\documentclass[fleqn,usenatbib]{mnras}

\usepackage[T1]{fontenc}
\usepackage{ae,aecompl}

\usepackage{graphicx}	
\usepackage{epstopdf}	
\usepackage{amsmath}	
\usepackage{amssymb}	

\usepackage{threeparttable}
\usepackage{placeins}
\usepackage{longtable}
\usepackage{lscape}

\title[Properties of supernova remnants in SIGNALS]{Properties of supernova remnants in SIGNALS galaxies - I . NGC 6822 and M33}

\author[S. Duarte Puertas et al.]
{Salvador Duarte Puertas$^{1,2,3,4}$,\thanks{E-mail: sduarte@ugr.es}
Laurent Drissen$^{3,4}$, Carmelle Robert$^{3,4}$, Laurie Rousseau-Nepton$^{5,6,7}$, \newauthor
 R. Pierre Martin$^8$, Philippe Amram$^9$, Thomas Martin$^{3,10}$\\ 
$^{1}$Departamento de F\'isica Te\'orica y del Cosmos, Universidad de Granada, 18071 Granada, Spain\\
$^{2}$Instituto Carlos I de Física Teórica y Computacional, Universidad de Granada, Granada, Spain\\
$^{3}$D\'epartement de physique, de g\'enie physique et d'optique, Universit\'e Laval, Qu\'ebec (QC), G1V 0A6, Canada\\
$^{4}$Centre de recherche en astrophysique du Qu\'ebec\\
$^{5}$Dunlap Institute of Astronomy and Astrophysics, University of Toronto, 50 St. George St, Toronto, ON, M5S 3H4, Canada\\
$^{6}$Canada-France-Hawaii Telescope, 65-1238 Mamalahoa Hwy, Kamuela, Hawaii 96743, USA\\
$^{7}$Department of Astronomy \& Astrophysics at the University of Toronto, 50 St. George St, Toronto, ON, M5S 3H4, Canada\\
$^{8}$Department of Physics and Astronomy, University of Hawai'i at Hilo, Hilo, HI 96720, USA\\
$^{9}$Aix Marseille Univ, CNRS, CNES, LAM, Marseille, France\\
$^{10}$C\'egep Garneau, 1660 Boulevard de l'Entente, Qu\'ebec, QC G1S 4S3, Canada\\
}
\date{Accepted XXX. Received YYY; in original form ZZZ}

\pubyear{2024}

\begin{document}
\label{firstpage}
\pagerange{\pageref{firstpage}--\pageref{lastpage}}
\maketitle

\begin{abstract}
We present a spatially resolved study of the kinematical properties of known supernova remnants (SNRs) in the nearest galaxies of the SIGNALS survey, namely NGC 6822 (one object) and M33 (163 objects), based on data obtained with the SITELLE Imaging Fourier Transform Spectrometer (iFTS) at the Canada-France-Hawaii Telescope. The purpose of this paper is to provide a better scheme of identification for extragalactic SNRs and, in particular, to distinguish between HII regions and SNRs. For that we have used diagrams which involve both the [SII]/H$\alpha$ ratio and the velocity dispersion ($\sigma$). We also introduce a new parameter, $\xi = {[SII] \over H\alpha} \times \sigma $, which enhances still the contrast between SNRs and the rest of the ionised gas. More than 90\% of the SNRs in our entire sample show an integrated [SII]/H$\alpha$ ratio larger than the canonical value (0.4). 86\% of the SNRs present in our field show a significant velocity dispersion. The spectral resolution of our observations allows us to observe the complex velocity structure of some SNRs.
\end{abstract}

\begin{keywords}
galaxies: individual: M33, NGC 6822 -- galaxies: ISM -- ISM: supernova remnants -- HII regions 
\end{keywords}



\section{Introduction}

A core-collapse supernova (SN) is the most dramatic stage in the life of a massive ($>$ 8 M$_\odot$) star, where it cannot maintain gravitational equilibrium and its core collapses, bounces back, thus generating a shock wave which, under favorable conditions \citep{2022MNRAS.515.1610G}, leads to the fast ejection of a large fraction of the stellar envelope. The result of the interaction between this rapidly ejected material and the surrounding circumstellar (in the very early stages) and interstellar medium (ISM), known as a supernova remnant (SNR), brightly shines for thousands of years with distinct signatures across the electromagnetic spectrum \citep{2017hsn..book.2005L}. SNRs are thus, in principle at least, relatively easy to identify in the Local Universe.  Because both Type Ia and core-collapse supernovae play a key role in the chemical evolution of galaxies and the dynamics of the interstellar medium, studying the properties of SNRs provides an essential element to our understanding of the global ISM in nearby galaxies.

In the visible part of the spectrum, the [SII]/H$\alpha$ ratio is traditionally used to distinguish between SNRs and HII regions: HII regions typically have values of [SII]/H$\alpha \sim$ 0.1 \citep[e.g.,][]{1976A&A....53..443D, 2010ApJS..187..495L} while SNRs have a value of at least 0.4. The diffuse ionized medium (DIG) however also presents high values of this ratio, complicating the unambiguous identification of SNRs. In addition to the traditional one, there are other non-empirical identification methods using different emission line ratios \citep[e.g.][]{2020MNRAS.491..889K,2023A&A...672A.148C}. \citet{2020MNRAS.491..889K} point out that a large number of shock models have [SII]/H$\alpha$ ratios lower than 0.4, and in particular that the use of 0.4 as a strict cutoff disfavours the identification of SNRs with slow shock velocities \citep[see however][]{2023ApJ...943...15W}.
 
\cite{2018ApJ...855..140L} noticed that a substantial number of SNR candidates in M33 show larger emission line widths than HII regions and suggests that velocity broadening could be a potential discriminant to appropriately select SNRs. Following this finding, \citet{2019ApJ...887...66P} proposed to use measurements of the kinematics of extragalactic SNR candidates in addition to the widely used [SII]/H$\alpha$ ratio as a means to ``purify" the surveys. They illustrate this by using long-slit spectra of well-known SNRs in the Large Magellanic Clouds and a sample in M83. 

In the present work, we follow in their footsteps by studying known SNRs in two Local Group galaxies, NGC 6822 \citep{2019AJ....157...50D} and M33 \citep{2010ApJS..187..495L,2014ApJ...793..134L}, using datacubes obtained with the imaging Fourier transform spectrometer (iFTS) SITELLE at the Canada-France-Hawaii telescope (CFHT). These data are part of the SIGNALS survey \citep{2019MNRAS.489.5530R}, one of CFHT's Large Programs, which aims at studying massive star formation and HII regions in a sample of local (distance $\leq$ 10 Mpc) extended galaxies by measuring bright line ratios and kinematics.

While most previous kinematics studies of SNRs in M33 are based on long-slit and fibre spectroscopy and thus do not sample the entire area of the objects, SITELLE \citep{2019MNRAS.485.3930D} allows us to perform a spatially resolved study of all the targets alongside that of their surrounding HII regions and DIG, thanks to its wide field-of-view (11$^\prime \times 11^\prime$). Moreover, in the H$\alpha$ spectral region\footnote{In the case of nearby galaxies, such as the complete sample of galaxies in the SIGNALS project, the H$\alpha$ spectral region refers to the SN3 bandpass.}, SIGNALS' nominal resolution of R = 5000 (see below) is very close to that suggested by \citet{2019ApJ...887...66P} for the purpose of discriminating SNRs from HII regions. 

The vast majority of SIGNALS target lie beyond the Local Group (median distance = 4 Mpc), but selected fields in nearby galaxies (M33, M31, NGC 6822, WLM, Sextans A and IC 1613) have also been targeted in order to serve as templates for the rest of the sample. Although the focus of SIGNALS is on HII regions and massive star formation, the sample is also very well suited for the study of SNRs. The initial goal of the present work was to combine the standard [SII]/H$\alpha$ criterion with the 2D and integrated velocity dispersion ($\sigma$) information in order to provide a framework allowing a clear and unambiguous identification of SNRs in the Local Group that could be used for more distant gas-rich galaxies. This is thus the first in a series of papers dedicated to this topic.

The structure of this paper is organised as follows: in Sect.~\ref{sec:2_observations} we describe the observations, the data reduction and the analysis of the data. In Sect.~\ref{sec:3_ngc6822} we present the morphological and kinematical results of Ho 12, the SNR in NGC 6822, and we show the scheme of identification proposed in this work. The application of this identification scheme to a sample of SNR candidates from M33 can be seen in Sect.~\ref{sec:4_m33}; Finally, in Sect.~\ref{sec:5_conclu} we present our conclusions. 

\section{Observations, data reduction and analysis}
\label{sec:2_observations}

\begin{table*}
        \centering
        \caption{Characteristics of the datacubes.}
        \label{tab:fields}
        \begin{tabular}{lccccccl}
                \hline
                Field &  RA & Dec & Filter & R& Exposure/step & Num. Steps& Date\\
                \hline
                NGC 6822 & 19:44:47.6 & -14:45:22 &SN3 & 3300& 21s & 590 & July 5, 2017 \\
                  &   &  &SN2 & 560 & 82s & 132 & June 30, 2017 \\
                  &   &  &SN1 & 400 & 139s & 69 & July 9, 2016 \\
                M33-F1 & 01:34:24.1 & +30:44:56 &SN3 & 2900 & 18s & 505 & October 12, 2017 \\
                 &  & &SN2 & 1020 & 38s & 219 & September 28, 2017 \\
                 &  &  &SN1 & 1020 & 49s & 172 & September 28, 2017 \\ 
                M33-F2 & 01:33:36.2 & +30:42:57 &SN3 & 2900 & 18s & 505 & October 13, 2017 \\
                 &  & &SN2 & 1020 & 38s & 219 & September 28, 2017 \\
                 &  &  &SN1 & 1020 & 49s & 172 & September 28, 2017 \\
                M33-F3 & 01:34:24.1 & +30:34:33 &SN3 & 2900 & 18s & 505 & October 17, 2017 \\
                 &  & &SN2 & 1020 & 38s & 219 & September 28, 2017 \\
                 &  &  &SN1 & 1020 & 49s & 172 & September 28, 2017 \\
                M33-F4 & 01:33:33 & +30:32:55 &SN3 & 2200 & 18s & 403 & October 13, 2017 \\
                 &  & &SN2 & 1020 & 38s & 219 & September 28, 2017 \\
                 &  &  &SN1 & 1020 & 49s & 172 & September 28, 2017 \\
                M33-F5 & 01:34:00.1 & +30:53:35 &SN3 & 5000 & 13s & 842 & October 2, 2021 \\
                 &  & &SN2 & 1020 & 46s & 219 & October 4, 2021 \\
                 &  &  &SN1 & 1020 & 59s & 171 & October 6, 2021 \\
                M33-F6 & 01:33:19.1 & +30:51:51 &SN3 & 5000 & 13s & 842 & September 30, 2021 \\
                 &  & &SN2 & 1020 & 46s& 219 & November 5, 2021 \\
                 &  &  &SN1 & 1020 & 59s & 171 & November 10, 2021 \\
                M33-F7 & 01:32:50.7 & +30:34:50 &SN3 & 5000 & 13s & 842 & October 4, 2018 \\
                 &  & &SN2 & 1020 & 46s & 220 & October 11, 2018 \\
                 &  &  &SN1 & 1000 & 59s & 171 & October 12, 2018 \\   
                M33-F8 & 01:32:51.1 & +30:22:34 &SN3 & 5000 & 13s & 842 & October 15, 2020 \\
                 &  & &SN2 & 1020 & 46s & 219 & December 8, 2020 \\
                 &01:32:47.1  &   +30:22:44&SN1 & 1000 & 59s & 171 & 17 November 2020 \\  
                M33-F9 & 01:33:41.6 & +30:18:47 &SN3 & 5000 & 13s & 842 & September 29, 2019 \\   
                 & 01:33:42.6 & +30:20:57&SN2 & 1020 & 46s & 220 & December 11, 2020 \\
                 & 01:33:38.6 & +30:21:08  &SN1 & 1000 & 59s & 171 & November 14, 2020 \\                            
                \hline
        \end{tabular}
\end{table*}

SITELLE provides the spatially resolved spectrum of the sources in an $11' \times 11'$ field of view with a sampling of $0.32''$/pixel, in selected bandpasses of the visible range, with a spectral resolution adapted to the need of the observer. Three bandpasses, centered on the HII regions' bright lines, are used in SIGNALS: SN3 (647 - 685 nm, R = 5000), SN2 (482 - 513 nm, R = 1000) and SN1 (363 - 386 nm, R = 1000). The SN3 datacubes (which include H$\alpha$, the [NII] $\lambda\lambda$ 6548,84 doublet and the [SII] $\lambda\lambda$ 6717,31 doublet) provide the required kinematics information as well as the [SII]/H$\alpha$ ratio traditionally used to identify SNRs. In this paper, we will also show images extracted from the SN1 and SN2 datacubes, although a detailed analysis of the line ratios will not be presented.

Half of the data cubes used in this paper (NGC 6822, M33 Fields 1 to 4) were observed as part of SIGNALS' proof of concept program; the spectral resolution and total exposure times of some of them therfore did not meet the SIGNALS standards (exemplified by Fields 5 to 9 in M33, see Table\,\ref{tab:fields}). Nevertheless, with a spectral resolution of R $\sim$3000 in the H$\alpha$ spectral region (SN3 filter), these data are of sufficient quality to start our investigation of SNRs in Local Group galaxies.

The raw data, consisting of two complementary 2D interferograms \citep[acquired using the two CCDs on the receiving end of the Michelson; see][]{2019MNRAS.485.3930D} were transformed into a single fully calibrated spectral datacube with ORBS, SITELLE's dedicated data reduction software \citep{2015ASPC..495..327M}. Photometric calibration is secured from images and data cubes of spectrophotometric standard stars such as G93-48, P330E and HZ21. Wavelength calibration is performed using a high spectral resolution laser data cube. It may show distortions due to aberrations and deformations in the optical structure, leading to zero-velocity offsets of $\sim 15 - 20$ km/s across the field of view. We thus refine the wavelength calibration by measuring the centroid positions of the night sky OH emission lines (in the SN3 filter) in the science data cubes following the procedure described in \citet{2018MNRAS.473.4130M}, reducing the uncertainty on the velocity calibration across the entire field to less than 2 km/s. We also applied a barycentric correction to the observed spectra.

\begin{figure}
	\includegraphics[width=\columnwidth]{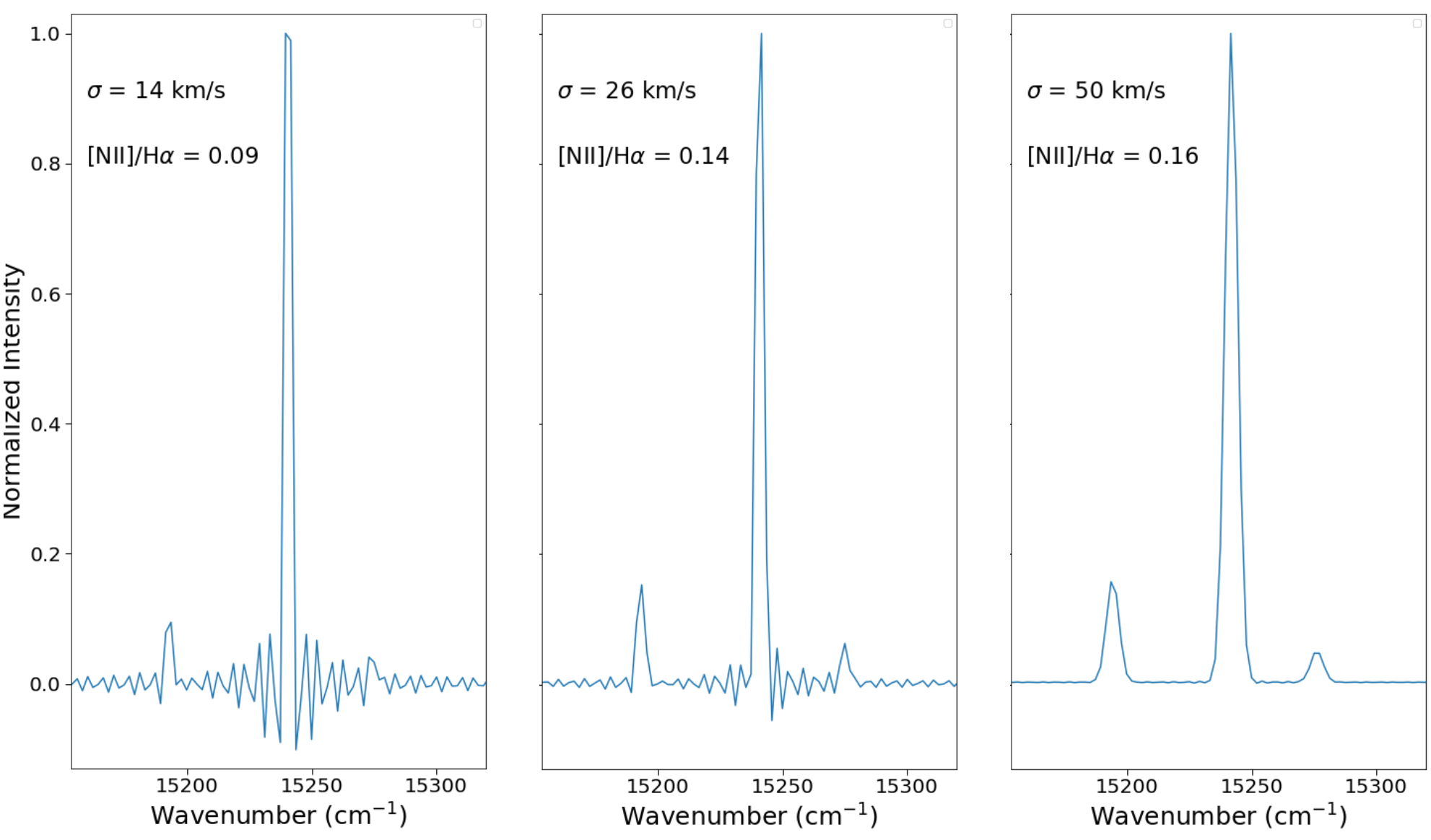}
    \caption{Normalized ORCS fits to the integrated spectrum of three regions in M33, showing the effects of the line broadening on a datacube with R = 5000 spectral resolution. See text for details.}
    \label{fig:lineshape}
\end{figure}

Spectral data from SITELLE (Fig. \ref{fig:lineshape}) can be a bit confusing for first-time users, for two reasons. First, the natural units of an FTS are wavenumbers (cm$^{-1}$): in the SN3 datacube, the [SII] doublet is therefore to the left of H$\alpha$. Secondly, each line is surrounded by oscillating sidelobes, alternating between positive and negative values.This is because, as described by \citet{2016MNRAS.463.4223M}, SITELLE's Instrument Line Shape (ILS) is a sinc function. However, any line broadening caused by turbulent motion, an expanding bubble or a velocity gradient along an extended aperture will transform the natural ILS into a so-called sincgauss function, the convolution of a sinc and a gaussian. ORCS, SITELLE's analysis software suite \citep{2020ASPC..522...41M}, is thus used to fit the observed line profiles for each pixel of the field of view (or an integrated region within it) and produce maps (or single values) of line intensities, radial velocity, and velocity dispersion (identified as $\sigma$) as well as their uncertainties. 

We note that since $\sigma$ characterizes the gaussian function which, convoluted with the ILS, reproduces the observed line shape, it represents the natural line broadening beyond the instrumental one. The minimum line broadening measurable with ORCS is a function of the cube's spectral resolution and the signal-to-noise ratio (S/N) of the data \citep[see Fig. 2 in][]{2016MNRAS.463.4223M}. With the current SN3 data cubes, this value is of the order of 10 - 15 km~s$^{-1}$. Figure \ref{fig:lineshape} shows fits to integrated (circles of 5 pixel radius) spectra of three regions in the M33-F9 field, illustrating the effect of line broadening on the spectral profile. On the left, a small ($\sim$ 25 pc diameter) HII region with a barely detectable broadening. Note the oscillating sidelobes of the iFTS sinc profile, which extend well beyond the [NII] doublet. The central spectrum is that of a part of a large HII complex with multiple expanding filaments along the line of sight, well discernable when moving through the spectral cube. With an easily measurable line broadening, we notice that the sidelobes display a significantly lower amplitude than the left-hand spectrum. On the right, a portion of one SNR in M33; the width of the central peak in indeed about twice that of the instrumental profile, and the sinc sidelobes are barely detectable between H$\alpha$ and the [NII] lines. These spectra are fits to real data, and are therefore noiseless and thus amenable to a pedagogical demonstration. But, as we shall see throughout this paper, actual data also clearly show this behaviour and ORCS's ability to fit them and extract the required properties of the source.

ORCS simultaneously fits all the lines within a given data cube, but allows the user to either force identical velocities and velocity dispersion for all the lines, to let them completely independent of each other, or to select groups of lines with the same values. The maps and integrated spectra presented in this paper were obtained with the first option, but we also looked at the impact of allowing the values of velocity dispersion for the [SII] doublet and H$\alpha$ to differ from each other (see section \ref{sec:4_2_nearby_str}).

To produce the 2D maps, we chose to subtract a common background to all pixels in a given cube, selected from a region without any obvious nebular contribution. For the integrated spectrum of individual objects, a nearby background devoid of bright HII regions was selected. This choice is also discussed in section \ref{sec:4_2_nearby_str}.

\begin{figure*} 
	\includegraphics[width=7.0truein]{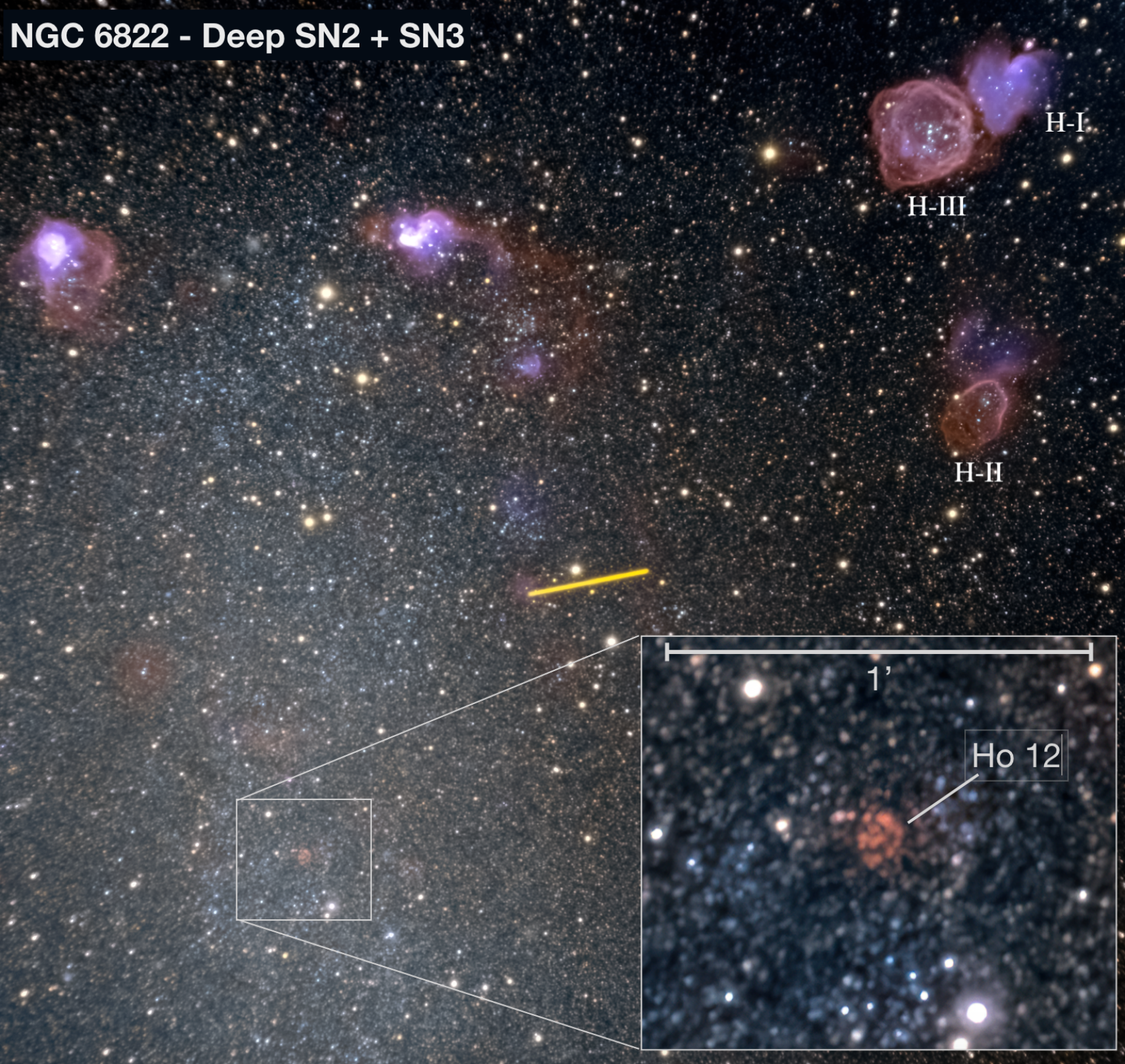}
    \caption{Composite image ($10' \times 10'$, with North up and East to the left) of NGC 6822 from the SN2 and SN3 deep frames (see text). HII regions Hubble-I, II and III (discussed later) are identified. The insert shows the location of Ho 12. Asteroid (654) Zelinda is seen crossing the field of view near the center of the image.}
    \label{fig:N6822all}
\end{figure*}

\section{Ho 12 in NGC 6822}
\label{sec:3_ngc6822}
Barnard's galaxy (NGC 6822) is the nearest target of the SIGNALS sample \citep[$\sim$0.52 Mpc, see Table A.1 in][]{2019MNRAS.489.5530R} and as such offers an excellent opportunity to study its nebular content with exquisite details. At the distance of NGC 6822, 1$'' \sim$ 2.5 pc. The oxygen abundance of NGC 6822 is 12+log(O/H) = 8.14 $\pm$ 0.08 \citep{2016ApJ...830...64B}. Only one SNR has been identified in this dwarf irregular galaxy, and it was first catalogued as an HII region (No. 12, hereafter Ho 12) by \citet{1969ApJS...18...73H}. Detailed analyses of this object have been performed by \citet{2004AJ....128.2783K} using X-ray and radio data as well as narrow-band visible images, and, more recently, by \citet{2019AJ....157...50D} using the integral field spectrograph WiFeS at the ANU 2.3 m telescope. The later concentrated on the analysis of Ho 12's line ratios but did not provide information on its kinematics.

\begin{figure}
	\includegraphics[width=\columnwidth]{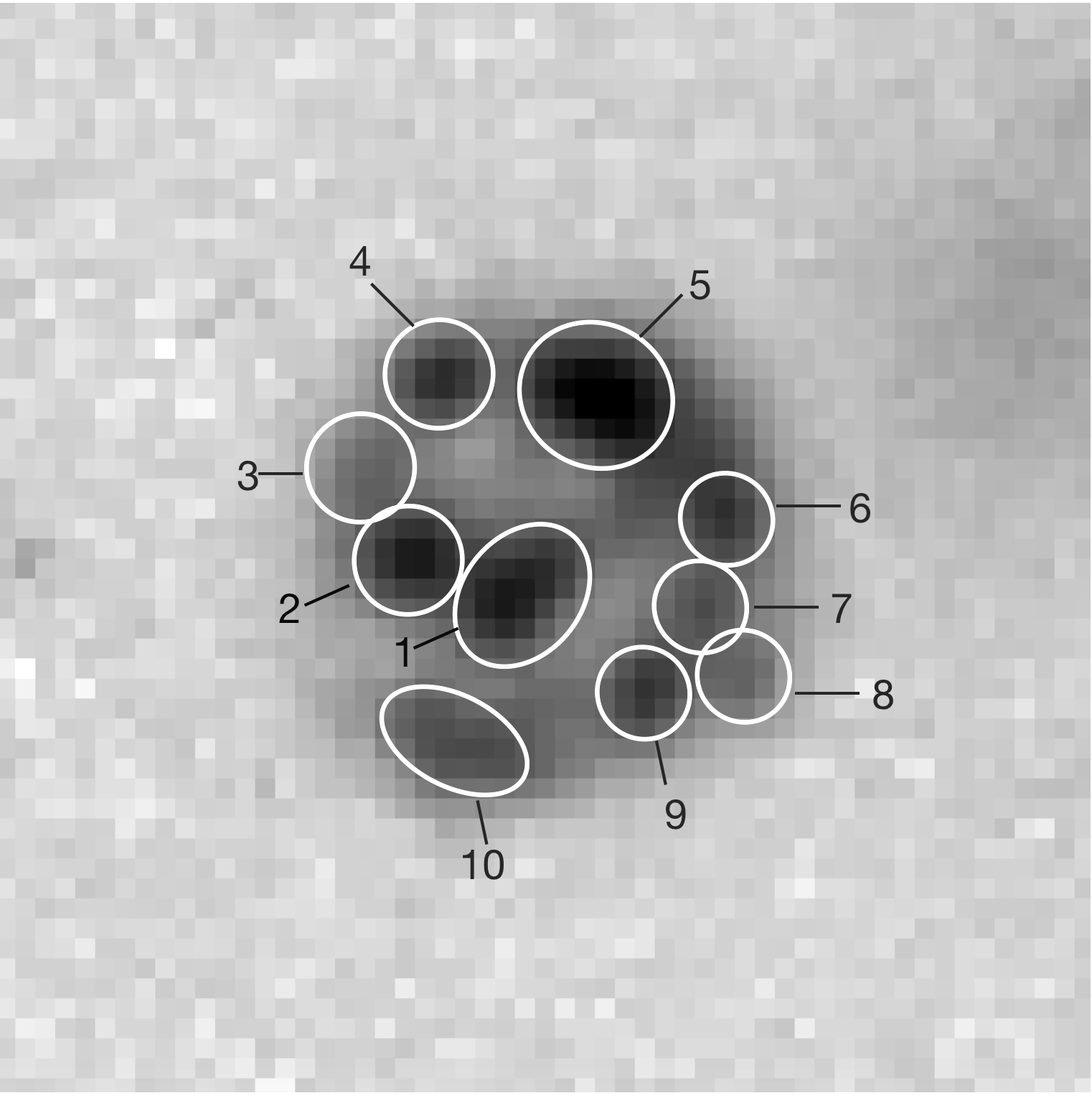}
        \includegraphics[width=\columnwidth]{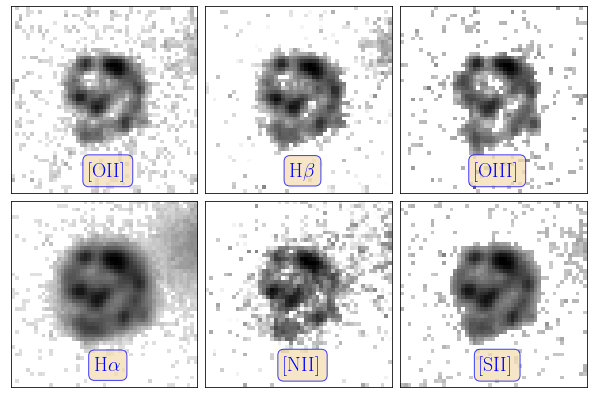}
    \caption{(upper panel) H$\alpha$ image of Ho 12 (17.3$''$, or 43 pc, on a side), with the knots labelled. (Lower panel) Emission line maps: [OII]$\lambda$3727 from the SN1 cube; H$\beta$ and [OIII]$\lambda$5007 from the SN2 cube; H$\alpha$, [NII]$\lambda$6584 and [SII]$\lambda\lambda$6717,6731 from the SN3 cube. North is at top, East to the left.}
    \label{fig:Ho12-knots}
\end{figure}

\begin{figure}
	\includegraphics[width=\columnwidth]{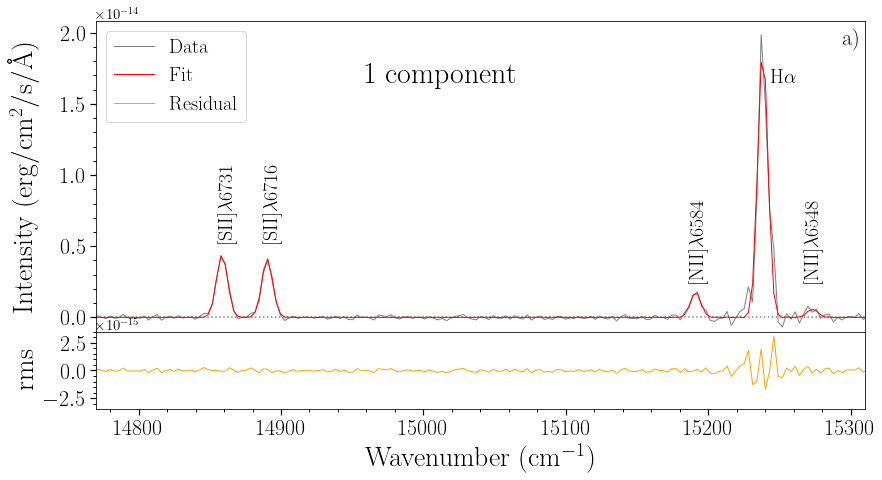}
	\includegraphics[width=\columnwidth]{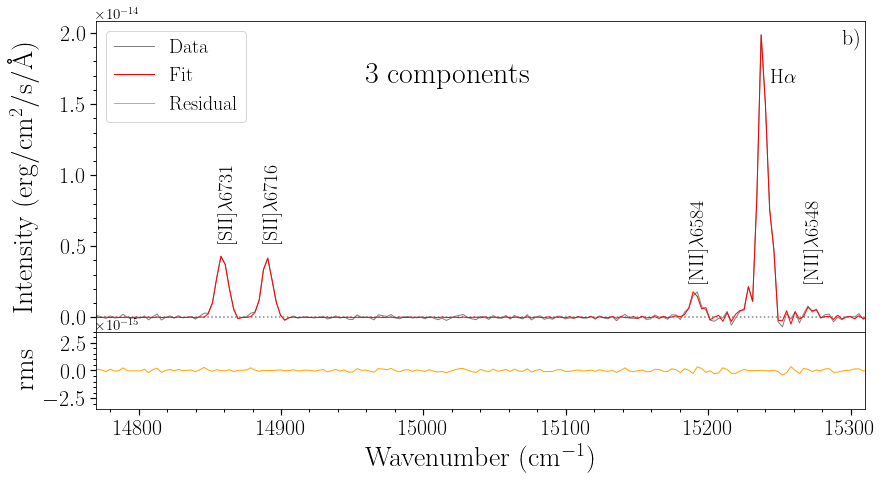}
    \includegraphics[width=\columnwidth]{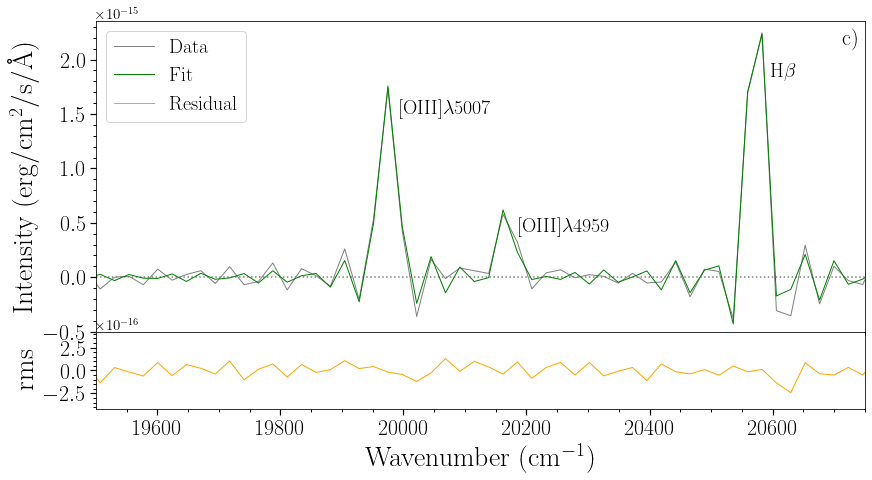}
    \includegraphics[width=\columnwidth]{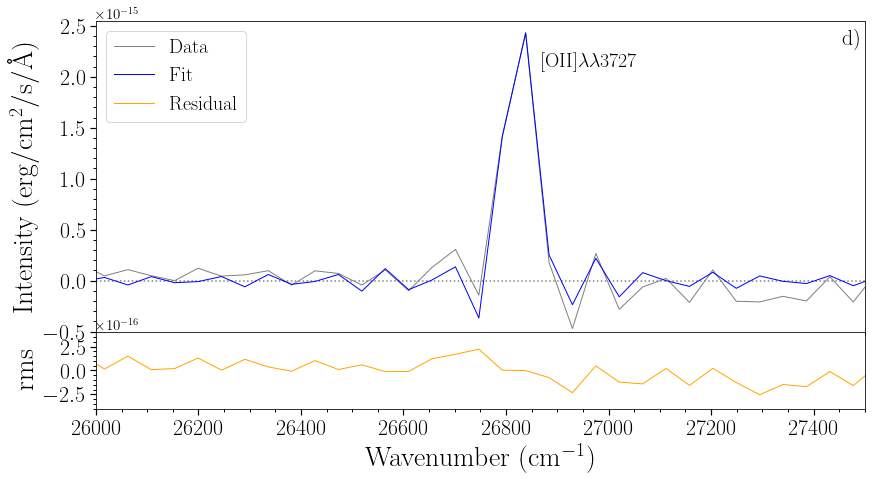}
    \caption{Integrated spectrum (r = 4.2$''$ = 10.5 pc) of NGC 6822-Ho 12, with fits using one velocity component (panel a) and three velocity components (panel b). Green and blue coloured lines show the fit obtained in the SN2 (panel c) and SN1 (panel d) filters, respectively; residuals from the fit are also shown in the lower panels.}
    \label{fig:N6822snr-fits}
\end{figure}

\subsection{Morphology and integrated spectrum}
\label{sec:3_1_ngc6822_morpho}
Figure \ref{fig:N6822all} presents Ho 12 in the context of SITELLE's data. While Fourier transforms of the interferograms collected by SITELLE lead to the spectral content of the source, adding all the individual frames from these interferograms provides a deep image in the bandpass of the filter. Figure \ref{fig:N6822all} is a color-coded rendering of the SN2 (blue) and SN3 (orange) deep images of the entire field, along with an insert enlarging the region around Ho 12 to emphasize the spatial resolution of the data. The yellow stripe in the center of this image is not an artefact, but rather the asteroid (654) Zelinda crossing the field of view. 

Composed of a series of knots seemingly aligned in a spiral shape, bathing in a more diffuse component, all within a diameter of $\sim$21 pc (Fig. \ref{fig:Ho12-knots}), Ho 12 shows a very peculiar morphology at all wavelengths. In the upper panel of Fig. \ref{fig:Ho12-knots}, we have identified 10 substructures for future reference. Knots 1, 5 and 10 are clearly elongated and could be further subdivided with higher resolution data. The H$\alpha$ flux of these knots, their average heliocentric velocity and velocity dispersion are presented in Table \ref{tab:knots}. The lower panel of fig. \ref{fig:Ho12-knots} presents Ho 12 in different emission lines: [OII]$\lambda$3727 from the SN1 cube; H$\beta$ and [OIII]$\lambda$5007 from the SN2 cube; H$\alpha$, [NII]$\lambda$6584, and [SII]$\lambda\lambda$6717,6731 from the SN3 cube. These images can be compared with fig. 3 from \citet{2019AJ....157...50D} and fig. 1 from \citet{2004AJ....128.2783K}. Images at [NII] and [OII] have, to our knowledge, never been published. As with all nebulae in NGC 6822, the [NII] lines are weak, reflecting the low metallicity of this galaxy.

Ho 12's background-subtracted, integrated spectrum in the SN3, SN2, and SN1 filters, using a circular aperture with a radius of 13 pixels (4.2$''$, or 10.5 pc), is shown in Fig. \ref{fig:N6822snr-fits}. The upper panel of this figure also presents a fit, using ORCS, considering only a single velocity component. Although the overall fit is relatively good, a close inspection of the H$\alpha$ line and its residuals reveals that the fit missed a weak red component, did not reproduce the peak of the line, as well as the blue wing. We therefore included three components in the fit, the result of which is shown in the lower panel of the same figure. The flux, central velocities and broadening of these individual components are:

\noindent F$_1$(H$\alpha$) = $5.77 \pm 0.31 \times 10^{-14}$ erg~cm$^{-2}$~s$^{-1}$, V$_1 = -63 \pm 2$ km~s$^{-1}$, $\sigma_1 = 58 \pm 3$ km~s$^{-1}$

\noindent F$_2$(H$\alpha$) = $0.89 \pm 0.15 \times 10^{-14}$ erg~cm$^{-2}$~s$^{-1}$, V$_2 = -194 \pm 2$ km~s$^{-1}$, $\sigma_2 = 24 \pm 8$ km~s$^{-1}$

\noindent F$_3$(H$\alpha$) = $0.46 \pm 0.12 \times 10^{-14}$ erg~cm$^{-2}$~s$^{-1}$, V$_3 = +138 \pm 2$ km~s$^{-1}$, $\sigma_3 = 59 \pm 14$ km~s$^{-1}$

The total flux, F$_{tot}$(H$\alpha$) = $7.12 \pm 0.36 \times 10^{-14}$ erg~cm$^{-2}$~s$^{-1}$, is in excellent agreement with the value  of $6.9 \times 10^{-14}$ erg~cm$^{-2}$~s$^{-1}$ (no uncertainty provided) from \citet{2004AJ....128.2783K}, obtained with a similar aperture in narrow-band images from the Local Group Survey \citep{2006AJ....131.2478M}. The global line ratios are [NII]/H$\alpha$ = 0.096 and [SII]/H$\alpha$ = 0.52, virtually identical to those obtained by \citet{2019AJ....157...50D}. This fit ensures a precise measurement of the line fluxes, but the real velocity structure of Ho 12 is much more complex than a three-component model would suggest, however good the fit might be, as we shall see in the next section.

\begin{table}
        \caption{Properties of the knots in Ho 12. Values are obtained within the limits of the contours shown in Fig.~\ref{fig:Ho12-knots}.}
        \label{tab:knots}
        \begin{tabular}{cccc}
                \hline
                Knot &  F(H$\alpha$) & V$_{helio}$ & $\sigma$ \\
                     &  10$^{-15}$erg~cm$^-2$~s$^{-1}$ & km/s & km/s \\
                \hline
                1 & 6.1 & -160 & 71 \\
                2 & 4.1 & -115 & 73 \\
                3 & 1.3 & -111 & 83 \\
                4 & 2.7 & -44 & 58 \\
                5 & 9.6 & -57 & 65 \\
                6 & 2.1 & -60 & 57 \\
                7 & 1.4 & -70 & 45 \\
                8 & 0.8 & -57 & 60 \\
                9 & 1.8 & -63 & 45 \\
                10 & 3.4 & -36 & 70 \\
                \hline
        \end{tabular}
\end{table}

\subsection{Kinematics}
\label{sec:3_2_ngc6822_kine}

\begin{figure}
	\includegraphics[width=\columnwidth]{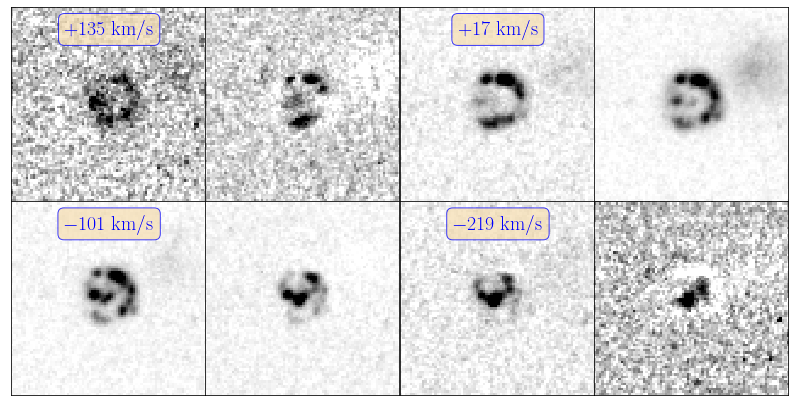}
    \caption{Consecutive frames centered on the H$\alpha$ line (from the SN3 data cube) showing the velocity structure of Ho 12.}
    \label{fig:N6822snr-frames}
\end{figure}

\begin{figure*}
	\includegraphics[width=0.44\textwidth]{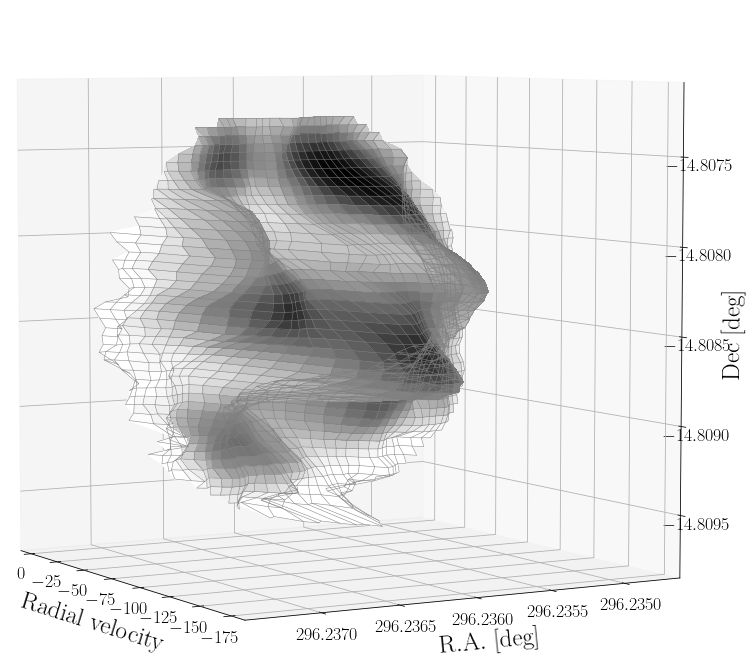}
	\includegraphics[width=0.51\textwidth]{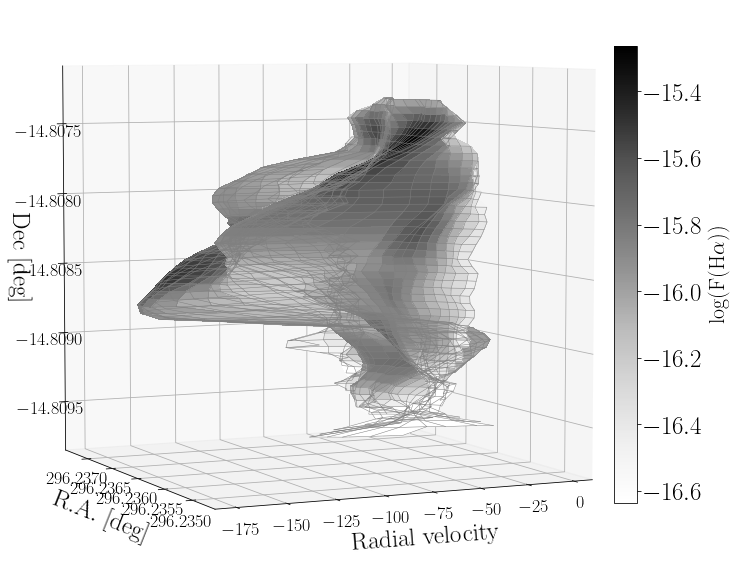}
    \caption{Three-dimensional view of Ho 12 (R.A., Dec, and radial velocity diagram). The greyscale indicates the H$\alpha$ flux.}
    \label{fig:ho12_ha_3d}
\end{figure*}

\begin{figure}
	\includegraphics[width=\columnwidth]{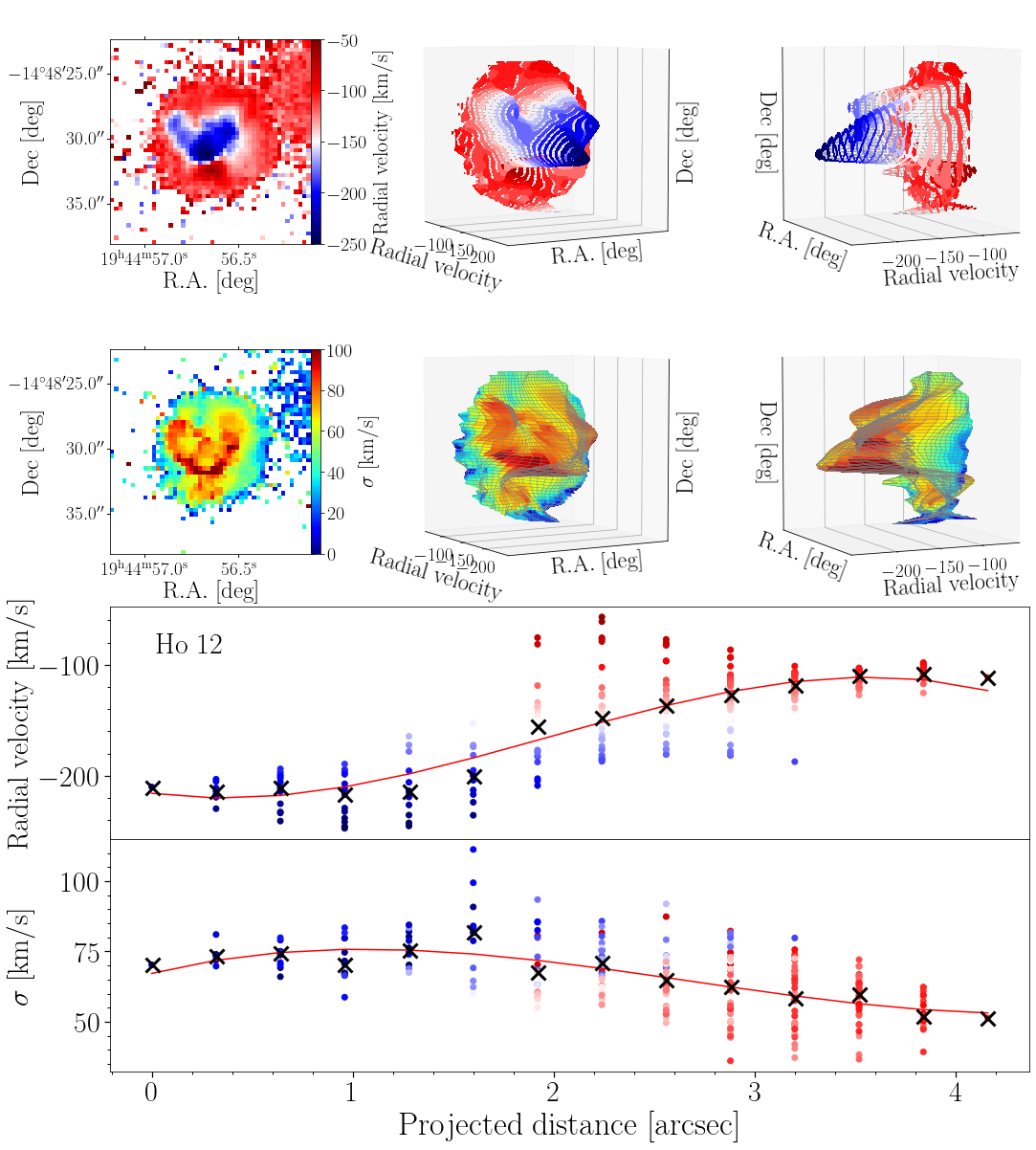}
    \caption{Map of radial velocity and three-dimensional view (R.A., Dec, and radial velocity)of Ho 12  color-coded by the radial velocity (upper panels), map of $\sigma$ and three dimensional view color-coded by the $\sigma$ (middle panels), and its radial profiles along Ho 12 (lower panels). Individual pixels in the radial profiles are color-coded according to their radial velocity. Black crosses and red solid lines represent the medians at a given projected distance and a polynomial fit to the data, respectively.}
    \label{fig:N6822snr-velo}
\end{figure}

Figure \ref{fig:N6822snr-frames} presents the individual frames within the SN3 datacube, centered on H$\alpha$, spanning nearly 400 km/s, showing the complex velocity structure of Ho 12. This is emphasized in the three-dimensional view (R.A., Dec, and radial velocity) shown in Fig.~\ref{fig:ho12_ha_3d} as well as the upper left panel of Fig.~\ref{fig:N6822snr-velo}, obtained by fitting a single component to the spectra of individual pixels. We did not attempt to fit two components (which would correspond to the receding and approaching caps of the expanding supernova remnant) because the spectral resolution was insufficient to clearly distinguish two peaks in the individual spectra; therefore the resulting velocity is the flux-averaged one along the line of sight, most likely favoring the blueshifted component. Indeed, while the average heliocentric velocity of HII regions in the surroundings of Ho 12 is $\sim -50$ km/s, that of Ho 12 is $\sim -75$ km/s. The brightest component of knot 1 shows the most blueshifted velocity (with an average of $\sim$-175 km/s) while the most redshifted region ($\sim$-5 km/s) corresponds to the diffuse zone between knots 9 and 10.

Figure \ref{fig:N6822snr-velo} shows the maps and three-dimensional views of the radial velocity (upper panel) and velocity dispersion ($\sigma$, middle panel), and their radial profiles along Ho 12. The outer, diffuse, component reaches velocities ($\sim -50$ km/s) virtually identical to those of neighboring HII regions. As expected for an expanding shell, the velocity dispersion decreases towards the outskirts of the supernova remnant\footnote{Geometrically, in the central part of an expanding shell, the distance between the H$\alpha$ emission line produced by the farthest and the nearest layers is maximum and the $\sigma$ will be maximum; on the contrary, as we move towards the outermost part of the SNR, the smaller the distance becomes the smaller the $\sigma$. This can be seen in Ho 12 (see Fig.~\ref{fig:N6822snr-velo}) and in the shell and compact SNRs in M33 (see Section \ref{sec:4_1_m33_indi}).}, but the complex knotty structure of Ho 12 is reflected in the broad range of values at a given distance. Not surprisingly, the largest velocity dispersion values (above 100 km/s, like nowhere else in the entire galaxy) are found in the diffuse regions between the central knots: their flux is not dominated by a single entity but reflects the kinematics of the gas from one side to the other along the line of sight. We come back to these properties of Ho 12 in comparison with the largest HII regions of NGC 6822 (Hubble I, II, and III) in Sect.~\ref{sec:3_4_ngc6822_iden}.

\begin{figure}
	\includegraphics[width=\columnwidth]{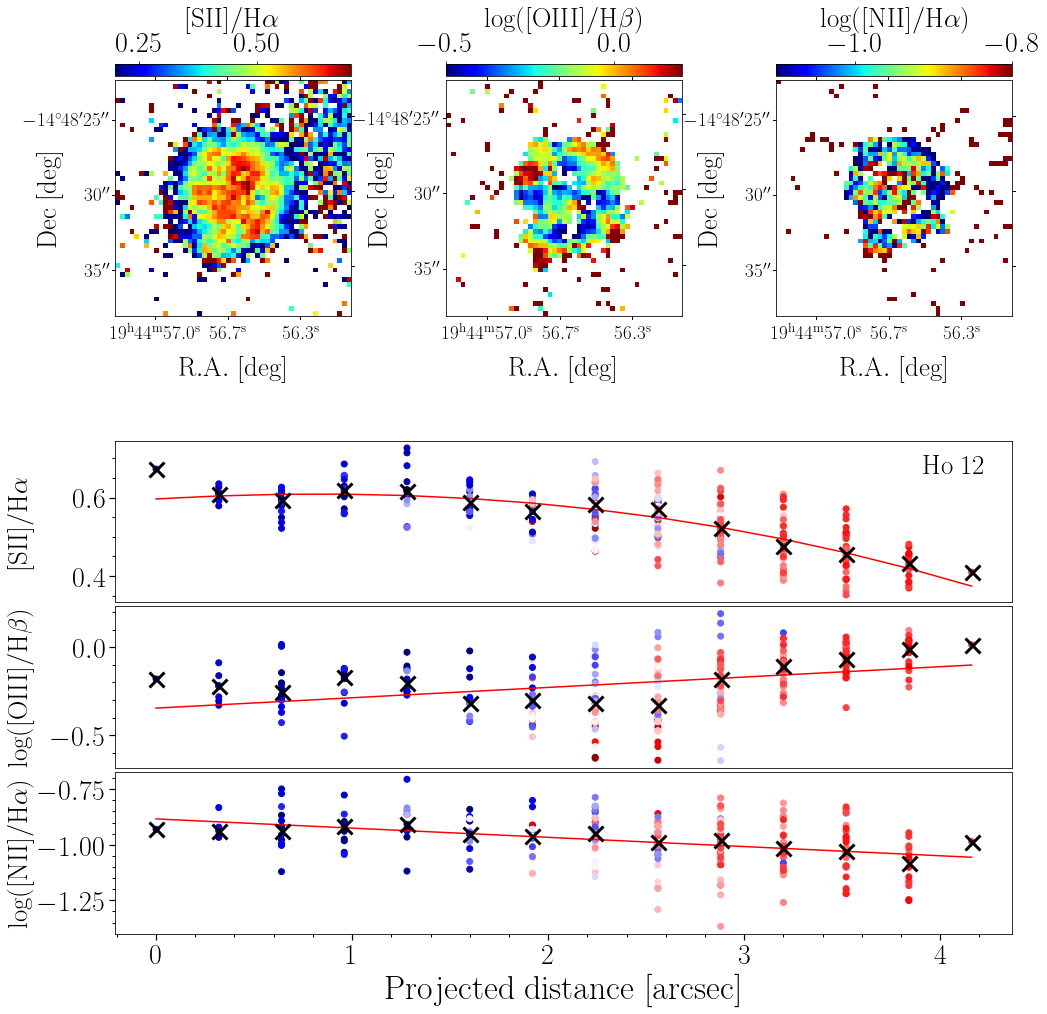}
    \caption{Maps of [SII]/H$\alpha$ (upper left panel), log([OIII]/H$\beta$) (upper middle panel), and log([NII]/H$\alpha$) (upper right panel) of Ho 12 and their corresponding radial profiles. All the pixels in the radial profiles are color-coded according to their radial velocity, from the reddest (+15 km/s) to the bluest color (-185 km/s). Black crosses and red solid lines represent the medians at a given projected distance and a polynomial fit to the data, respectively.}
    \label{fig:N6822snr-morpho}
\end{figure}

\subsection{Line ratios and global trends}
\label{sec:3_3_ngc6822_lineratio}

Figure~\ref{fig:N6822snr-morpho} shows the following line ratio maps of Ho 12: [SII]/H$\alpha$, log([OIII]/H$\beta$), and log([NII]/H$\alpha$) and their corresponding radial profiles. Throughout Ho 12, the vast majority of pixels ($>$ 95\%) show an [SII]/H$\alpha$ ratio larger than 0.4, where the highest values are reached in the inner part. The log([NII]/H$\alpha$) ratio also decreases outwards, with an average value of -1.01, typical for a low-metallicity galaxy. Adjacent individual knots often have very different line ratios, and the log([OIII]/H$\beta$) map is very patchy, which is reflected in its corresponding radial profile with no clear radial trend. A detailed study of individual knots in Ho 12, as well as that of the structure of M33 supernova remnants, is certainly warranted but beyond the scope of this paper.

\subsection{Simultaneous use of [SII]/H\texorpdfstring{$\alpha$}{alpha} and \texorpdfstring{$\sigma$}{sigma}}
\label{sec:3_4_ngc6822_iden}
Supernova remnants are not the only objects showing large values of [SII]/H$\alpha$ or velocity dispersion in galaxies. Outside of AGNs (which often display both concurrently), the central parts of giant HII regions show significant velocity dispersion due to the effects of stellar winds and supernovae \citep{2015MNRAS.451.3001T}, while their outskirts, like the diffuse ionised gas, show large values of [SII]/H$\alpha$ \citep{2020MNRAS.496..339P}.
The purpose of this paper being to provide a better scheme of identification for extragalactic SNRs and in particular to distinguish them from HII regions using both the [SII]/H$\alpha$ ratio and $\sigma$ from SITELLE data cubes, we will now make use of the product of these two properties, $\xi$ = ([SII]/H$\alpha) \times \sigma$, on a pixel-by-pixel basis and in the integrated spectrum \citep[see also][]{2023MNRAS.524.3623V}. We will also compare these properties in Ho 12 and in three large HII regions in NGC 6822 as a testbed for M33 and more distant galaxies. 

\begin{figure}
	\includegraphics[width=\columnwidth]{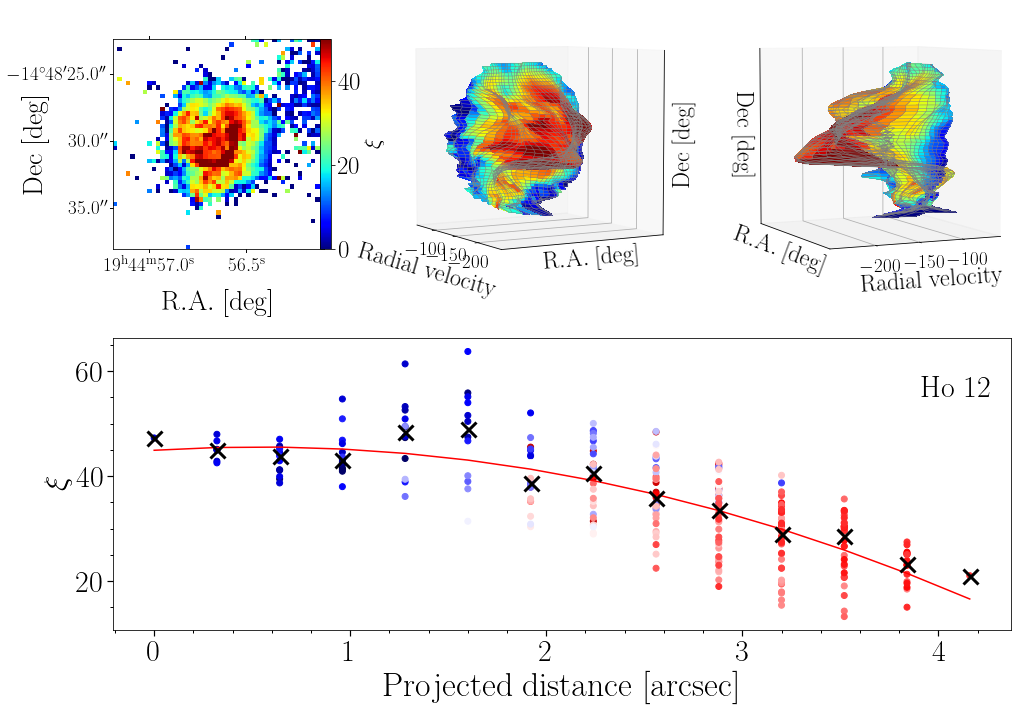}
 \caption{Map of $\xi$ and three-dimensional view (R.A., Dec, and radial velocity) of Ho 12  color-coded by $\xi$ (upper panels), and its radial profile along Ho 12 (lower panels). All the points in the radial profiles are color-coded by its radial velocity, red higher and blue lower. Black crosses and red solid lines represent the medians and polynomial fit of each property and the projected distance, respectively.}
    \label{fig:N6822snr-siihasig}
\end{figure}

\begin{figure}
	\includegraphics[width=\columnwidth]{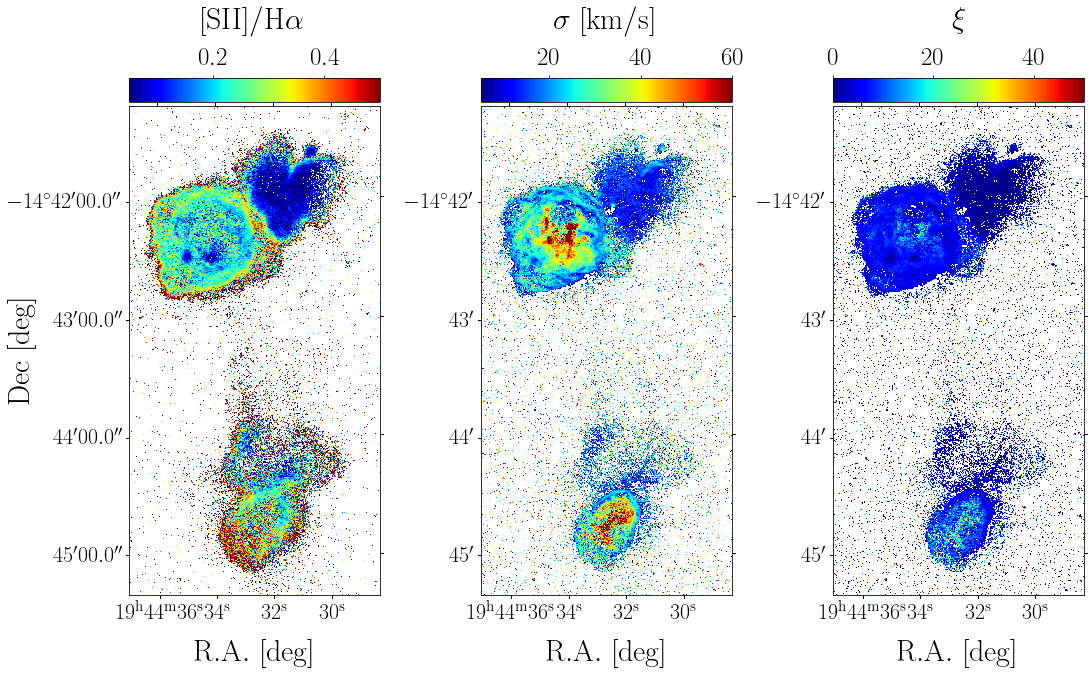}\\
	\includegraphics[width=\columnwidth]{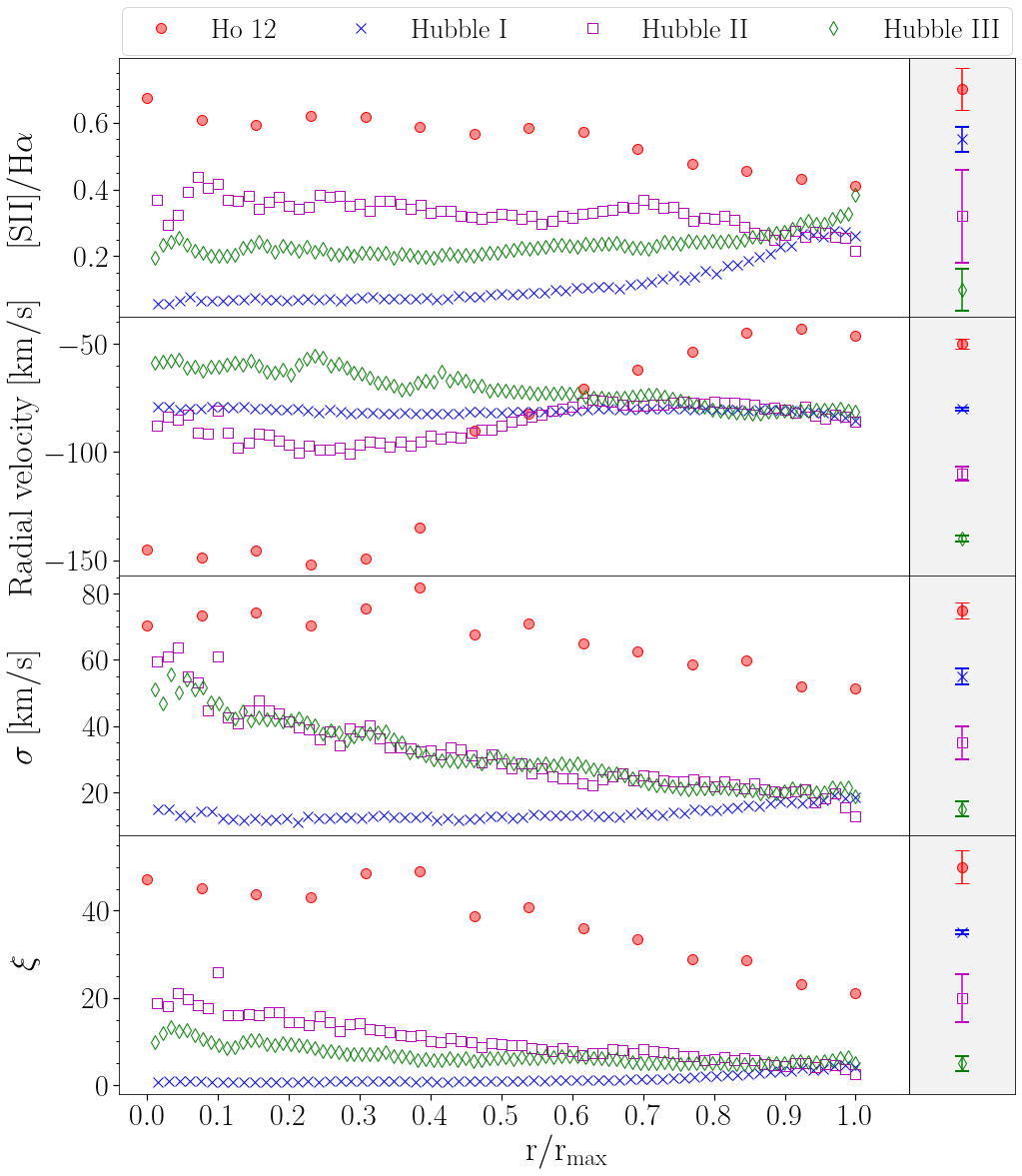}
    \caption{[SII]/H$\alpha$ ratio, $\sigma$, and $\xi$ maps of the large HII regions Hubble I (region at north-west), Hubble III (region at north-east) and Hubble II (region at south) in NGC 6822. The lower panels show, from top to bottom, the normalized radial profiles, r/r$\rm _{max}$, of [SII]/H$\alpha$, log([OIII]/H$\beta$), log([NII]/H$\alpha$), radial velocity, $\sigma$, and $\xi$. The median of each properties along the normalized radius in bins of 0.32 arcsec is shown for the SNR Ho 12 (red circles) and the HII regions Hubble I (blue crosses), Hubble II (magenta squares) and Hubble III (green diamonds). To derive the medians of each HII regions we considered all pixels with signal-to-noise ratio in [SII] and H$\alpha$ greater than 1, while for Ho 12 greater than 5 in both [SII] and H$\alpha$. Right panels: in the vertical grey band, the typical uncertainties of each parameter and region are represented.}
    \label{fig:N6822ghr}
\end{figure}

\begin{figure}
	\includegraphics[width=\columnwidth]{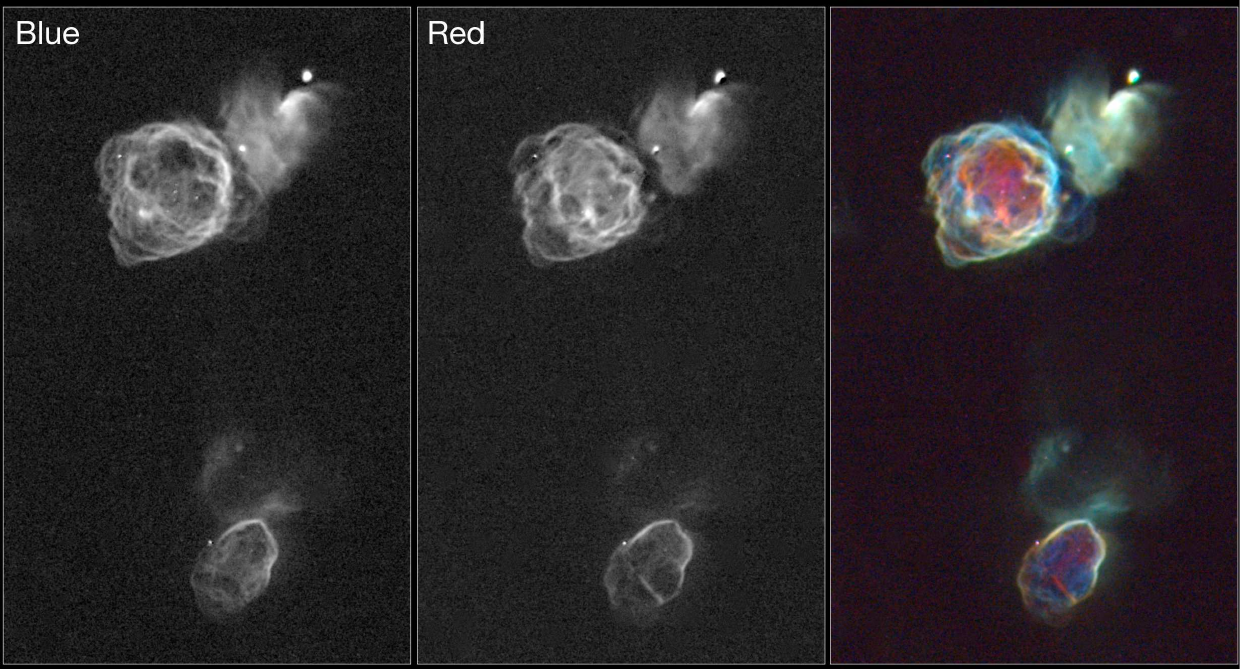}
    \caption{Blue and red velocity components for Hubble I, II and III extracted from the SN3 data cube and separated by 115 km/s. The right panel is a color composite, or Doppler image.}
    \label{fig:Hubble-Dop}
\end{figure}

In Fig.~\ref{fig:N6822snr-siihasig} we show the map and three-dimensional view of $\xi$, and its radial profile along Ho 12, color-coded by the radial velocity. As expected, the product of these two properties emphasizes the contrast between the inner and outer parts of the SNR. But its usefulness is event more obvious in Fig.~\ref{fig:N6822ghr}, where Ho 12 is compared with three large HII regions first identified by \cite{1925ApJ....62..409H}, i.e. Hubble I, II, and III. The upper panel of this figure presents the maps of the [SII]/H$\alpha$ ratio, the velocity dispersion, and the product of these two properties, while their radial profiles (along the normalized radius in bins of 0.32 arcsec) are shown in the lower panels. The right panels show the typical uncertainties of each parameter (i.e. [SII]/H$\alpha$, radial velocity, $\sigma$, and $\xi$) for each region. Hubble I, II, and III have larger uncertainties than Ho 12 because we have considered pixels with lower S/N ($\geq$1) than in the case of Ho 12 ($\geq$5), standing out in the case of the [SII]/H$\alpha$ parameter for the Hubble II region. However, the radial distributions shown for Hubble I, II and III evolve smoothly for each of the properties considered in Fig.~\ref{fig:N6822ghr}.

Let's first consider the [SII]/H$\alpha$ ratio: the behaviour of Hubble I and III is typical of photoionised regions, with values lower than the standard 0.4 threshold everywhere and increasing towards the outer edge. Ho 12's [SII]/H$\alpha$ ratio is larger than 0.4 throughout and decreasing outwards, which is expected from a typical SNR. But Hubble II is intriguing: its [SII]/H$\alpha$ ratio is very close to 0.4 in the inner regions and, contrary to the other HII regions, it decreases outwards. Hubble II also displays relatively high values of velocity dispersion (50 - 70 km/s) in its inner regions, like Hubble III. Going through the data cube indeed reveals both caps of their expanding shell, as shown in Fig.~\ref{fig:Hubble-Dop}. Hubble I is much more quiet in this respect and its [SII]/H$\alpha$ ratio is on average very low. So the innermost parts of some large HII regions such as Hubble II and III (r/r$\rm _{max} < 0.1$) almost rival those of Ho 12 in terms of velocity dispersion. 
It is thus in the $\xi$ radial distribution that the distinction between Ho 12 and all three HII regions becomes obvious: the SNR shows systematically higher values than the HII regions. The average value of $\xi$ in Ho 12 is above 40 until r/r$\rm _{max} \sim 0.4$, while for the HII regions it is almost always smaller than 20. We also note that the overall difference in radial velocity (i.e., maximum - minimum value) in Ho 12 ($\sim$150 km/s) is much larger than that of the HII regions ($\sim$30 km/s). This is compatible with the high velocity ejected material associated with the SN explosion itself, and contributes to the significantly larger value of $\sigma$ ($\sim$70 km/s) in the {\it integrated} spectrum of Ho 12 compared to that of the HII regions ($\sim$30 km/s for Hubble II and III, 15 km/s for Hubble I): see Table~\ref{tab:integrated_ngc6822} where the integrated characteristics of Ho 12, Hubble I, Hubble II, and Hubble III are presented.

\begin{table*}
\setlength{\tabcolsep}{4pt}
        \caption{Integrated properties of Ho 12, Hubble I, II, and III.}
        \label{tab:integrated_ngc6822}
        \begin{tabular}{ccccccc}
                \hline
                Region & [SII]/H$\alpha$ & [OIII]/H$\beta$ & [NII]/H$\alpha$ & V$_{helio}$ & $\sigma$ & $\xi$ \\
                     & & & & km/s & km/s & km/s\\
                \hline
                Ho 12 & 0.52$\pm$0.03 & 0.75$\pm$0.08 & 0.10$\pm$0.02 & -75$\pm$1 & 69$\pm$1 & 36$\pm$2 \\
                Hubble I & 0.11$\pm$0.02 & 3.59$\pm$0.54 & 0.03$\pm$0.01 & -82$\pm$1 & 15$\pm$1 & 2$\pm$1 \\
                Hubble II & 0.29$\pm$0.02 & 1.14$\pm$0.09 & 0.08$\pm$0.01 & -82$\pm$1 & 30$\pm$1 & 9$\pm$1 \\
                Hubble III & 0.23$\pm$0.02 & 1.70$\pm$0.16 & 0.06$\pm$0.01 & -78$\pm$1 & 28$\pm$1 & 6$\pm$1\\
                \hline
        \end{tabular}
\end{table*}

\begin{figure}
	\includegraphics[width=\columnwidth]{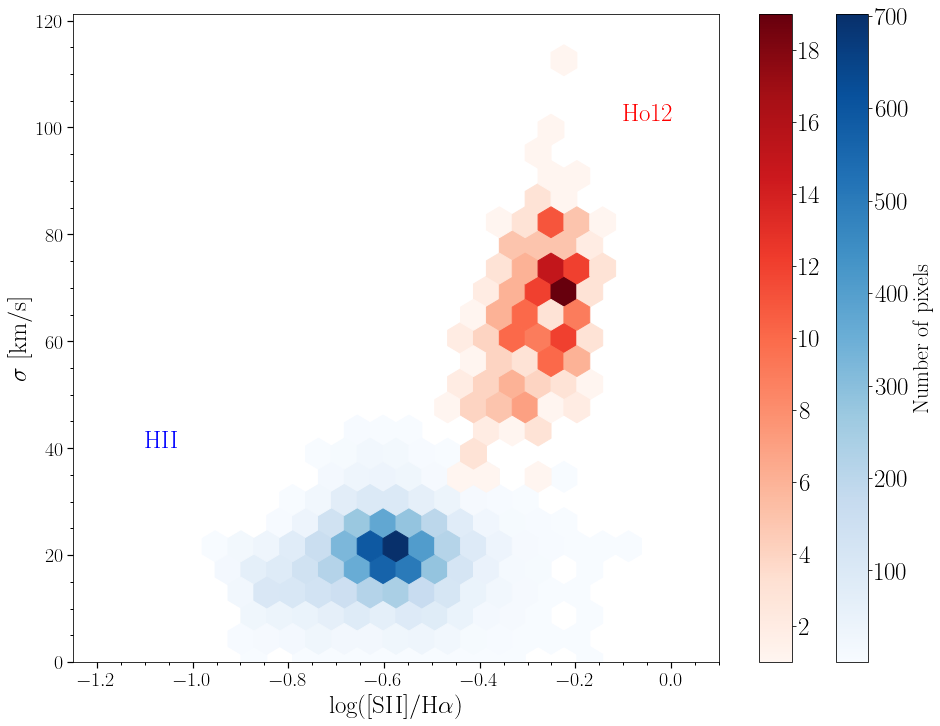}
    \caption{Velocity dispersion as a function of the [SII]/H$\alpha$ ratio. Only those pixels with S/N $\geq$ 5 for H$\alpha$ and [SII] have been considered. The red hexbins correspond to the pixels of SNR Ho 12 while the blue hexbins correspond to all pixels coming from the HII regions of NGC 6822.}
    \label{fig:N6822sigma}
\end{figure}

\begin{figure}
	\includegraphics[width=\columnwidth]{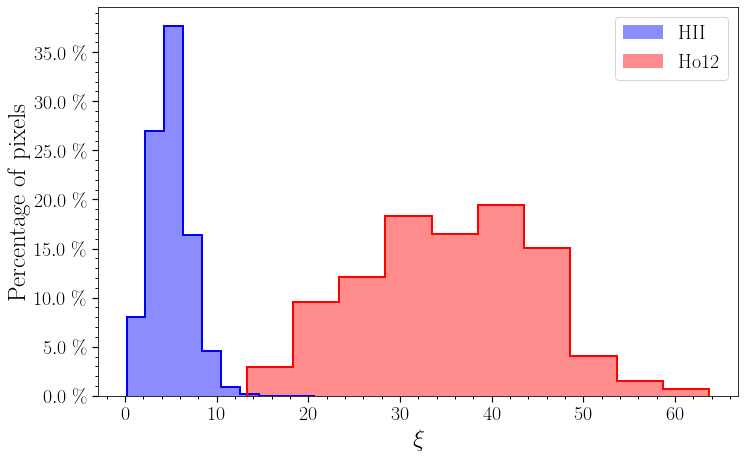}
    \caption{Distribution of $\xi$ for the same pixels as in Fig.~\ref{fig:N6822sigma}.}
    \label{fig:siihasigma}
\end{figure}

Figure~\ref{fig:N6822sigma} presents $\sigma$ as a function of the [SII]/H$\alpha$ ratio, for all the pixels having a S/N ratio larger than 5 in both [SII] and H$\alpha$ in the entire field of view for Ho 12 (red hexbin) and all the HII regions (blue hexbin). This diagram presents a very clear separation between the SNR and the HII regions. 
HII regions in NGC 6822 can be summarized by the following characteristics: i) most pixels have lower $\sigma$ and [SII]/H$\alpha$ values than Ho 12 values; ii) pixels with [SII]/H$\alpha$ values $\geq$ 0.4 show $\sigma$ values lower than Ho 12; and iii) pixels with $\sigma$ values greater than 30 km s$^{-1}$ show values of [SII]/H$\alpha$ $<$ 0.4.
The same pixels are also represented in Fig.~\ref{fig:siihasigma}, which depicts their distribution of the $\xi$ parameter. Again, the distribution of the HII regions is clearly different from that of the SNR.

\begin{figure}
	\includegraphics[width=\columnwidth]{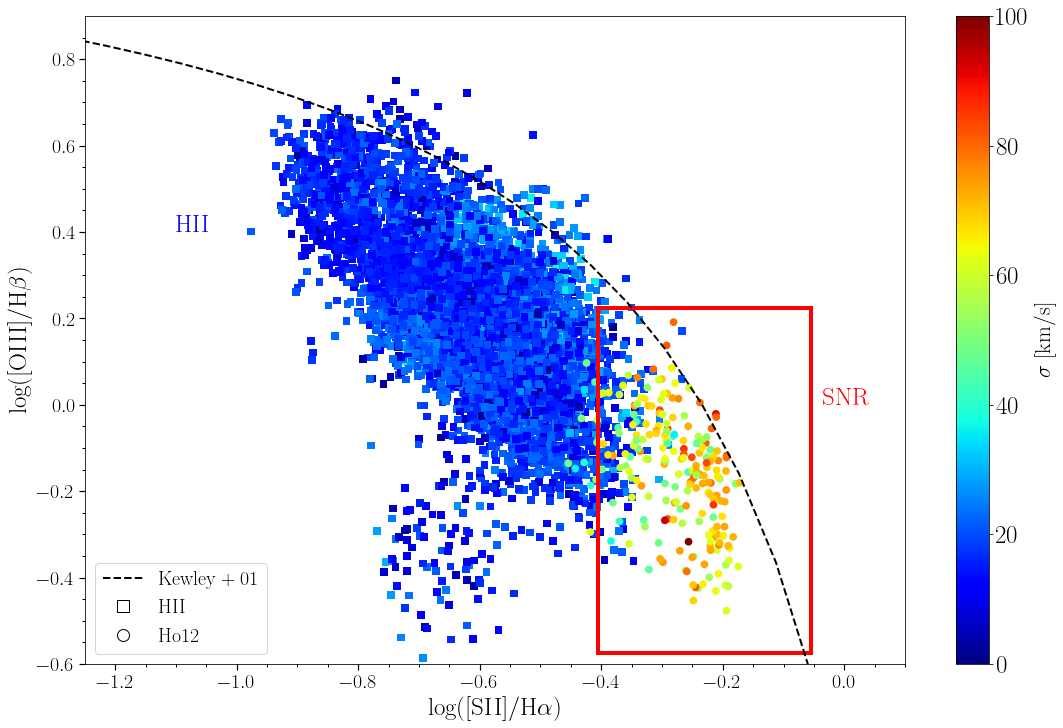}
    \caption{Composite BPT diagram of all HII regions identified in NGC 6822 and the SNR Ho 12. Only those pixels with S/N $\geq$ 5 for H$\alpha$, H$\beta$ and [SII], and [OIII] have been considered. Each point and square represents a pixel coming from Ho 12 and HII regions, respectively, and is color-coded according to its $\sigma$. The red rectangle shows the theoretical area corresponding to the SNRs.}
    \label{fig:N6822bpt}
\end{figure}

We conclude this section with a standard a diagnostic diagram, log([OIII]/H$\beta$) versus log([SII]/H$\alpha$) (Fig.~\ref{fig:N6822bpt}), but with a twist since each pixel is now color-coded according to its velocity dispersion, $\sigma$. In this figure we take into account all the ionised gas in NGC 6822. The SNR (identified with filled circles) stands out in the lower right part of this diagram, but there is a significant overlap between the SNR and HII regions: the number of pixels from HII regions for which the [SII]/H$\alpha$ ratio exceeds the threshold value used to detect SNRs (0.4) is not negligible. As shown in Fig.~\ref{fig:N6822ghr}, these pixels tend to concentrate on the outskirts of the HII regions. On the other hand, pixels with a significant $\sigma$ within HII regions have [SII]/H$\alpha \leq 0.3$. 

We have shown that, in NGC 6822 at least, the addition of the velocity dispersion parameter to the standard [SII]/H$\alpha$ ratio, all based on the same data cube, is an excellent asset to discriminate between HII regions and supernova remnants. There are however some potential caveats that we should be aware of:

\begin{itemize}
\item Only one SNR, Ho 12, is known in NGC 6822; its morphology is very peculiar, and thus may not be representative of all SNRs in nearby galaxies.

\item Ho 12 has a high surface brightness, is very well isolated from any HII region and the DIG, and belongs to the nearest galaxy in the SIGNALS survey. This obviously helps in its unambiguous identification as an SNR. In the next section, we explore M33 and we will see that, even though it is also a member of the Local Group, its large number of HII regions and the ubiquity of the DIG complicate our case.

\item Some zones in Hubble II, a large HII region, simultaneously present both a high [SII]/H$\alpha$ ratio (0.35 - 0.45) and a relatively high velocity  dispersion (50 - 65 km/s). This might suggest that it has it been the site of a supernova blast in the recent past.
\end{itemize}

\section{M33}
\label{sec:4_m33}
\begin{figure*}
	\includegraphics[width=7truein]{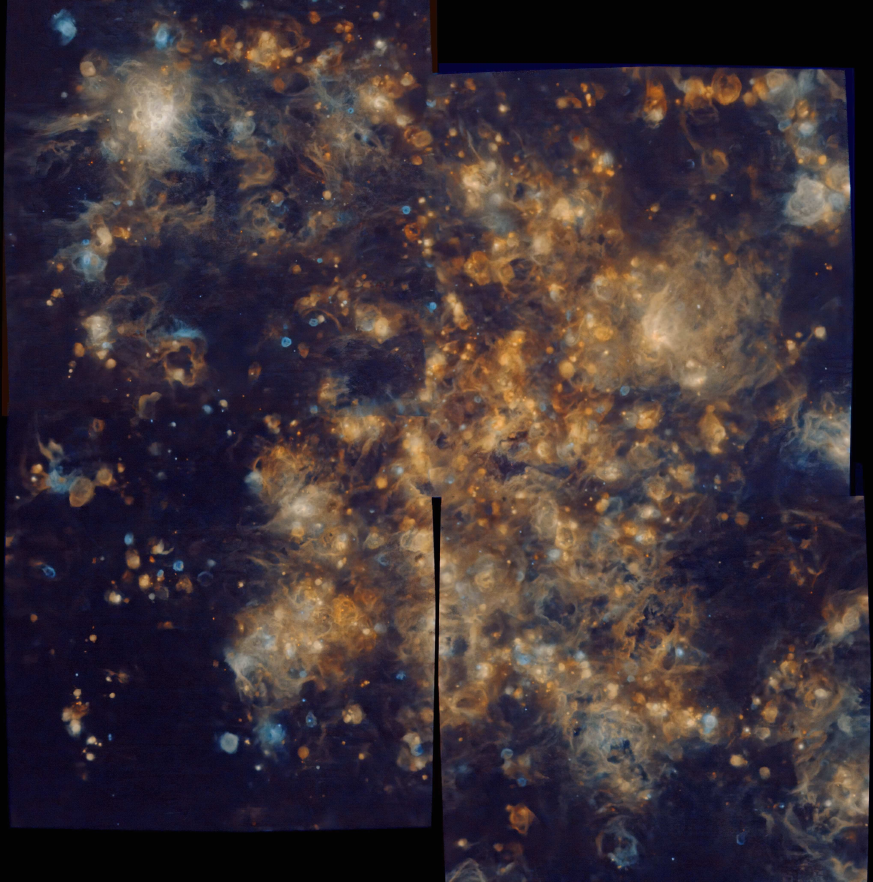}
    \caption{Color-coded mosaic of the four central fields of M33 (fields 1, 2, 3, and 4 in Table\,\ref{tab:fields}). H$\alpha$ in orange and a combination of [OII] and [OIII] in blue. The field of view is $\sim 22' \times 22'$, with North at the top and East to the left.}
    \label{fig:M33color}
\end{figure*}

\begin{figure*}
	\includegraphics[width=7truein]{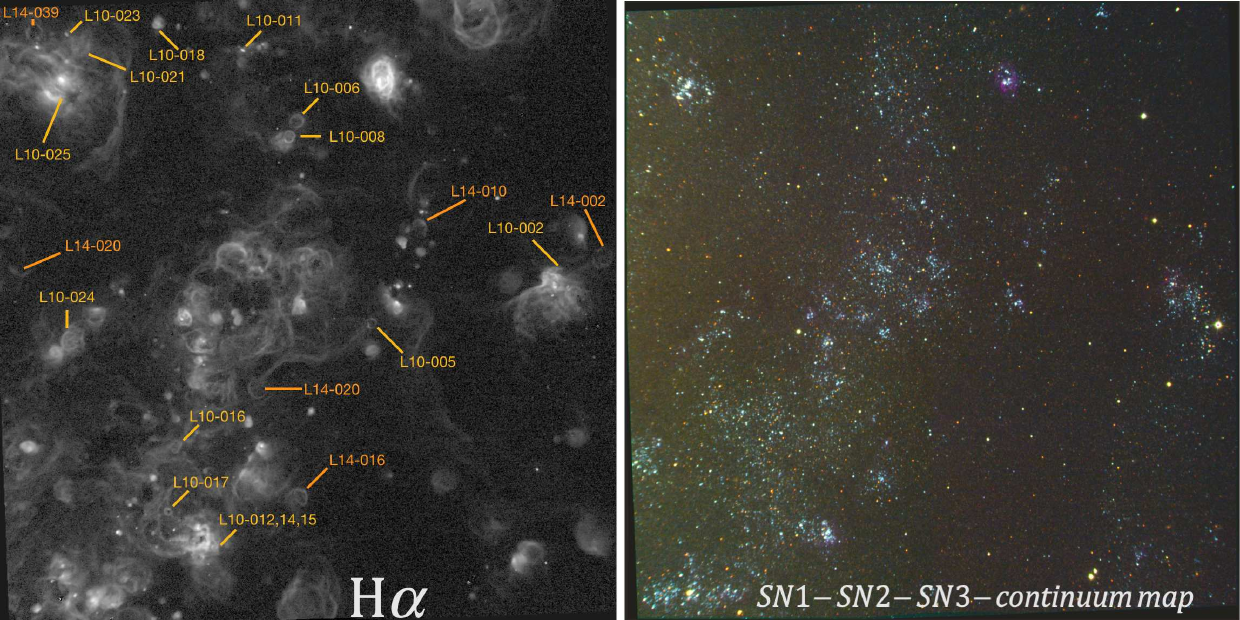}
    \caption{H$\alpha$ map (left panel) and continuum map (right panel, composition of the continuum maps in SN1, SN2, and SN3) of Field 7 in M33. In the left panel, we highlight the position of all the SNRs known in this field. FOV is $11' \times 11'$, with North at the top and East to the left.}
    \label{fig:M33F7}
\end{figure*}

The actively star-forming Triangulum galaxy hosts $\sim$200 SNR candidates, most of them spectroscopically confirmed and well studied, both in the visible band, the radio and X-rays \citep[e.g.,][and references therein]{1978A&A....63...63D,1998ApJS..117...89G,1999ApJS..120..247G,2010ApJS..187..495L, 2014ApJ...793..134L,2018ApJ...855..140L}. The central oxygen abundance of M33, as determined with the direct method, is 12+log(O/H) = 8.59 $\pm$ 0.02, with a galactocentric gradient of -0.037 $\pm$ 0.007 dex kpc$^{-1}$\citep{2022ApJ...939...44R}. The combination of the proximity of this galaxy \citep[D = 0.86 Mpc,][]{2014AJ....148...17D} and the large number of known SNRs make M33 a perfect candidate to study the suitability of the [SII]/H$\alpha$ vs. $\sigma$ diagram proposed in Sect.~\ref{sec:3_4_ngc6822_iden} and expand our analysis beyond the single case of Ho 12 in NGC 6822. In this second part of the paper we analyse a sample of 163 among the 217 known SNR candidates in M33 provided by \cite{2010ApJS..187..495L} - herafter L10 - and \cite{2014ApJ...793..134L} - hereafter L14, which are included in nine pointings observed with SITELLE (see Table~\ref{tab:fields}) for the SIGNALS survey. In Fig.~\ref{fig:M33color} we present a color-coded mosaic of the four central fields of M33 (Fields 1, 2, 3, and 4 in Table~\ref{tab:fields}). In Fig.~\ref{fig:M33F7} we show the H$\alpha$ map of Field 7 (left panel, highlighting the location of known SNR candidates) and its corresponding continuum map.

\begin{figure}
	\includegraphics[width=\columnwidth]{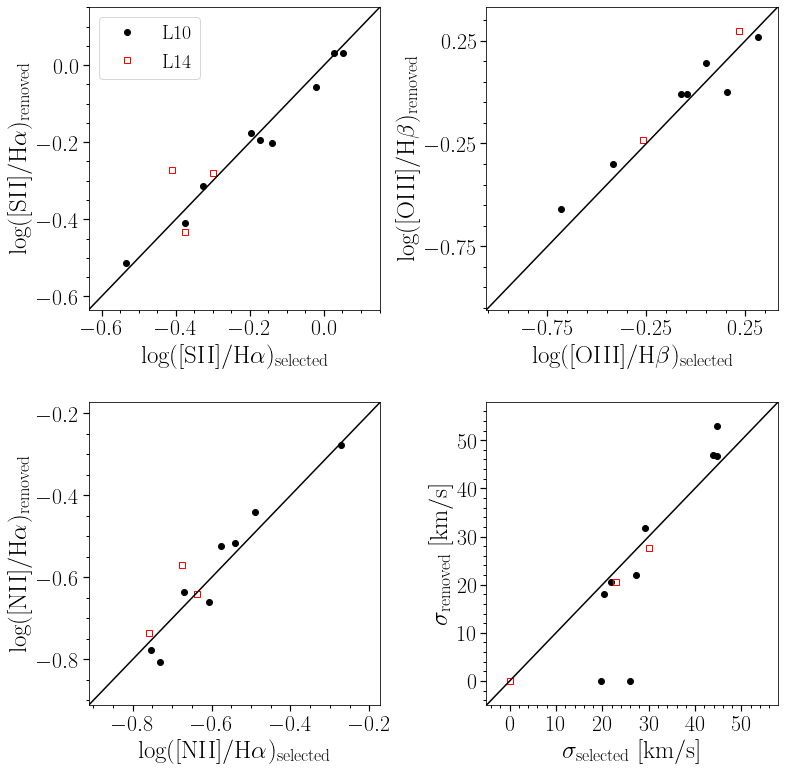}
    \caption{Comparison of line ratios and velocity dispersions for all SNRs present in two data cubes. All the emission lines have a signal-to-noise ratio higher than 5. The X axis represents the values for the selected cube while the Y axis represents those of the other cube. Objects shown in black dots come from \citet{2018ApJ...855..140L} catalogue while the red squares come from \citet{2014ApJ...793..134L}. The black solid line indicates the 1:1 relation.}
    \label{fig:compa_repe}
\end{figure}

As explained in Sect.~\ref{sec:2_observations}, we fitted the [OII]$\lambda$3727, [OIII]$\lambda\lambda$ 4959,5007, H$\beta$, [NII]$\lambda\lambda$6548,84, H$\alpha$, and [SII]$\lambda\lambda$6717,6731 emission lines for each of the 163 SNR candidates considering only one velocity component. Among those, 12 are located in overlapping regions of some fields (the duplicates). We thus kept for subsequent analysis the data from the cube having the best spectral resolution and S/N ratio. Figure \ref{fig:compa_repe} shows the comparison between the properties (line ratios and $\sigma$) of the selected and discarded duplicates: the values are in good agreement regardless of the spectral resolution, with the exception of two objects for which the spectral resolution of one of the data cubes was insufficient to distinguish the observed line width from that of the ILS (and were therefore attributed a value of $\sigma = 0$ in this dataset).  

\begin{figure}
	\includegraphics[width=\columnwidth]{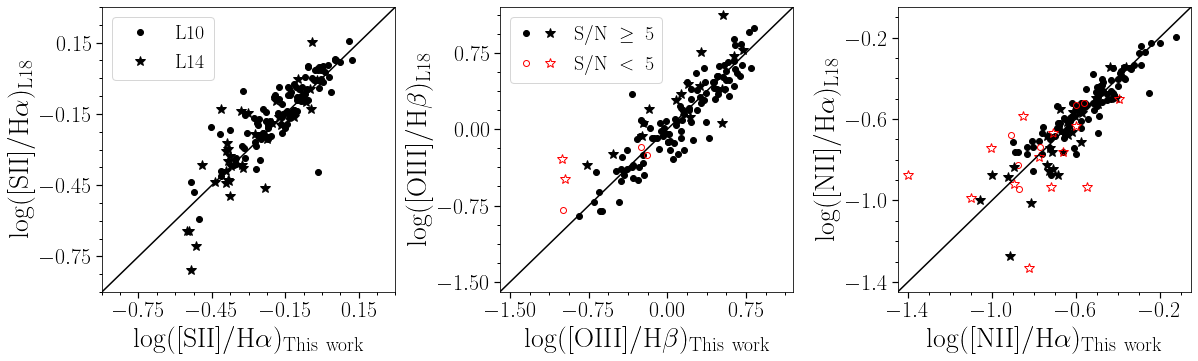}
    \caption{Comparison of the line ratios obtained in this work (x-axis) with those of \citet{2018ApJ...855..140L} (y-axis). SNRs with S/N $\geq$ 5 (in the emission lines shown in each panel) are shown in black dots while those with S/N < 5 are represented by red squares. The black solid line indicates the 1:1 relation.}
    \label{fig:compa_L18}
\end{figure}

Figure \ref{fig:compa_L18} presents a comparison between our {\it spatially integrated} properties of the individual objects in terms of line ratios (log([SII]/H$\alpha$) on the left panel, log([OIII]/H$\beta$) on the central panel, and log([NII]/H$\alpha$) on the right panel) with the work of \cite{2018ApJ...855..140L}. In general, the agreement is good, with the largest discrepancies occurring for the SNR candidates presented by \cite{2014ApJ...793..134L}, which tend to be more extended and diffuse than those found previously. But the large scatter in this figure is not surprising, given the noticeable spatial variations in line ratios within a given object and the fact that neither of the previous studies completely cover the whole SNR, as  slits and fibres (with 1.5 arcsec radius) were used, whereas our line ratios represent the integrated spectrum of each object.

\subsection{Morphology and properties of the sample}
\label{sec:4_1_m33_indi}

In this section we study the implication of the morphology of the SNRs in the [SII]/H$\alpha$ vs. $\sigma$ diagram. For this purpose, we only considered the SNRs  and the morphology defined by \cite{2010ApJS..187..495L}; therefore, the sample decreases to 118 objects. We did not consider \cite{2014ApJ...793..134L} SNR morphologies as they are not defined in the same way than \cite{2010ApJS..187..495L}.\footnote{The SNRs in \cite{2014ApJ...793..134L} are divided into more morphological subcategories than \cite{2010ApJS..187..495L}. We may add uncertainties to our results by erroneously including them in the morphological categories of \cite{2010ApJS..187..495L}.}

\cite{2010ApJS..187..495L} defined four morphological types in their sample of SNR: i) Type A, practically complete shells; ii) type A$^\prime$ (A2 in this article), small and bright objects, which are not shell-like but are well defined (they present an elongated shape); iii) type B, partial shells; and iv) type C, amorphous or poorly defined objects that may be a single filament. We define an additional morphological type, A3, corresponding to compact and bright objects. According to this classification there are 67 SNRs in type A, 5 in type A2, 4 in type A3, 22 in type B, and 20 in type C.
Table~\ref{tab:snrprop} presents the median and standard deviation values of the properties [SII]/H$\alpha$, $\sigma$, and $\xi$ when grouping all the SNRs according to their morphological classification. 

\begin{figure*}
	\includegraphics[width=0.49\textwidth]{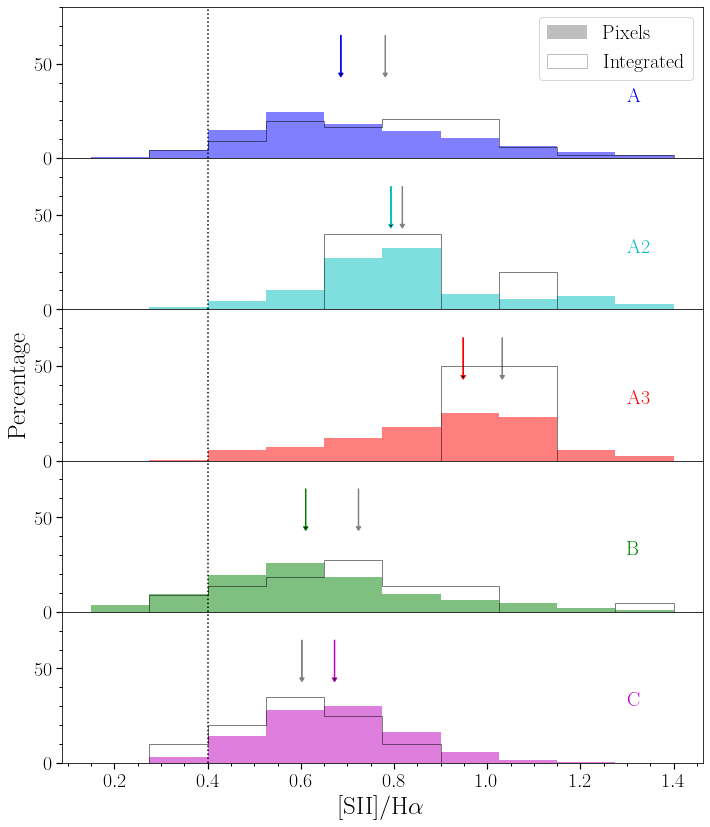}
	\includegraphics[width=0.49\textwidth]{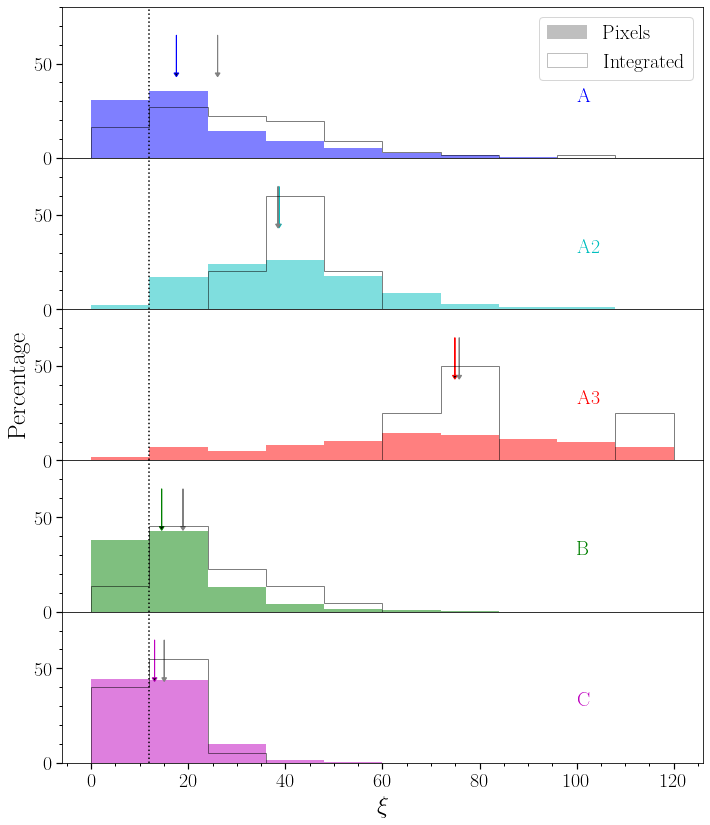}
    \caption{Distributions of [SII]/H$\alpha$ (left panel) and $\xi$ (right panel) for each morphological type (from top to bottom: type A, A2, A3, B, and C). In both panels, the distributions of the integrated values and for all the pixels of each SNR are shown as grey line and coloured histogram, respectively. The grey and coloured arrows indicate the position of the median value for each morphological type of the integrated value and the pixels contained in each SNR. The dotted vertical lines show the limit value of [SII]/H$\alpha$=0.4 (left panel) and $\xi$=12 (right panel).}
    \label{fig:props_morpho}
\end{figure*}

Figure~\ref{fig:props_morpho} shows the distributions of [SII]/H$\alpha$ (left panel) and $\xi$ (right panel) for each morphological type. We first note that while the [SII]/H$\alpha$ values span a wide range of values regardless of the morphological type, the most compact ones (A2 and A3) show higher median values (0.82 and 1.03, respectively) than the more extended types A and B (0.77 and 0.72, respectively). On the other hand, type C objects (poorly defined structure) show the lowest [SII]/H$\alpha$ values on average (0.6), as expected. 
 
Differences among the morphological types are more pronounced for $\xi$ than for [SII]/H$\alpha$ (right-hand panels of Fig.~\ref{fig:props_morpho}). Type A3 shows the highest median values ($\sim$76, with a wide distribution), followed by type A2 ($\sim$38). No SNR of the A2 and A3 types (and only $\sim$1\% of the pixels) shows values of $\xi$ lower than 12. When considering the integrated values, 10\% of the candidates of type A presents values of $\xi$ lower than 12. It is tempting to suggest an evolutionary link between the compact A3 type to type A2 and then the type A shells. The average values of $\xi$ determined for types B and C are significantly lower ($\sim$19 and $\sim$13, respectively) and could represent the final stages of detectable SNRs before a complete dissolution into the ISM. The range of $\xi$ spanned in both cases is similar, but only $\sim$10\% of type B candidates also show values lower than 12, while in type C they are $\sim$45\%. This is possibly due to the fact that type B objects evolve into type C. 

\begin{figure*}
	\includegraphics[width=0.43\textwidth]{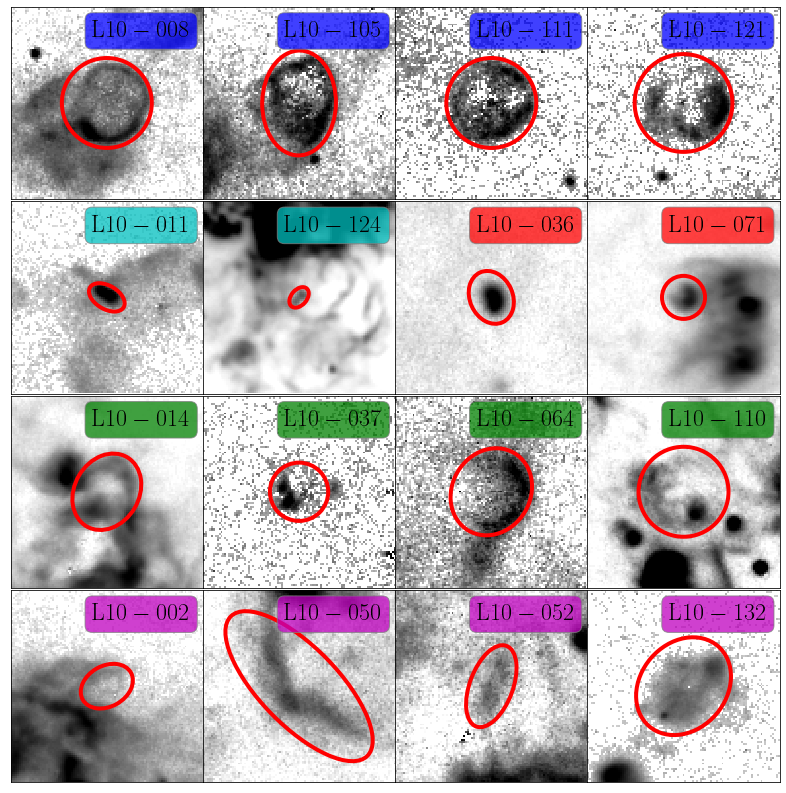}
	\includegraphics[width=0.5\textwidth]{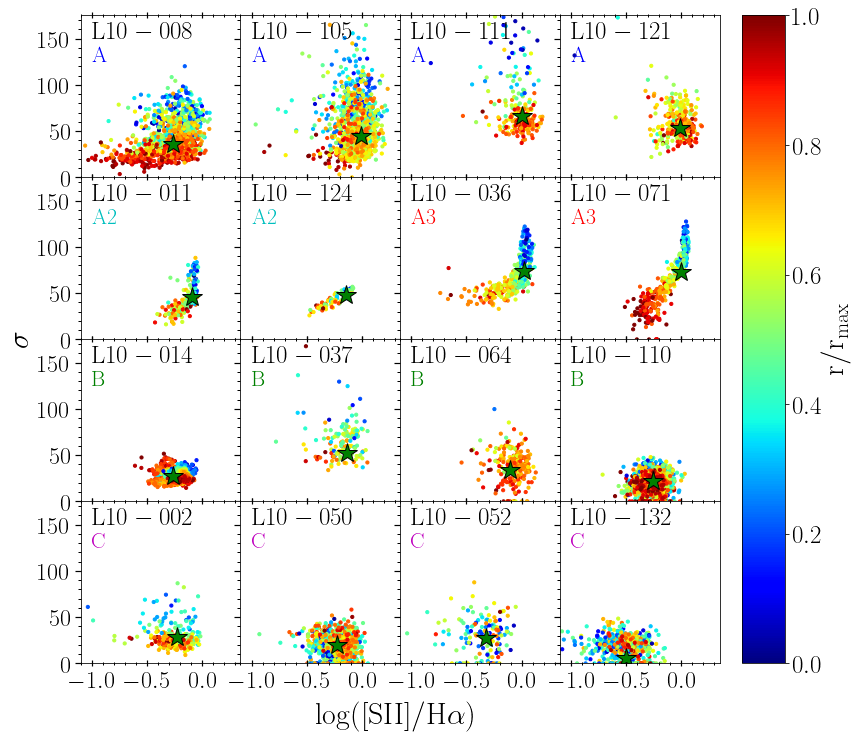}
    \caption{Left panel: H$\alpha$ flux images of SNRs illustrating the different morphological types. The region used to fit the integrated spectrum is shown as a red ellipse. Right panel: $\sigma$ vs. [SII]/H$\alpha$ diagram for the same objects; identification numbers are from \citet{2010ApJS..187..495L}. The green stars in each plot represents the integrated value. The coloured dots correspond to all pixels with F(H$\alpha$) $\geq$ 3$\times$10$^{-17}$ erg s$^{-1}$ cm$^{-2}$ enclosed inside the red ellipse in the left panel and have been colour-coded by their radial distance (r/r$_{max}$).}
    \label{fig:props_morpho_selec}
\end{figure*}

In Fig.~\ref{fig:props_morpho_selec} we show the H$\alpha$ flux maps (left panel), and the [SII]/H$\alpha$ vs. $\sigma$ diagram (right panel) for a subsample of 16 selected SNRs in M33 representative of all morphological types. SNRs with A, A2, or A3 morphological types present higher values of $\sigma$ in the central parts, decreasing as we move towards the outer parts of the SNR, as in the case of Ho 12 (see Sect.~\ref{sec:3_2_ngc6822_kine}). In the case of the A2 and A3 types shown in this section, except for L10-121, we see that the value of [SII]/H$\alpha$ remain constant as $\sigma$ decreases until r/r$_{max}\ \sim$0.5, from which it begins to decrease with $\sigma$. In the case of B and C types, this dependence is less evident, although in L10-002 and L10-037 it is. Moreover, it can be seen that the scatter of [SII]/H$\alpha$ at a constant $\sigma$ value is lower in the A2 and A3 types than in the rest of the morphological types.

\begin{figure}
	\includegraphics[width=\columnwidth]{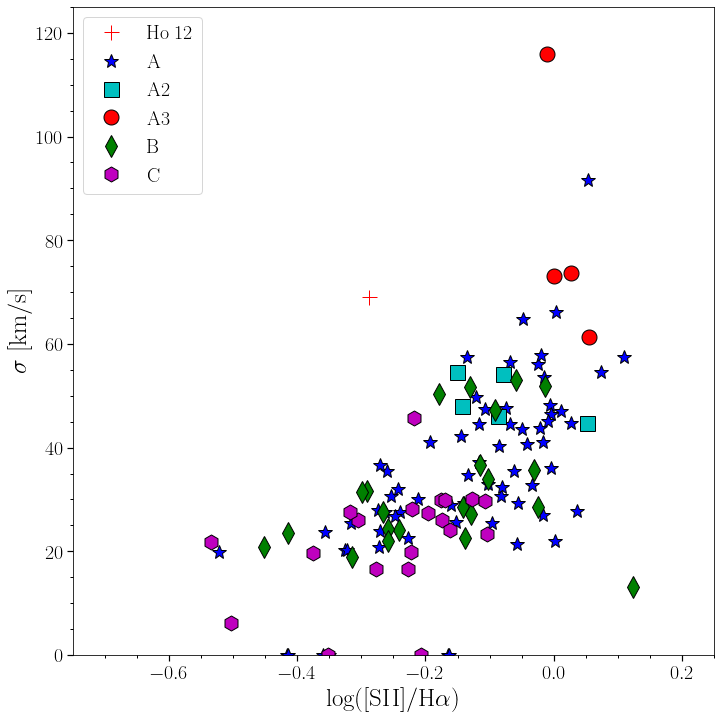}
    \caption{Velocity dispersion as a function of the [SII]/H$\alpha$ ratio. Red dots, blue stars, cyan squares, green diamonds, and magenta hexagons represent A3, A, A2, B, and C morphological type of SNR, respectively. The red cross shows the position in this diagram for Ho 12 in NGC 6822.
    }
    \label{fig:sig_siiha_morpho}
\end{figure}

\begin{figure}
	\includegraphics[width=\columnwidth]{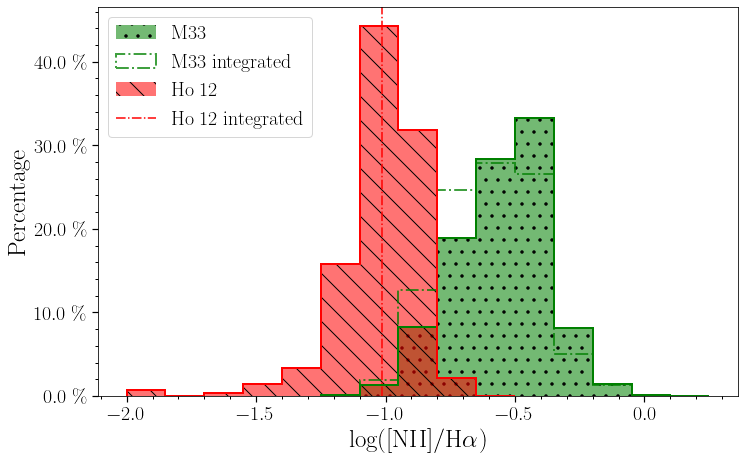}
    \caption{Distribution of log([NII]/H$\alpha$) for the SNR candidates in M33 and Ho 12. Solid green histogram (with diagonal line hatch) and dashed green histogram (with dotted hatch) represent the distribution for all the pixels and the integrated values from the 163 SNR candidates in M33, respectively. Solid red histogram and vertical dashed red line show the distribution of all the pixels and the integrated values from Ho 12, respectively.}
    \label{fig:hist_niiha}
\end{figure}

Figure~\ref{fig:sig_siiha_morpho} shows the $\sigma$ vs. [SII]/H$\alpha$ diagram taking into account the morphology of each SNR candidate. According to this diagram, the evolution of a SNR could be interpreted as follows: compact objects (type A3) have very high values of [SII]/H$\alpha$ and $\sigma$, following a well-defined path, with a small dispersion in [SII]/H$\alpha$ at constant $\sigma$. These SNRs would evolve into more elongated type A2 objects, where their $\sigma$ would decrease as they expand through the interstellar medium. As they expand, the shell-like structure (type A) becomes more evident. The central parts of type A SNRs can still show high $\sigma$ values, more likely because the velocity components of the approaching and receding parts of the shell are detected simultaneously, see Sect.~\ref{sec:4_3_m33_kine}. The expansion and interaction with the ISM cause $\sigma$ and [SII]/H$\alpha$ to decrease. We observe that the [SII]/H$\alpha$ vs. $\sigma$ relation is no longer as well defined as the more compact objects (i.e., the dispersion is larger), as expected. As they continue expanding, the structures become less obvious, the type B SNRs have not yet fully mixed with the ISM and retain sufficiently high values of [SII]/H$\alpha$ and $\sigma$ in their innermost parts to allow their detection. Type C objects have already evolved much further, it is difficult to identify the structure of the SNR clearly, and we are often only able to observe a filament. The value of $\sigma$ is low, although some of them still maintain values of [SII]/H$\alpha$ higher than 0.4. We note that Ho 12 in NGC 6822 does not follow the global trend defined by the M33 SNRs in the $\sigma$ vs. [SII]/H$\alpha$ diagram. This may be related to the metallicity of the host galaxies. Taking into account that [NII] emission depends on metallicity \citep[e.g.][]{2004MNRAS.348L..59P,2009MNRAS.398..949P} we show the distribution of log([NII]/H$\alpha$) in Fig.\ref{fig:hist_niiha}. In this Figure we notice that the log([NII]/H$\alpha$) values for Ho 12 are lower than those of the SNR candidates of M33. The most likely explanation is that the global metallicity of M33 is higher than what has been found in NGC6822, although the [NII]/H$\alpha$ ratio is also sensitive to other nebular properties like the ionization parameter and the abundance of secondary nitrogen. We will explore further this topic with a larger number of SNRs in galaxies of the SIGNALS survey (Duarte Puertas et al., in prep.).

\begin{table*}
        \centering
        \caption{Morphological properties of SNRs from L10 in M33.}
        \label{tab:snrprop}
        \begin{tabular}{ccccccc}
                \hline
                 &  A & A2 & A3 & B & C & All \\
                \hline
                [SII]/H$\alpha$ & 0.78$\pm$0.22 & 0.82$\pm$0.15 & 1.03$\pm$0.06 & 0.72$\pm$0.22 & 0.60$\pm$0.14 & 0.73$\pm$0.22 \\
                $\sigma$ & 33$\pm$17 & 48$\pm$4 & 73$\pm$21 & 28$\pm$12 & 25$\pm$11 & 30$\pm$18 \\
                $\xi$ & 26$\pm$20 & 38$\pm$4 & 76$\pm$17 & 19$\pm$12 & 15$\pm$8 & 22$\pm$20 \\
                Number of objects & 67 & 5 & 4 & 22 & 20 & 118 \\
                      
                \hline
        \end{tabular}
\end{table*}

\begin{figure*}
	\includegraphics[width=\textwidth]{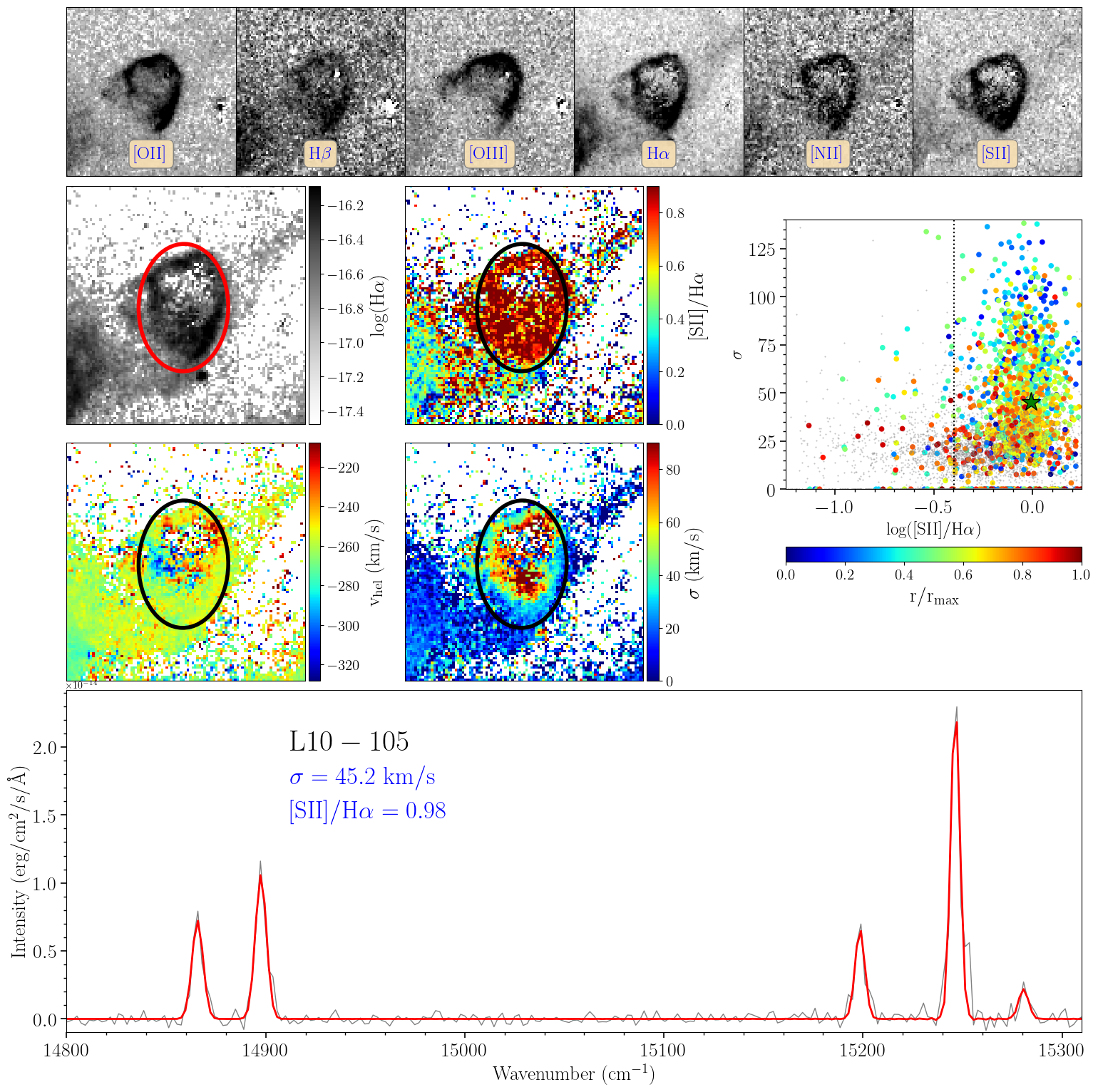}
    \caption{Properties of L10-105. Top panel: emission line maps (32$'' \times 32''$), from left to right :  [OII], H$\beta$, [OIII], H$\alpha$, [NII], and [SII]. Middle panel, left: log(H$\alpha$) flux, [SII]/H$\alpha$ ratio, heliocentric radial velocity, and $\sigma$.Middle panel, right: $\sigma$ vs. [SII]/H$\alpha$ diagram for all pixels in the maps, with the coloured dots corresponding to all pixels with F(H$\alpha$) $\geq$ 1$\times$10$^{-17}$ erg s$^{-1}$ cm$^{-2}$ enclosed by the red ellipse in the left panel and have been colour-coded by r/r$_{max}$. The green star shows the position of the integrated value within the ellipse. Lower panel: Continuum-subtracted integrated spectrum of the region shown as a red (or black) ellipse above. The grey line shows the real spectrum and the red coloured line show the fit obtained with ORCS in the SN3 bandpass.}
    \label{fig:l10-105}
\end{figure*}

\begin{figure*}
	\includegraphics[width=\textwidth]{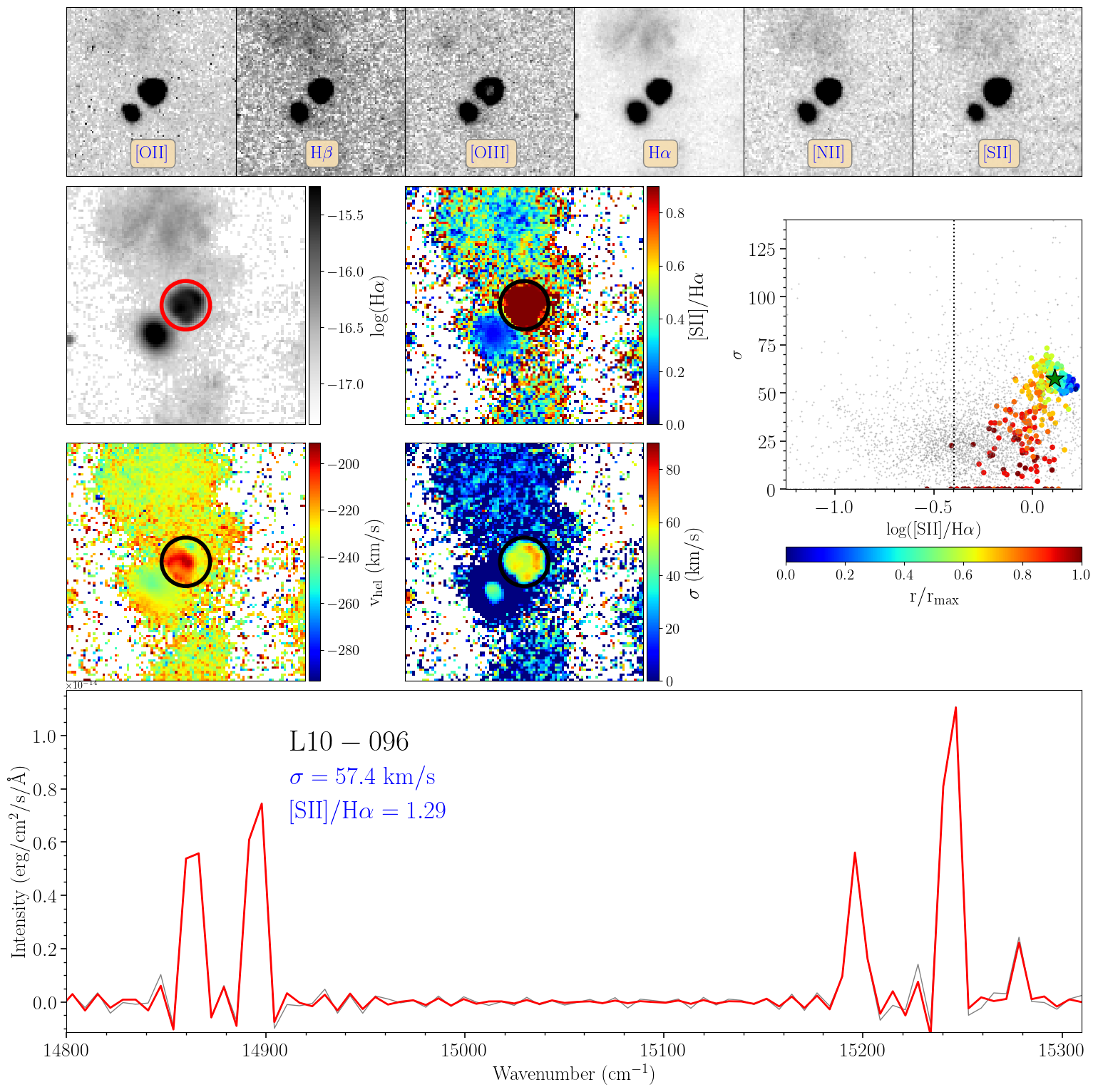}
    \caption{Same as Fig.~\ref{fig:l10-105} for L10-096.}
    \label{fig:l10-096}
\end{figure*}

\begin{figure*}
	\includegraphics[width=\textwidth]{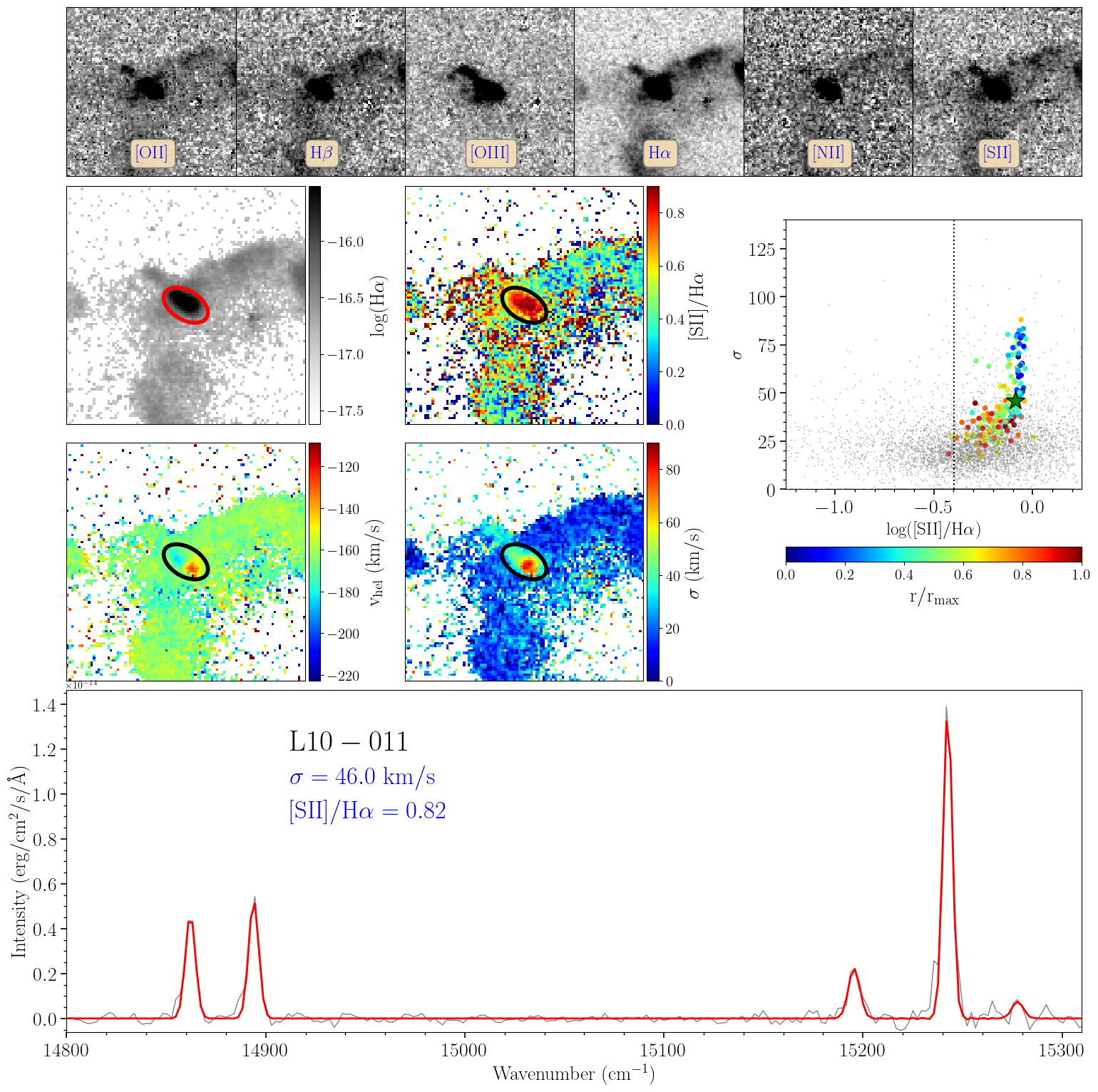}
    \caption{Same as Fig.~\ref{fig:l10-105} for L10-011.}
    \label{fig:l10-011}
\end{figure*}

\begin{figure*}
	\includegraphics[width=\textwidth]{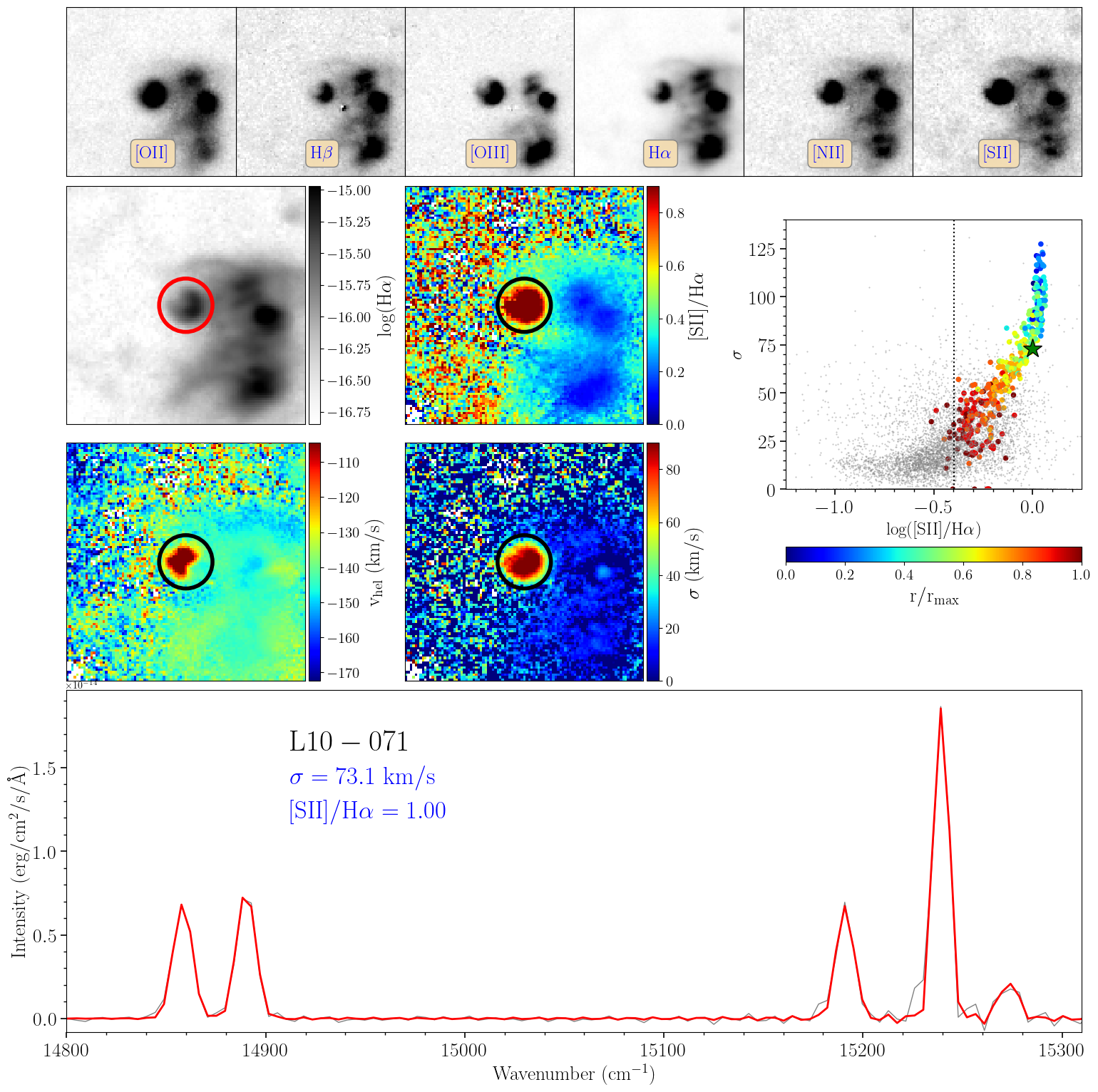}
    \caption{Same as Fig.~\ref{fig:l10-105} for L10-071.}
    \label{fig:l10-071}
\end{figure*}

\begin{figure*}
 	\includegraphics[width=\textwidth]{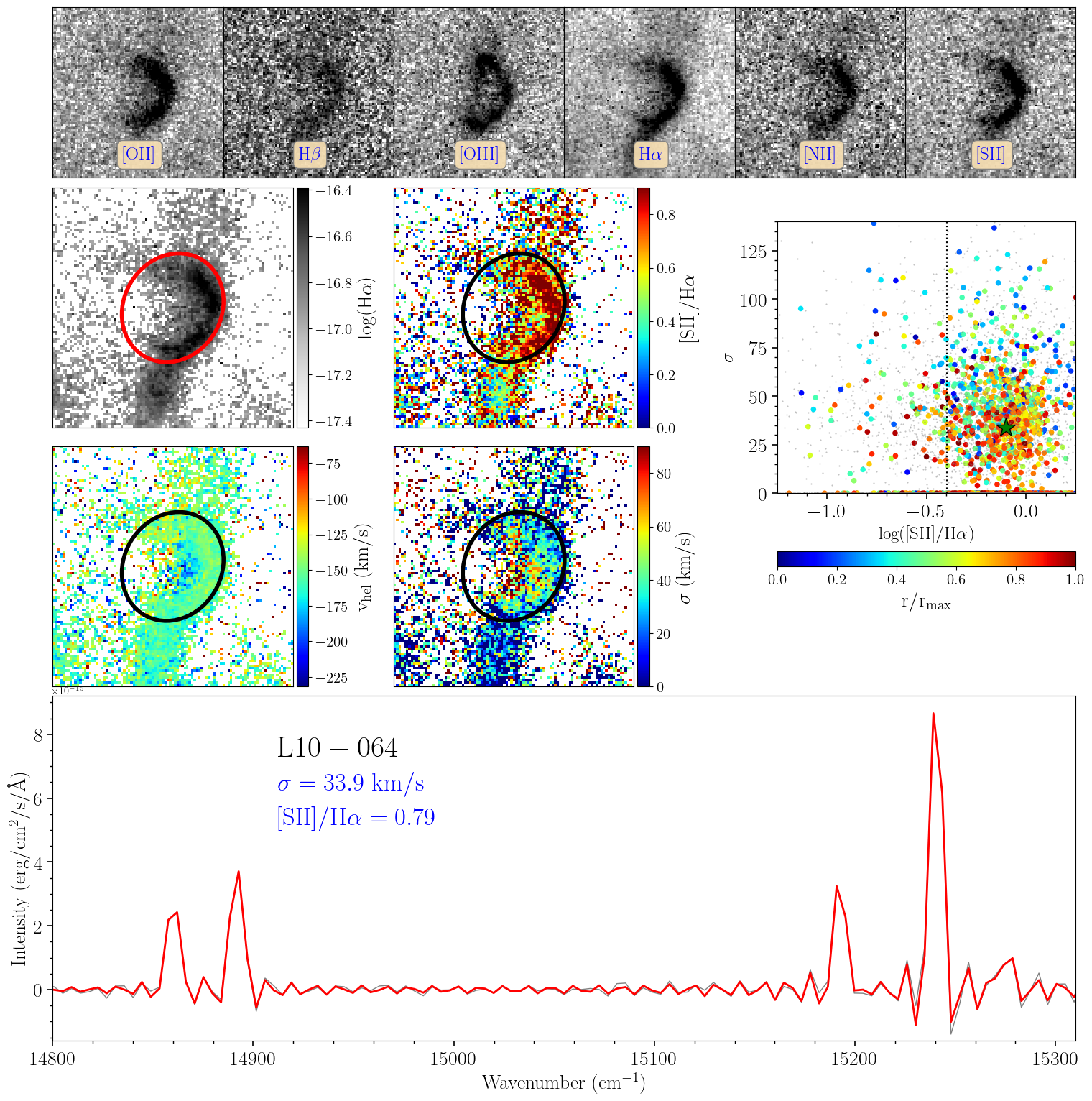}
     \caption{Same as Fig.~\ref{fig:l10-105} for L10-064.}
     \label{fig:l10-064}
 \end{figure*}

\begin{figure*}
	\includegraphics[width=\textwidth]{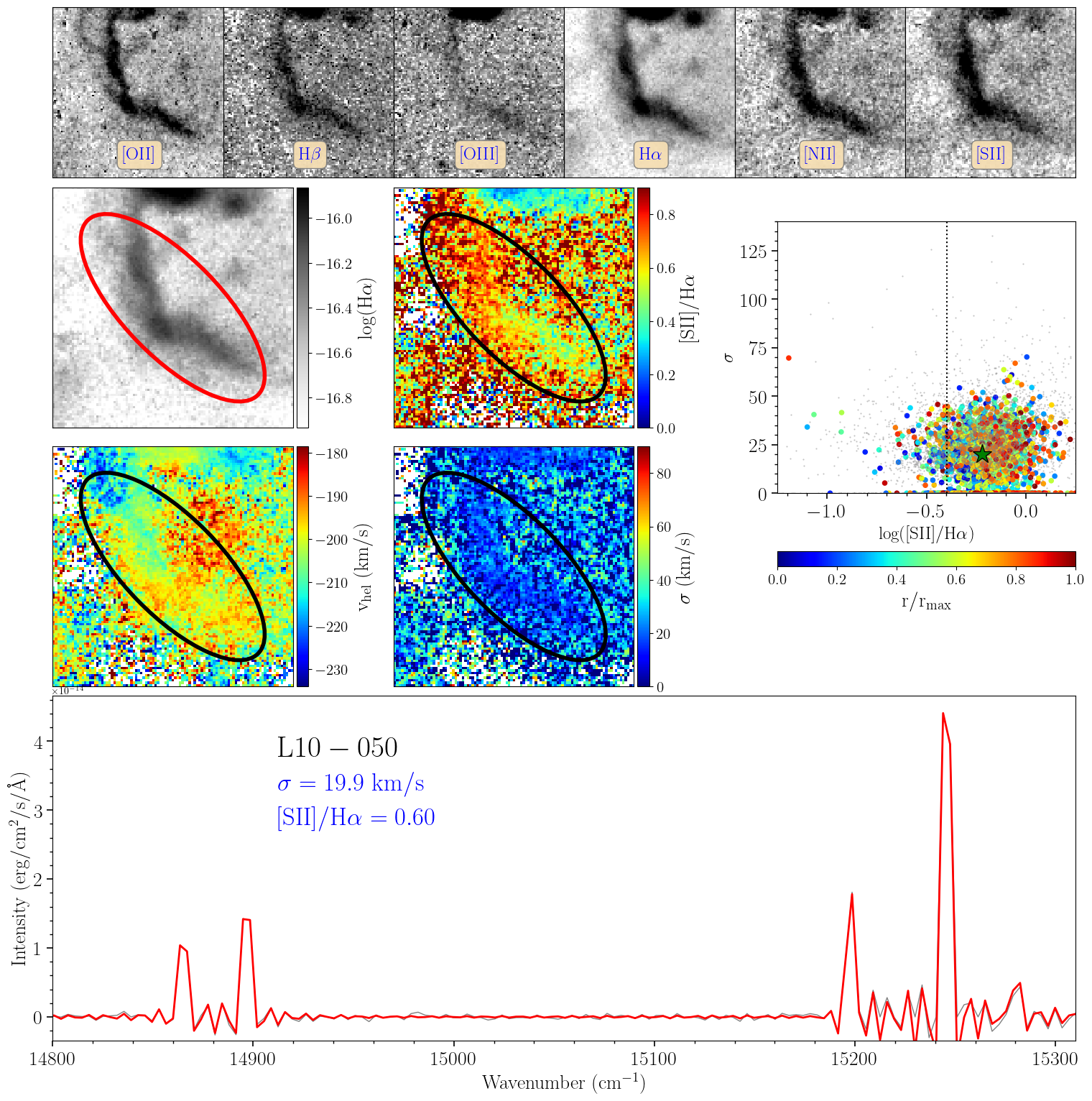}
    \caption{Same as Fig.~\ref{fig:l10-105} for L10-050.}
    \label{fig:l10-050}
\end{figure*}

Figures \ref{fig:l10-105}, \ref{fig:l10-096}, \ref{fig:l10-011}, \ref{fig:l10-071}, \ref{fig:l10-064}, and \ref{fig:l10-050} present some examples of SNRs with different morphologies (e.g., L10-105, L10-096, L10-096, L10-071, and L10-050). In each figure we show: i) the flux maps for the [OII], H$\beta$, [OIII], H$\alpha$, [NII] and [SII] emission lines; ii) the maps of [SII]/H$\alpha$, radial velocity, and $\sigma$; iii) the log([SII]/H$\alpha$) vs. $\sigma$ diagram; and iv) the integrated SNR spectrum and the fit obtained with ORCS in the SN3 bandpass. The remaining SNRs are shown in  Sect.~\ref{sec:4_2_nearby_str} and in the Appendix (Supplementary material). 

Figure \ref{fig:l10-105} shows L10-105. According to \cite{2010ApJS..187..495L}, the morphological type for this SNR is A. It has a similar shape in H$\alpha$ and [SII], with a prominent south-eastern wall, which is however almost invisible in [OIII]. The [SII]/H$\alpha$ ratio is high in the whole SNR, however $\sigma$ presents higher values in the central part than in the outermost part, as expected from an expanding shell. The integrated spectrum hints at at least one additional velocity component, which could be attributed to the ISM filament crossing the field of view.

Figure \ref{fig:l10-096} presents L10-096. This object is in some ways similar to Ho 12 in NGC 6822, showing a series of knots along the outer ring. The morphology at the different wavelengths is similar, except that the strongest [OIII] emission is seen slightly further out the ring. We also note that the largest values of $\sigma$ are seen in the knots of the outer ring. A small expanding (with larger values of $\sigma$ at its core), purely photoionized, HII region is seen at its lower left.

L10-011 shows an A2 morphology (see Fig.~\ref{fig:l10-011}). Its kinematics show a bipolar behaviour, with lower radial velocities in the northwest and higher in the southeast. This southeastern part shows higher values of $\sigma$ than the northwestern part. An HII region to the west and the SNR candidate L14-023 to the south can be observed close to L10-011. There is a filament with a large value of [SII]/H$\alpha$ that crosses the SNR from behind, connecting both the HII region and the SNR candidate. This filament has a low $\sigma$ value ($\sim$20 km/s). Its average radial velocity is $\sim$-160 km/s and has a [SII]/H$\alpha$ value larger than 0.4, being larger on the outside of the filament than on the inside. Moreover, taking into account the $\sigma$ map, it can be seen how to the northeast of this SNR there is a zone with higher $\sigma$ values than 20 km/s. Taking into account the radial velocity of this zone and comparing it with the velocity of L10-011, it is compatible with being material associated with this SNR.

L10-071 is a compact SNR (Fig.~\ref{fig:l10-071}); its outer ring is barely resolved. Morphologically it is similar in the different emission lines. The behaviour in the log([SII]/H$\alpha$) vs. $\sigma$ diagram has already been discussed in the previous section. 

Figure \ref{fig:l10-064} presents L10-064. It is a type B SNR. Its western part shows a similar morphology in all emission lines, while the eastern part is open. A filament is visible from north to south, the southern part being the brightest. As in the cases of L10-011 and L10-105, the filament has high [SII]/H$\alpha$ and low $\sigma$ values. L10-064 presents high [SII]/H$\alpha$ values and tentatively higher values of $\sigma$ in the central part ($\sim$120 km/s) than in the outer part ($\lesssim$40 km/s).

In Fig.~\ref{fig:l10-050} we show L10-050, of type C: it is fragmented and stands out as a filament. The morphology is similar in all visible emission lines, although [OIII] is rather faint. [SII]/H$\alpha$ is high but $\sigma$ is low ($\sim$20 km/s).

\subsection{Detectability, spectral characteristics of SNRs and their immediate environment}
\label{sec:4_2_nearby_str}

The detectability of SNRs in galaxies outside the Local Group obviously depends on their evolutionary status and location with respect to bright ISM regions: an old SNR within a dense HII region is less likely to be detected with optical data than an isolated younger one. In addition, the diffuse ionized gas (DIG) is known to harbor large values of [SII]/H$\alpha$ \citep{2021MNRAS.508.1582M}, albeit with small velocity dispersion. Without any information on the velocity dispersion, some filaments of the DIG could thus be mistaken for SNRs \citep[see for instance the case of NGC 4214: ][]{2013MNRAS.429..189L,2023MNRAS.524.3623V}. In M33, L10-011 and L10-064 are examples of SNRs which are seen superimposed on large DIG filaments with values of [SII]/H$\alpha$ larger than 0.4. 

For the same reasons, the spectral characteristics of SNRs, in terms of line ratios and kinematics, depend on the choice of the background used to extract their net spectrum; this choice is not obvious in cases where the SNR lies in a highly inhomogeneous nebular structure of filament. As we mentioned before, we have deliberately chosen to subtract a general background as much devoid of nebular features as possible to produce the maps and the integrated spectra of SNRs presented here and forced ORCS to use the same velocity dispersion for all the lines in the SN3 cube. This has little impact for the vast majority of objects, since the contrast between the SNRs and their surroundings is high. 

\begin{figure}
	\includegraphics[width=\columnwidth]{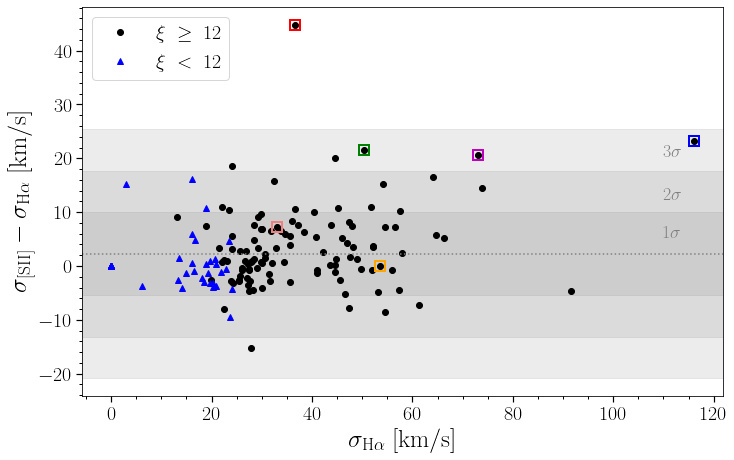}
    \caption{Difference between the velocity dispersions of the H$\alpha$ line ($\sigma_{H\alpha}$) and the [SII] doublet ($\sigma_{[SII]}$) for our sample of 163 SNRs in M33. Coloured squares highlight the position of a few selected objects SNRs, the spectrum of which are displayed in Fig.~\ref{fig:diff_sigma_sii_ha_spec}: red, blue, green, magenta, orange, and salmon for L10-025, L10-039, L10-034, L10-071, L10-121, and L10-049, respectively.} 
    \label{fig:diff_sigma_sii_ha}
\end{figure}

\begin{figure}
	\includegraphics[width=\columnwidth]{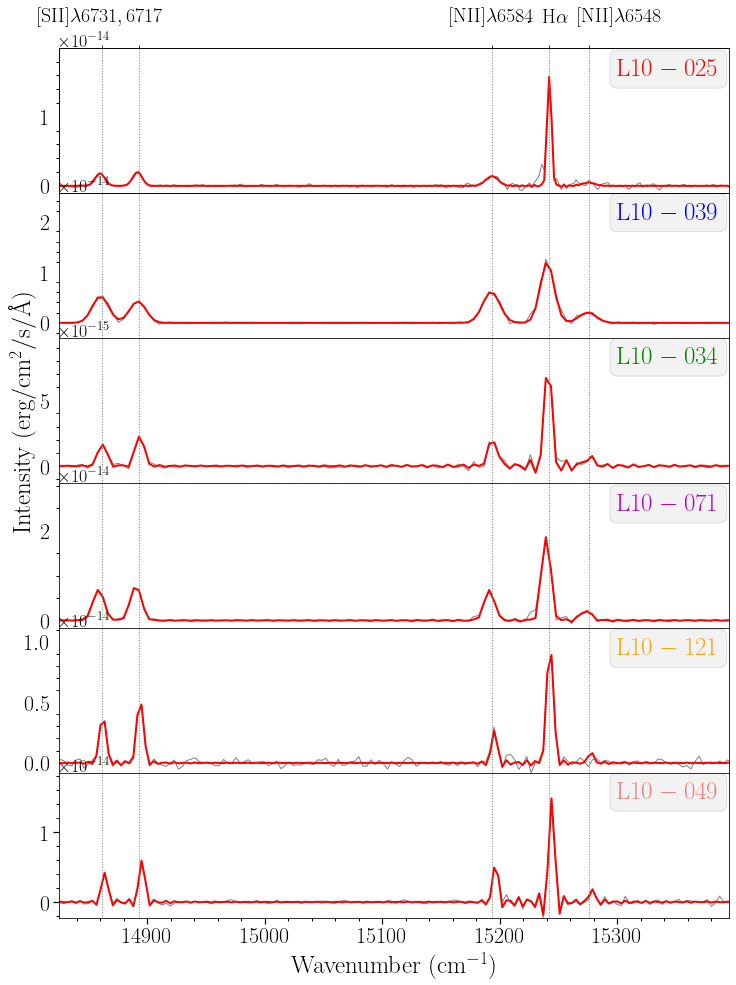}
    \caption{Spectra of the SNRs identified in Fig.~\ref{fig:diff_sigma_sii_ha}.  In each panel, the grey line shows the real spectrum and the red lines show the fit obtained with ORCS. Grey dotted lines show the locations of the emission lines from SN3 filter. The SNR label is indicated in each panel (upper right part).}
    \label{fig:diff_sigma_sii_ha_spec}
\end{figure}

\begin{figure*}
	\includegraphics[width=\textwidth]{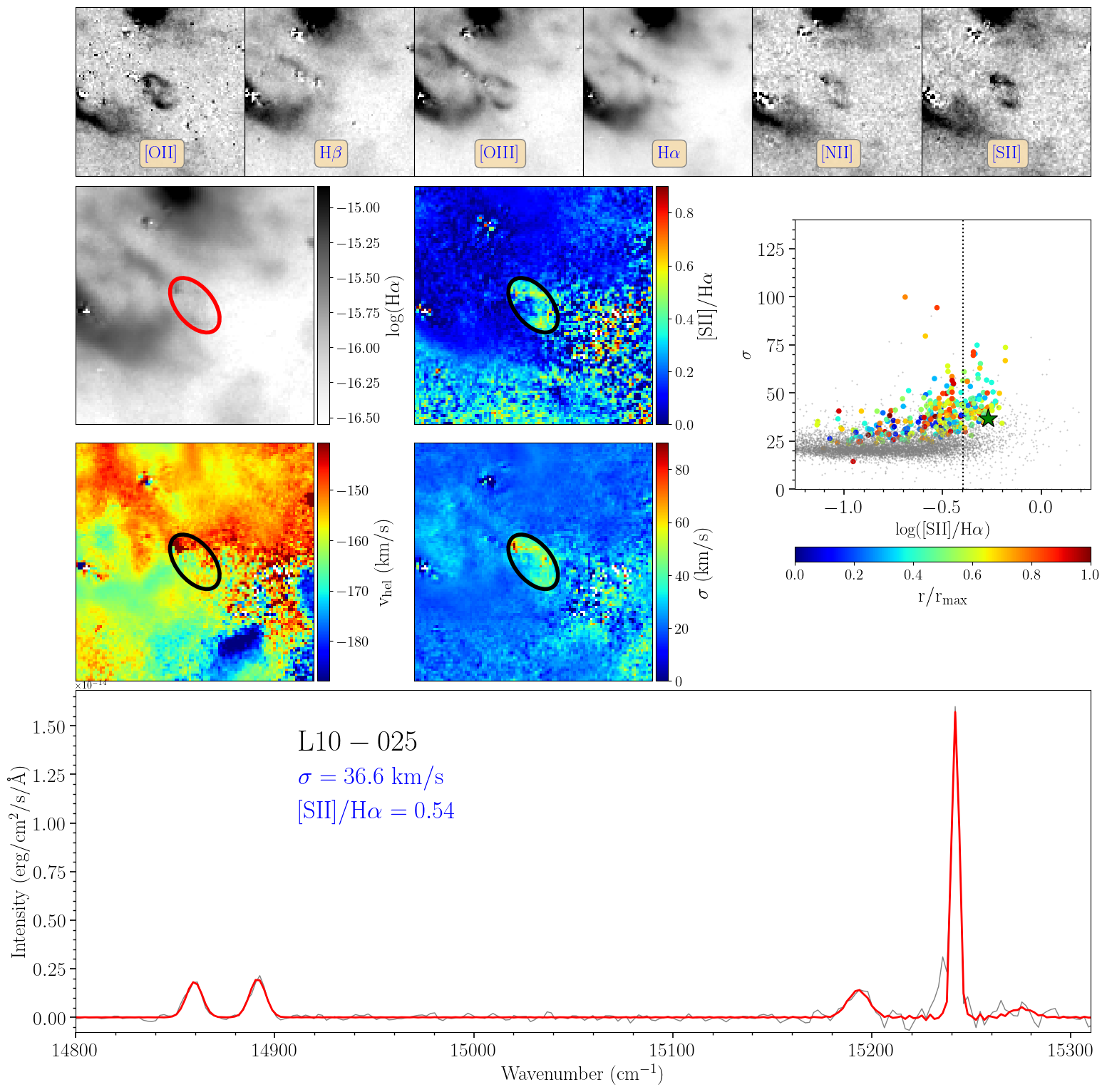}
    \caption{Same as Fig.~\ref{fig:l10-105} for L10-025.}
    \label{fig:l10-025}
\end{figure*}

We have however visually noticed a few cases for which the width of the H$\alpha$ line was noticeably smaller than that of the [SII] doublet. We have therefore performed another fit on the integrated spectrum of all SNRs with ORCS, but this time allowing it to fit different velocity dispersions for H$\alpha$ and the forbidden lines. The values of $\sigma_{H\alpha}$ and $\sigma_{[SII]}$ are compiled in Table 5 and plotted in Fig.~\ref{fig:diff_sigma_sii_ha}; examples of spectra from selected objects are shown in Fig.~\ref{fig:diff_sigma_sii_ha_spec}. Although there is a small offset globally (the average difference is 2 km/s), about 87\% of the SNRs show no significant difference between the two values of velocity dispersion. Twelve objects (7\% of the sample) have a difference larger than 15 km/s in favor of [SII], the most obvious one being L10-025 (upper panel of Fig.~\ref{fig:diff_sigma_sii_ha_spec} and Fig.~\ref{fig:l10-025}), which is indeed surrounded by a bright HII region. We infer that in more distant galaxies, a much larger fraction of SNRs will display a significant difference between the two values of velocity dispersion: extra care will then need to be taken to subtract a proper background.

\subsection{Velocity structure: SNRs with multiple components}
\label{sec:4_3_m33_kine}

\begin{figure*}
	\includegraphics[width=\textwidth]{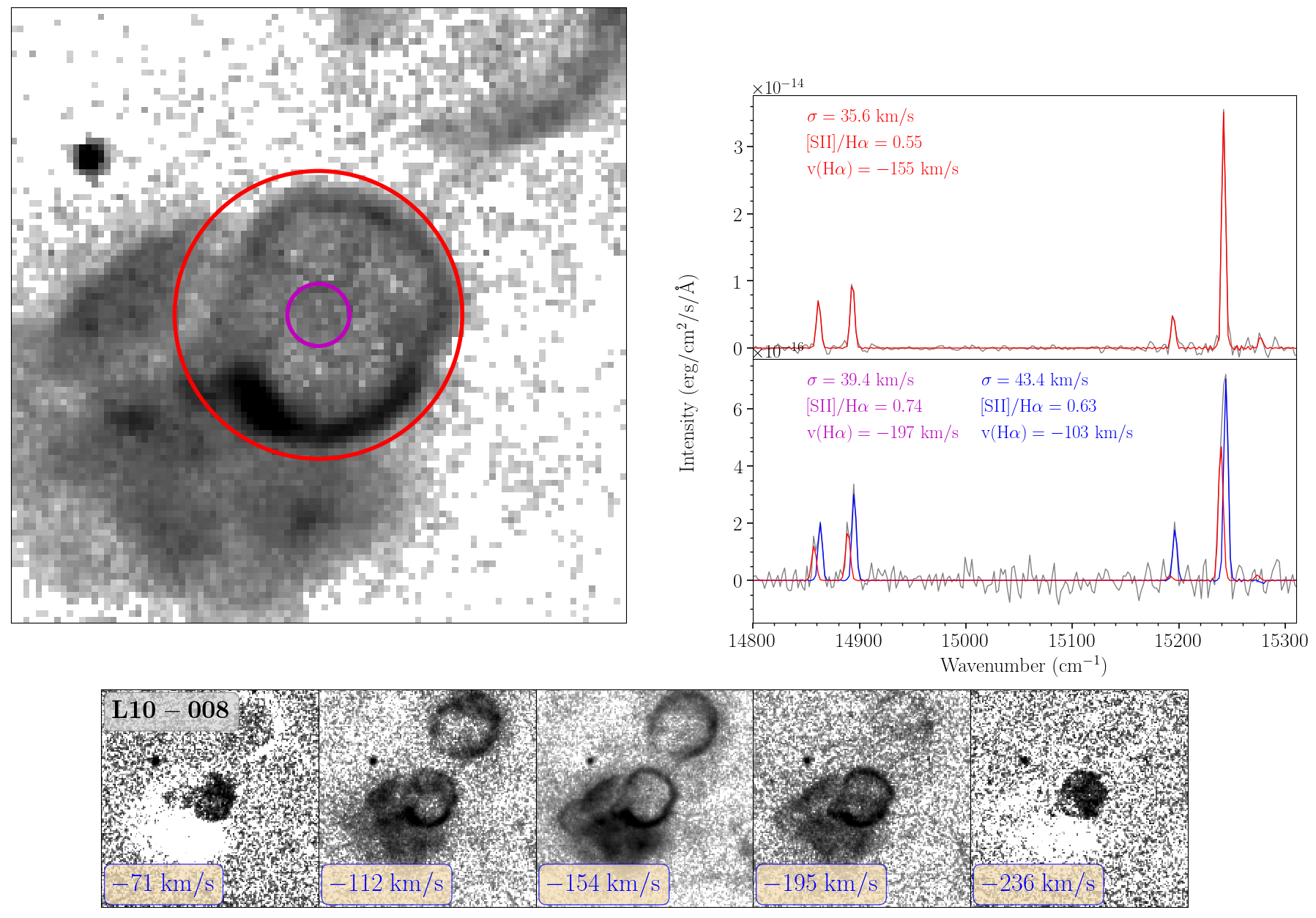}
    \caption{Spatial and velocity structure of L10-008. The upper left panel (35$''$ on a side) shows the integrated H$\alpha$ flux map with ellipses corresponding to regions chosen to represent the integrated spectrum (red) and a selected region of the interior (magenta). The corresponding spectra and their fits are shown on the upper right panel. The lower panel (50$''$ on a side) shows the individual frames within the datacube, centred on H$\alpha$, where the transition in radial velocity between the two outermost layers of the central part of the SNRs can be seen. }
    \label{fig:l10-008_2cv}
\end{figure*}

\begin{figure*}
	\includegraphics[width=\textwidth]{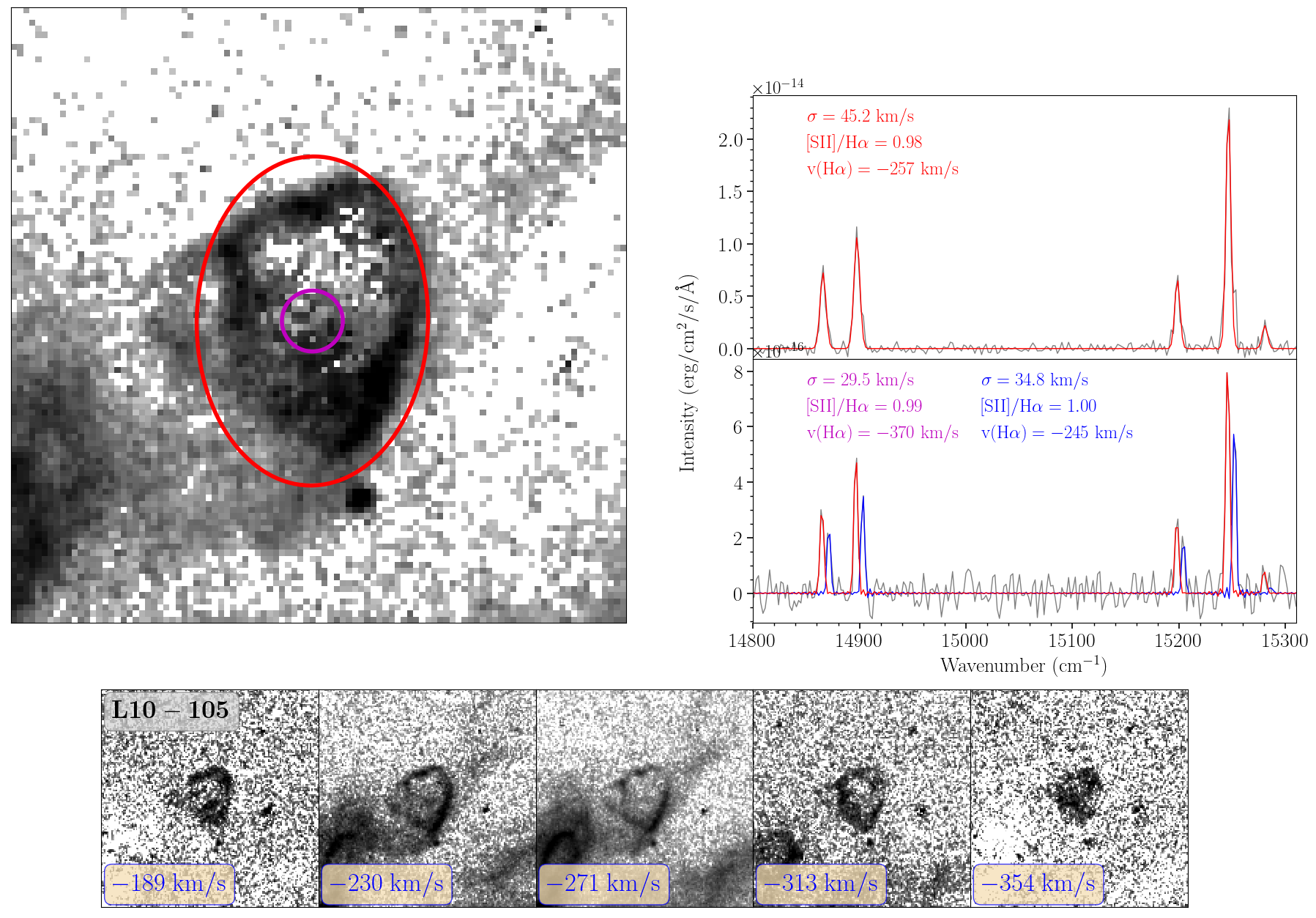}
    \caption{Same as Fig.~\ref{fig:l10-008_2cv}, for L10-105.}
    \label{fig:l10-105_2cv}
\end{figure*}

While a large velocity dispersion is a clear indication of an expansion in SNRs, a combination of a proper morphology, significant expansion velocity and appropriate spectral resolution can allow us to distinguish the two layers (the farthest and the closest) of these objects. A few SNRs in our fields could be studied this way, and we present here the best examples, namely L10-008 (Fig.~\ref{fig:l10-008_2cv}) and L10-105 (Fig.~\ref{fig:l10-105_2cv}), for which the approaching and receding velocity components can be clearly separated, both spectroscopically and spatially. The figures present the integrated H$\alpha$ map, the integrated spectrum (dominated by the outer ring), a spectrum of a selected inner region where the two distinct components are seen, as well as a series of images corresponding to the complete range of velocities seen in the data cube. The emission line profile of the integrated spectrum in the central part of the SNR can be resolved and shows two velocity components (separated by 94 km/s for L10-008 and 125 km/s for L10-105). Both the approaching and the receding caps of the expanding shell are clearly visible.

\subsection{Global properties of SNRs and the threshold for detection}
\label{sec:4_4_m33_doubt_det}

Table~\ref{tab:Table5} presents the values of some properties of SNRs in M33 from this work. Only 15 objects in our sample of 163 (9\%) have an integrated [SII]/H$\alpha$ ratio inferior to the canonical value of 0.4 (six objects have a ratio between 0.37 and 0.4). Such a high ''success rate'' is expected given that this ratio was originally used to select the SNRs (although not on an integrated value basis). But the situation is not as unequivocal when the integrated velocity dispersion of the objects is considered, nor the product of these two parameters, $\xi$. We also note significant differences between the L10 and L14 samples (we consider here in the L14 samples only the new SNRs which were not previously detected by L10). 

What would be the ideal discriminant between SNRs and HII regions (or superbubbles driven by stellar winds) if both the velocity dispersion and the [SII]/H$\alpha$ ratio are available? In NGC 6822, we saw (Fig. \ref{fig:siihasigma}) that $\xi$ = 12 clearly separates the individual pixels in Ho12 from those of the largest HII regions; this is also true for their integrated properties (Table \ref{tab:integrated_ngc6822}). But the sample is very small in this galaxy, and the HII regions studied here can be considered as giant, harbouring a large number of massive stars. The M33 data provide a better framework for this purpose. We have thus measured the integrated values of $\sigma$ and $\xi$ for dozens of HII regions and ring nebulae in our fields (Drissen et al., in prep.): excluding the central parts of giant HII regions NGC 588, NGC 592, NGC 595 and NGC 604 and all SNRs, $\sigma$ varies between 8 km/s and 20 km/s and $\xi$ between 1.5 and 9. The value of $\xi = 12$ based on the NGC 6822 data alone thus seems a bit too conservative and could be slightly reduced. Before examining the impact of varying the threshold value for $\xi$, we must emphasize important differences between the L10 and L14 samples.

While the number of SNRs in the L10 sample with no measurable integrated velocity dispersion in our data ($\sigma$ = 0) is low (8, or 7\%), this fraction rises to 33\% in the L14 sample. The average value of velocity dispersion in the L10 sample ($\overline{\sigma}_{L10}$ = 34 km/s) is in fact twice as large as that of the new candidates from the L14 sample ($\overline{\sigma}_{L14}$ = 17 km/s), although this criterion was never considered in either study. We find no correlation between the values of the [SII]/H$\alpha$ ratio (nor the velocity dispersion) and the origin of the supernova (core collapse or type Ia) as determined by \cite{2014ApJ...793..134L} based on the distance between the SNR and groups of OB stars. 

Table \ref{tab:xivar} presents the impact of the choice of $\xi$ on the fraction of SNRs in the L10 and L14 samples that do not meet the threshold value. Again, the dichotomy between the two samples is clear, with a large fraction of L14 candidates being excluded, even with a low threshold value. But the vast majority of these objects nevertheless have a [SII]/H$\alpha$ ratio larger than 0.4. They could be older SNRs.

\begin{landscape}
\begin{table}
\setlength{\tabcolsep}{2pt}
\scriptsize
\caption{Properties of SNRs in M33. Reddening corrected line fluxes relative to H$\beta$ = 100.}
\centering
\label{tab:Table5}
\begin{tabular}{l c c c c c c c c c c c c c c c}
\hline\hline  \\
(1) & (2) & (3) & (4) & (5) & (6) & (7) & (8) & (9) & (10) & (11) & (12) & (13) & (14) & (15) & (16) \\
   Name &  Field &          RA &          Dec & [SII]/H$\alpha$ & $\sigma$ & $\sigma$([SII]) & $\xi$ & Morphology &       F(H$\alpha$) &            [OII]3727 &         [OIII]5007 &            H$\alpha$ &        [NII]6584 &        [SII]6717 &       [SII]6717 \\
 ID & & & & & km\, s$^{-1}$ & km\, s$^{-1}$ & & &  10$^{-15}$ erg cm$^{-2}$  s$^{-1}$ &  & & & & &\\
\hline\\[-2ex]
L10-001 & Field8 & 01:32:30.41 & +30:27:41.00 &            0.76 &       50 &              49 &    38 &          A &  $\rm 71.32\pm 2.21$ &    $\rm 468.8\pm 28.4$ &    $\rm 51.0\pm 4.3$ &  $\rm 286.0\pm 14.2$ &  $\rm 13.3\pm 3.0$ &  $\rm 44.4\pm 3.1$ & $\rm 31.2\pm 2.5$ \\
L10-002 & Field7 & 01:32:31.26 & +30:35:30.79 &            0.60 &       28 &              29 &    17 &          C &  $\rm 17.56\pm 0.29$ &    $\rm 402.7\pm 24.2$ &  $\rm 278.6\pm 15.1$ &  $\rm 286.0\pm 15.4$ &  $\rm 14.7\pm 1.6$ &  $\rm 34.7\pm 1.6$ & $\rm 25.3\pm 1.4$ \\
L10-003 & Field8 & 01:32:42.54 & +30:20:58.00 &            0.56 &       31 &              33 &    17 &          A & $\rm 101.90\pm 1.40$ &    $\rm 663.0\pm 44.8$ &    $\rm 57.9\pm 4.0$ &  $\rm 286.0\pm 10.3$ &  $\rm 15.8\pm 1.4$ &  $\rm 32.7\pm 1.4$ & $\rm 23.0\pm 1.2$ \\
L10-004 & Field8 & 01:32:44.72 & +30:22:14.60 &            0.72 &       42 &              45 &    30 &          A &  $\rm 14.02\pm 0.42$ &    $\rm 805.4\pm 77.8$ &    $\rm 36.5\pm 4.6$ &  $\rm 286.0\pm 15.1$ &  $\rm 13.5\pm 2.8$ &  $\rm 41.9\pm 3.1$ & $\rm 29.9\pm 2.6$ \\
L10-005 & Field7 & 01:32:46.73 & +30:34:37.80 &            0.73 &       57 &              68 &    42 &          A &  $\rm 20.53\pm 1.00$ &    $\rm 999.8\pm 87.8$ &  $\rm 350.3\pm 27.4$ &  $\rm 286.0\pm 25.7$ &  $\rm 16.9\pm 4.4$ &  $\rm 42.4\pm 5.2$ & $\rm 31.0\pm 4.3$ \\
L10-006 & Field7 & 01:32:52.71 & +30:38:12.60 &            0.54 &       24 &              21 &    13 &          A &  $\rm 42.72\pm 0.74$ &    $\rm 883.0\pm 33.1$ &  $\rm 303.9\pm 11.0$ &  $\rm 286.0\pm 11.0$ &  $\rm 17.3\pm 1.5$ &  $\rm 30.4\pm 1.6$ & $\rm 23.2\pm 1.4$ \\
L10-008 & Field7 & 01:32:53.40 & +30:37:56.90 &            0.55 &       36 &              40 &    20 &          A &  $\rm 70.91\pm 1.04$ &    $\rm 559.7\pm 21.6$ &    $\rm 52.9\pm 2.4$ &   $\rm 286.0\pm 7.4$ &  $\rm 14.7\pm 1.4$ &  $\rm 32.1\pm 1.5$ & $\rm 23.0\pm 1.3$ \\
L10-009 & Field8 & 01:32:54.10 & +30:25:31.80 &            0.75 &       30 &              31 &    22 &          C &  $\rm 16.36\pm 0.37$ &    $\rm 832.1\pm 60.8$ &  $\rm 280.3\pm 16.4$ &  $\rm 286.0\pm 17.0$ &  $\rm 22.3\pm 2.4$ &  $\rm 43.9\pm 2.3$ & $\rm 30.7\pm 1.9$ \\
L10-011 & Field7 & 01:32:57.02 & +30:39:27.10 &            0.82 &       46 &              51 &    38 &         A2 &  $\rm 34.94\pm 0.78$ &    $\rm 632.9\pm 16.9$ &   $\rm 144.5\pm 2.7$ &   $\rm 286.0\pm 7.7$ &  $\rm 19.9\pm 2.5$ &  $\rm 43.4\pm 2.3$ & $\rm 38.3\pm 2.1$ \\
L10-012 & Field7 & 01:33:00.15 & +30:30:46.20 &            0.35 &       21 &              21 &     7 &          B & $\rm 447.08\pm 3.56$ &    $\rm 419.5\pm 14.2$ &   $\rm 175.1\pm 5.8$ &   $\rm 286.0\pm 8.4$ &  $\rm 12.6\pm 0.7$ &  $\rm 20.7\pm 0.8$ & $\rm 14.6\pm 0.7$ \\
L10-014 & Field7 & 01:33:00.57 & +30:31:00.67 &            0.54 &       27 &              27 &    15 &          B & $\rm 247.57\pm 2.31$ &    $\rm 483.7\pm 14.1$ &   $\rm 100.3\pm 3.3$ &   $\rm 286.0\pm 7.1$ &  $\rm 15.6\pm 0.8$ &  $\rm 31.5\pm 0.9$ & $\rm 22.7\pm 0.8$ \\
L10-015 & Field7 & 01:33:01.51 & +30:30:49.60 &            0.51 &       32 &              29 &    16 &          B &  $\rm 50.06\pm 0.63$ &    $\rm 586.2\pm 25.0$ &   $\rm 203.3\pm 8.3$ &  $\rm 286.0\pm 11.1$ &  $\rm 16.8\pm 1.1$ &  $\rm 29.8\pm 1.1$ & $\rm 21.4\pm 1.0$ \\
L10-016 & Field7 & 01:33:02.93 & +30:32:29.20 &            0.70 &       26 &              29 &    18 &          A &  $\rm 48.17\pm 0.73$ &    $\rm 639.3\pm 46.6$ &  $\rm 191.4\pm 10.1$ &  $\rm 286.0\pm 14.0$ &  $\rm 21.2\pm 1.4$ &  $\rm 41.7\pm 1.6$ & $\rm 28.6\pm 1.4$ \\
L10-017 & Field7 & 01:33:03.52 & +30:31:20.90 &            0.89 &       44 &              44 &    39 &          A &  $\rm 27.50\pm 0.59$ &    $\rm 823.6\pm 44.7$ &   $\rm 153.7\pm 7.8$ &  $\rm 286.0\pm 13.5$ &  $\rm 21.2\pm 2.0$ &  $\rm 53.2\pm 2.3$ & $\rm 36.0\pm 1.8$ \\
L10-018 & Field7 & 01:33:03.95 & +30:39:53.70 &            0.61 &       30 &              37 &    19 &          A &  $\rm 70.43\pm 0.95$ &    $\rm 620.9\pm 20.1$ &    $\rm 92.7\pm 1.6$ &   $\rm 286.0\pm 5.2$ &  $\rm 20.4\pm 1.4$ &  $\rm 35.1\pm 1.5$ & $\rm 26.3\pm 1.3$ \\
L10-020 & Field8 & 01:33:08.98 & +30:26:58.90 &            0.74 &       35 &              41 &    26 &          A &  $\rm 26.45\pm 0.61$ &    $\rm 619.9\pm 39.2$ &  $\rm 187.0\pm 12.1$ &  $\rm 286.0\pm 17.4$ &  $\rm 25.4\pm 2.6$ &  $\rm 43.4\pm 2.6$ & $\rm 30.1\pm 2.1$ \\
L10-021 & Field7 & 01:33:09.87 & +30:39:34.90 &            0.53 &       28 &              29 &    15 &          A & $\rm 111.47\pm 1.74$ &    $\rm 660.7\pm 38.3$ &   $\rm 149.5\pm 3.2$ &   $\rm 286.0\pm 6.8$ &  $\rm 24.4\pm 1.6$ &  $\rm 31.1\pm 1.6$ & $\rm 22.1\pm 1.3$ \\
L10-023 & Field7 & 01:33:11.19 & +30:39:42.96 &            0.71 &       55 &              62 &    38 &         A2 &  $\rm 13.12\pm 0.37$ &    $\rm 887.5\pm 46.4$ &   $\rm 235.2\pm 8.4$ &  $\rm 286.0\pm 12.4$ &  $\rm 23.2\pm 3.0$ &  $\rm 37.0\pm 2.8$ & $\rm 33.5\pm 2.6$ \\
L10-024 & Field7 & 01:33:11.28 & +30:34:23.50 &            0.47 &       20 &              16 &    10 &          A & $\rm 195.51\pm 2.29$ &    $\rm 464.5\pm 13.4$ &    $\rm 20.8\pm 1.7$ &   $\rm 286.0\pm 5.8$ &  $\rm 21.3\pm 1.1$ &  $\rm 27.5\pm 1.1$ & $\rm 19.7\pm 0.9$ \\
L10-025 & Field7 & 01:33:11.58 & +30:38:41.39 &            0.54 &       37 &              81 &    20 &          A &  $\rm 32.50\pm 0.71$ &    $\rm 496.1\pm 31.2$ &  $\rm 297.0\pm 12.0$ &  $\rm 286.0\pm 12.6$ &  $\rm 25.1\pm 4.0$ &  $\rm 27.8\pm 3.1$ & $\rm 25.7\pm 3.0$ \\
L10-026 & Field2 & 01:33:16.73 & +30:46:10.30 &            0.57 &       24 &              27 &    14 &          B &  $\rm 78.22\pm 0.82$ &    $\rm 464.0\pm 14.6$ &   $\rm 119.9\pm 4.6$ &   $\rm 286.0\pm 8.9$ &  $\rm 23.6\pm 1.0$ &  $\rm 33.3\pm 1.1$ & $\rm 24.0\pm 0.9$ \\
L10-027 & Field4 & 01:33:17.44 & +30:31:28.50 &            0.78 &       30 &              31 &    23 &          C &  $\rm 19.52\pm 0.44$ &    $\rm 211.2\pm 12.3$ &    $\rm 47.5\pm 5.9$ &  $\rm 267.0\pm 15.7$ &  $\rm 22.0\pm 2.1$ &  $\rm 46.5\pm 2.5$ & $\rm 31.7\pm 2.1$ \\
L10-029 & Field6 & 01:33:18.94 & +30:46:50.04 &            0.95 &       44 &              52 &    42 &          A &  $\rm 39.21\pm 1.48$ &    $\rm 464.2\pm 49.2$ &  $\rm 113.0\pm 13.2$ &  $\rm 286.0\pm 26.8$ &  $\rm 28.8\pm 3.8$ &  $\rm 57.2\pm 4.4$ & $\rm 38.0\pm 3.4$ \\
L10-030 & Field4 & 01:33:21.64 & +30:31:31.10 &            0.44 &       24 &              14 &    10 &          A &  $\rm 63.52\pm 1.37$ &     $\rm 232.7\pm 9.7$ &    $\rm 41.3\pm 4.0$ &  $\rm 245.1\pm 11.0$ &  $\rm 23.3\pm 2.0$ &  $\rm 26.4\pm 2.2$ & $\rm 17.6\pm 1.8$ \\
L10-033 & Field6 & 01:33:27.07 & +30:47:48.60 &            0.67 &       26 &              26 &    17 &          C &  $\rm 56.99\pm 0.89$ &    $\rm 242.6\pm 25.5$ &                   -- &  $\rm 286.0\pm 18.0$ &  $\rm 18.6\pm 1.5$ &  $\rm 39.4\pm 1.6$ & $\rm 27.6\pm 1.3$ \\
L10-034 & Field4 & 01:33:28.08 & +30:31:35.00 &            0.66 &       50 &              72 &    33 &          B &  $\rm 25.54\pm 0.49$ &    $\rm 391.9\pm 25.5$ &  $\rm 429.5\pm 27.2$ &  $\rm 259.3\pm 16.7$ &  $\rm 33.0\pm 2.4$ &  $\rm 38.3\pm 2.4$ & $\rm 27.9\pm 2.0$ \\
L10-035$^\dagger$ & Field6 & 01:33:28.99 & +30:47:42.80 &            0.29 &       22 &              21 &     6 &          C &   $\rm 5.38\pm 0.30$ &                     -- &                   -- &     $\rm 286.0\pm 96.7$ &  $\rm 15.4\pm 5.7$ &  $\rm 17.7\pm 5.3$ & $\rm 11.5\pm 4.5$ \\
L10-036 & Field2 & 01:33:29.05 & +30:42:17.00 &            1.06 &       74 &              88 &    78 &         A3 &  $\rm 67.35\pm 1.30$ &    $\rm 453.2\pm 17.1$ &   $\rm 160.0\pm 6.6$ &  $\rm 286.0\pm 11.5$ &  $\rm 47.7\pm 2.3$ &  $\rm 52.0\pm 2.0$ & $\rm 54.2\pm 2.1$ \\
L10-037 & Field6 & 01:33:29.45 & +30:49:10.80 &            0.74 &       52 &              63 &    38 &          B &  $\rm 13.33\pm 0.60$ &    $\rm 574.4\pm 71.3$ &  $\rm 305.8\pm 36.5$ &  $\rm 286.0\pm 34.7$ &  $\rm 32.3\pm 4.9$ &  $\rm 41.3\pm 4.8$ & $\rm 32.8\pm 4.2$ \\
L10-038 & Field6 & 01:33:30.21 & +30:47:43.80 &            0.42 &       20 &              21 &     8 &          C &  $\rm 55.88\pm 0.55$ &    $\rm 204.2\pm 18.7$ &    $\rm 90.0\pm 7.0$ &  $\rm 286.0\pm 15.3$ &  $\rm 17.6\pm 0.9$ &  $\rm 25.1\pm 1.0$ & $\rm 17.0\pm 0.8$ \\
L10-039 & Field4 & 01:33:31.20 & +30:33:33.40 &            0.98 &      116 &             139 &   113 &         A3 &  $\rm 76.43\pm 1.60$ &    $\rm 443.7\pm 23.2$ &  $\rm 445.9\pm 23.5$ &  $\rm 286.0\pm 16.0$ &  $\rm 63.4\pm 2.6$ &  $\rm 43.6\pm 2.0$ & $\rm 54.1\pm 2.3$ \\
L10-040 & Field2 & 01:33:31.34 & +30:42:18.30 &            0.30 &       20 &              17 &     6 &          A &  $\rm 63.29\pm 0.90$ &     $\rm 256.5\pm 9.5$ &   $\rm 104.6\pm 5.0$ &  $\rm 286.0\pm 10.8$ &  $\rm 22.2\pm 1.4$ &  $\rm 17.3\pm 1.4$ & $\rm 12.7\pm 1.2$ \\
L10-041 & Field4 & 01:33:31.80 & +30:31:01.10 &            0.96 &       27 &              25 &    26 &          A &  $\rm 22.82\pm 1.50$ &    $\rm 272.6\pm 36.0$ &   $\rm 74.3\pm 12.0$ &  $\rm 244.9\pm 29.1$ &  $\rm 25.1\pm 6.2$ &  $\rm 57.0\pm 7.6$ & $\rm 39.1\pm 6.2$ \\
L10-043 & Field2 & 01:33:35.39 & +30:42:32.10 &            0.49 &       26 &              25 &    13 &          C &  $\rm 29.29\pm 0.35$ &    $\rm 318.8\pm 12.3$ &    $\rm 59.0\pm 4.1$ &  $\rm 286.0\pm 10.6$ &  $\rm 26.6\pm 1.1$ &  $\rm 28.1\pm 1.2$ & $\rm 21.4\pm 1.1$ \\
L10-044 & Field6 & 01:33:35.63 & +30:49:22.63 &            0.95 &       56 &              55 &    53 &          A &  $\rm 22.82\pm 0.60$ &    $\rm 617.8\pm 50.4$ &   $\rm 100.7\pm 9.7$ &  $\rm 286.0\pm 20.9$ &  $\rm 31.9\pm 2.8$ &  $\rm 54.0\pm 2.7$ & $\rm 40.6\pm 2.3$ \\
L10-045 & Field4 & 01:33:36.09 & +30:36:27.01 &            0.83 &       54 &              70 &    45 &         A2 & $\rm 144.35\pm 1.83$ &     $\rm 352.1\pm 7.9$ &   $\rm 221.1\pm 5.0$ &   $\rm 286.0\pm 6.9$ &  $\rm 41.3\pm 1.5$ &  $\rm 46.0\pm 1.5$ & $\rm 37.4\pm 1.3$ \\
L10-046 & Field4 & 01:33:37.09 & +30:32:53.50 &            0.99 &       48 &              52 &    48 &          A &  $\rm 18.32\pm 0.24$ &    $\rm 475.6\pm 27.3$ &   $\rm 105.5\pm 7.3$ &  $\rm 286.0\pm 14.7$ &  $\rm 38.2\pm 1.5$ &  $\rm 58.2\pm 1.5$ & $\rm 40.6\pm 1.3$ \\
L10-047 & Field2 & 01:33:37.75 & +30:40:09.20 &            0.96 &       41 &              40 &    39 &          A &  $\rm 22.45\pm 0.69$ &  $\rm 1279.4\pm 136.6$ &  $\rm 236.6\pm 27.0$ &  $\rm 286.0\pm 31.3$ &  $\rm 41.9\pm 3.4$ &  $\rm 55.7\pm 3.4$ & $\rm 40.4\pm 2.8$ \\
L10-048 & Field2 & 01:33:38.01 & +30:42:18.20 &            0.53 &       21 &              17 &    11 &          A &  $\rm 16.70\pm 0.22$ &    $\rm 233.2\pm 10.1$ &    $\rm 22.1\pm 4.2$ &  $\rm 286.0\pm 11.6$ &  $\rm 26.8\pm 1.2$ &  $\rm 31.2\pm 1.3$ & $\rm 22.2\pm 1.1$ \\
L10-049 & Field2 & 01:33:40.66 & +30:39:40.80 &            0.79 &       33 &              40 &    26 &          A &  $\rm 34.53\pm 0.68$ &    $\rm 887.9\pm 71.3$ &  $\rm 154.8\pm 14.3$ &  $\rm 286.0\pm 22.8$ &  $\rm 37.8\pm 2.0$ &  $\rm 46.0\pm 2.2$ & $\rm 32.9\pm 1.9$ \\
L10-050 & Field2 & 01:33:40.73 & +30:42:35.70 &            0.60 &       20 &              17 &    12 &          C & $\rm 110.80\pm 1.10$ &    $\rm 343.6\pm 10.3$ &    $\rm 41.4\pm 3.0$ &   $\rm 286.0\pm 8.5$ &  $\rm 31.8\pm 1.0$ &  $\rm 35.0\pm 1.0$ & $\rm 25.0\pm 0.9$ \\
L10-051 & Field6 & 01:33:40.91 & +30:52:14.49 &            0.72 &       29 &              33 &    21 &          A &  $\rm 39.25\pm 0.92$ &    $\rm 346.7\pm 38.2$ &   $\rm 84.5\pm 11.5$ &  $\rm 286.0\pm 27.1$ &  $\rm 24.7\pm 2.4$ &  $\rm 42.5\pm 2.5$ & $\rm 30.0\pm 2.1$ \\
L10-052 & Field4 & 01:33:41.30 & +30:32:28.40 &            0.48 &       27 &              23 &    13 &          C &  $\rm 15.99\pm 0.26$ &    $\rm 329.8\pm 23.2$ &    $\rm 53.9\pm 6.3$ &  $\rm 286.0\pm 16.4$ &  $\rm 23.5\pm 1.5$ &  $\rm 28.5\pm 1.7$ & $\rm 19.5\pm 1.4$ \\
L10-053 & Field9 & 01:33:41.69 & +30:21:04.10 &            0.57 &       32 &              32 &    18 &          A &  $\rm 73.98\pm 1.07$ &    $\rm 637.5\pm 19.9$ &   $\rm 114.5\pm 4.3$ &   $\rm 286.0\pm 9.2$ &  $\rm 17.3\pm 1.4$ &  $\rm 33.9\pm 1.4$ & $\rm 23.3\pm 1.2$ \\
L10-054 & Field9 & 01:33:42.20 & +30:20:58.00 &            0.77 &       37 &              45 &    29 &          A &  $\rm 43.42\pm 1.13$ &    $\rm 587.2\pm 26.5$ &    $\rm 82.3\pm 5.4$ &  $\rm 286.0\pm 14.3$ &  $\rm 18.9\pm 3.0$ &  $\rm 45.8\pm 3.0$ & $\rm 30.8\pm 2.4$ \\
L10-055 & Field2 & 01:33:42.91 & +30:41:49.50 &            0.53 &       17 &              16 &     9 &          C &  $\rm 51.99\pm 0.53$ &    $\rm 267.6\pm 11.3$ &    $\rm 23.5\pm 4.2$ &  $\rm 286.0\pm 11.0$ &  $\rm 25.2\pm 1.0$ &  $\rm 30.7\pm 1.1$ & $\rm 22.2\pm 0.9$ \\
L10-056 & Field2 & 01:33:43.49 & +30:41:03.80 &            0.72 &       29 &              36 &    21 &          B &  $\rm 15.54\pm 0.25$ &    $\rm 606.0\pm 37.6$ &    $\rm 59.4\pm 6.3$ &  $\rm 286.0\pm 16.1$ &  $\rm 36.0\pm 1.8$ &  $\rm 41.6\pm 1.8$ & $\rm 30.7\pm 1.5$ \\
L10-057 & Field4 & 01:33:43.70 & +30:36:11.50 &            0.59 &       17 &              21 &    10 &          C &  $\rm 36.48\pm 0.38$ &     $\rm 259.6\pm 7.8$ &    $\rm 38.9\pm 2.8$ &   $\rm 286.0\pm 7.9$ &  $\rm 27.8\pm 1.1$ &  $\rm 35.0\pm 1.2$ & $\rm 24.3\pm 1.0$ \\
L10-058 & Field4 & 01:33:45.26 & +30:32:20.10 &            0.59 &       23 &              14 &    13 &          A &  $\rm 86.78\pm 0.70$ &     $\rm 269.4\pm 6.8$ &    $\rm 14.4\pm 2.2$ &   $\rm 286.0\pm 6.5$ &  $\rm 27.9\pm 0.8$ &  $\rm 34.7\pm 0.8$ & $\rm 24.5\pm 0.7$ \\
L10-059$^\dagger$ & Field2 & 01:33:47.46 & +30:39:44.70 & 0.99 &       36 &              44 &    36 &          A &  $\rm 29.75\pm 0.86$ &      $\rm < 1016.6 \pm 341.0$ & $\rm < 200.0 \pm 65.7$ & $\rm 286.0 \pm 95.7$ &  $\rm 51.1\pm 3.2$ &  $\rm 56.4\pm 3.5$ & $\rm 42.7\pm 3.0$ \\
L10-060 & Field2 & 01:33:48.35 & +30:39:28.40 &            0.67 &       30 &              40 &    20 &          C &  $\rm 20.92\pm 0.27$ &    $\rm 484.3\pm 26.8$ &    $\rm 80.3\pm 7.2$ &  $\rm 286.0\pm 15.5$ &  $\rm 40.9\pm 1.5$ &  $\rm 38.3\pm 1.5$ & $\rm 28.5\pm 1.3$ \\
L10-061 & Field4 & 01:33:48.50 & +30:33:07.90 &            0.73 &       22 &              24 &    16 &          B & $\rm 125.10\pm 1.13$ &     $\rm 285.7\pm 6.1$ &    $\rm 59.9\pm 2.3$ &   $\rm 286.0\pm 6.4$ &  $\rm 31.6\pm 0.9$ &  $\rm 42.8\pm 1.0$ & $\rm 29.8\pm 0.8$ \\
L10-062 & Field4 & 01:33:49.75 & +30:30:49.70 &            0.68 &        0 &               0 &     0 &          A &  $\rm 23.22\pm 0.42$ &    $\rm 289.1\pm 17.7$ &    $\rm 36.0\pm 5.7$ &  $\rm 267.0\pm 15.1$ &  $\rm 24.0\pm 1.8$ &  $\rm 40.5\pm 2.0$ & $\rm 27.9\pm 2.0$ \\
\hline\\[-2ex]
\end{tabular}
\begin{tablenotes}
       \item Continued.
\end{tablenotes}
\end{table}
\end{landscape}

\newpage
\clearpage

\begin{landscape}
\begin{table}
\setlength{\tabcolsep}{2pt}
\scriptsize
\centering
\begin{tabular}{l c c c c c c c c c c c c c c c}
\hline\hline  \\
(1) & (2) & (3) & (4) & (5) & (6) & (7) & (8) & (9) & (10) & (11) & (12) & (13) & (14) & (15) & (16) \\
   Name &  Field &          RA &          Dec & [SII]/H$\alpha$ & $\sigma$ & $\sigma$([SII]) & $\xi$ & Morphology &       F(H$\alpha$) &            [OII]3727 &         [OIII]5007 &            H$\alpha$ &        [NII]6584 &        [SII]6717 &       [SII]6717 \\
 ID & & & & & km\, s$^{-1}$ & km\, s$^{-1}$ & & &  10$^{-15}$ erg cm$^{-2}$  s$^{-1}$ &  & & & & &\\
\hline\\[-2ex]
L10-063 & Field4 & 01:33:49.90 & +30:30:16.70 &            1.09 &       28 &              13 &    30 &          A &   $\rm 4.57\pm 0.40$ &   $\rm 561.3\pm 115.9$ &  $\rm 249.4\pm 53.1$ &  $\rm 236.8\pm 49.9$ & $\rm 55.9\pm 11.1$ & $\rm 65.9\pm 10.2$ & $\rm 42.7\pm 7.9$ \\
L10-064 & Field4 & 01:33:50.12 & +30:35:28.60 &            0.79 &       34 &              40 &    27 &          B &  $\rm 27.23\pm 0.48$ &    $\rm 465.9\pm 25.8$ &  $\rm 224.8\pm 13.2$ &  $\rm 286.0\pm 16.0$ &  $\rm 37.7\pm 1.9$ &  $\rm 45.4\pm 2.0$ & $\rm 33.6\pm 1.7$ \\
L10-065 & Field2 & 01:33:51.06 & +30:43:56.20 &            0.55 &       24 &              21 &    13 &          B &  $\rm 90.09\pm 0.63$ &     $\rm 348.1\pm 7.1$ &    $\rm 32.2\pm 1.8$ &   $\rm 286.0\pm 5.4$ &  $\rm 31.7\pm 0.7$ &  $\rm 32.0\pm 0.7$ & $\rm 23.2\pm 0.6$ \\
L10-066 & Field4 & 01:33:51.67 & +30:30:59.60 &            1.19 &       55 &              46 &    65 &          A &  $\rm 13.49\pm 0.52$ &    $\rm 457.5\pm 51.1$ &  $\rm 135.7\pm 19.0$ &  $\rm 286.0\pm 33.2$ &  $\rm 33.6\pm 3.8$ &  $\rm 67.4\pm 4.4$ & $\rm 51.3\pm 3.7$ \\
L10-067 & Field4 & 01:33:51.61 & +30:30:42.64 &            0.79 &       23 &              34 &    18 &          C &   $\rm 9.76\pm 0.25$ &    $\rm 344.3\pm 23.5$ &    $\rm 55.1\pm 5.9$ &  $\rm 244.2\pm 14.0$ &  $\rm 33.6\pm 2.9$ &  $\rm 48.3\pm 3.1$ & $\rm 30.4\pm 2.5$ \\
L10-068 & Field5 & 01:33:52.15 & +30:56:33.40 &            0.48 &       20 &              17 &    10 &          A & $\rm 175.07\pm 2.03$ &    $\rm 270.6\pm 16.2$ &  $\rm 384.5\pm 21.5$ &  $\rm 286.0\pm 15.8$ &  $\rm 20.4\pm 1.1$ &  $\rm 27.9\pm 1.1$ & $\rm 19.7\pm 0.9$ \\
L10-069 & Field4 & 01:33:54.28 & +30:33:47.90 &            0.83 &       31 &              32 &    25 &          A &  $\rm 52.43\pm 0.64$ &     $\rm 313.3\pm 8.1$ &    $\rm 37.2\pm 2.4$ &   $\rm 286.0\pm 7.3$ &  $\rm 28.5\pm 1.3$ &  $\rm 48.8\pm 1.4$ & $\rm 34.0\pm 1.2$ \\
L10-070 & Field2 & 01:33:54.49 & +30:45:18.70 &            0.77 &       37 &              47 &    28 &          B &  $\rm 88.19\pm 0.71$ &     $\rm 402.5\pm 7.0$ &    $\rm 71.1\pm 2.0$ &   $\rm 286.0\pm 5.2$ &  $\rm 37.0\pm 0.9$ &  $\rm 44.5\pm 0.9$ & $\rm 32.4\pm 0.8$ \\
L10-071 & Field4 & 01:33:54.91 & +30:33:11.00 &            1.00 &       73 &              94 &    73 &         A3 &  $\rm 78.17\pm 1.34$ &    $\rm 439.8\pm 12.8$ &   $\rm 203.9\pm 6.3$ &   $\rm 267.1\pm 8.7$ &  $\rm 41.2\pm 2.0$ &  $\rm 52.9\pm 2.0$ & $\rm 47.3\pm 1.9$ \\
L10-072$^\dagger$ & Field2 & 01:33:55.01 & +30:39:57.30 &  1.33 &       13 &              22 &    17 &          B &  $\rm 11.28\pm 0.73$ &    $\rm < 813.6 \pm 273.3$ & $\rm < 97.4 \pm 36.3$ & $\rm 286.0 \pm 97.1$ &  $\rm 75.8\pm 9.0$ &  $\rm 75.8\pm 8.3$ & $\rm 57.0\pm 7.0$ \\
L10-073 & Field9 & 01:33:56.49 & +30:21:27.00 &            0.84 &       48 &              49 &    40 &          A &  $\rm 27.85\pm 0.78$ &    $\rm 890.5\pm 54.4$ &   $\rm 118.7\pm 7.9$ &  $\rm 286.0\pm 16.6$ &  $\rm 20.9\pm 2.5$ &  $\rm 50.2\pm 3.0$ & $\rm 34.2\pm 2.3$ \\
L10-074 & Field4 & 01:33:56.97 & +30:34:58.70 &            0.93 &       36 &              41 &    33 &          B &  $\rm 24.72\pm 0.49$ &    $\rm 419.9\pm 13.3$ &   $\rm 106.8\pm 4.3$ &  $\rm 286.0\pm 10.1$ &  $\rm 33.2\pm 2.0$ &  $\rm 53.9\pm 2.3$ & $\rm 39.4\pm 2.0$ \\
L10-075 & Field2 & 01:33:57.13 & +30:40:48.50 &            1.01 &       22 &              33 &    22 &          A &  $\rm 24.29\pm 1.05$ &  $\rm 1584.4\pm 324.7$ &   $\rm 64.3\pm 23.8$ &  $\rm 286.0\pm 58.3$ &  $\rm 45.9\pm 5.3$ &  $\rm 58.5\pm 5.3$ & $\rm 42.1\pm 4.5$ \\
L10-076 & Field4 & 01:33:57.13 & +30:35:06.10 &            0.49 &       19 &              30 &     9 &          B &  $\rm 19.74\pm 0.27$ &    $\rm 410.9\pm 10.7$ &   $\rm 166.8\pm 4.8$ &   $\rm 286.0\pm 8.1$ &  $\rm 33.9\pm 1.4$ &  $\rm 28.6\pm 1.5$ & $\rm 20.0\pm 1.3$ \\
L10-077 & Field4 & 01:33:58.06 & +30:32:09.60 &            0.44 &        0 &               0 &     0 &          C &  $\rm 26.19\pm 0.21$ &     $\rm 190.5\pm 5.3$ &    $\rm 20.0\pm 2.6$ &   $\rm 286.0\pm 7.1$ &  $\rm 22.8\pm 0.8$ &  $\rm 26.5\pm 0.9$ & $\rm 17.9\pm 0.9$ \\
L10-078 & Field2 & 01:33:58.13 & +30:37:53.04 &            1.13 &       45 &              46 &    50 &         A2 &  $\rm 18.01\pm 0.45$ &   $\rm 1017.9\pm 70.1$ &  $\rm 206.9\pm 16.0$ &  $\rm 286.0\pm 20.8$ &  $\rm 53.5\pm 2.9$ &  $\rm 64.2\pm 2.9$ & $\rm 48.6\pm 2.4$ \\
L10-079 & Field5 & 01:33:58.15 & +30:48:36.40 &            0.55 &       28 &              27 &    15 &          A &  $\rm 10.23\pm 0.24$ &    $\rm 247.1\pm 14.5$ &    $\rm 22.8\pm 4.3$ &  $\rm 286.0\pm 13.7$ &  $\rm 26.5\pm 2.4$ &  $\rm 32.4\pm 2.2$ & $\rm 22.1\pm 1.9$ \\
L10-080 & Field4 & 01:33:58.42 & +30:36:24.30 &            1.14 &       61 &              54 &    70 &         A3 &   $\rm 3.65\pm 0.17$ &  $\rm 1108.7\pm 138.8$ &  $\rm 281.7\pm 31.7$ &  $\rm 286.0\pm 33.1$ &  $\rm 50.4\pm 4.9$ &  $\rm 62.9\pm 5.1$ & $\rm 50.8\pm 4.4$ \\
L10-081 & Field4 & 01:33:58.51 & +30:33:32.30 &            0.68 &       30 &              37 &    20 &          C &  $\rm 13.87\pm 0.32$ &    $\rm 554.1\pm 29.9$ &   $\rm 167.8\pm 8.9$ &  $\rm 286.0\pm 14.6$ &  $\rm 31.9\pm 2.4$ &  $\rm 40.3\pm 2.6$ & $\rm 27.4\pm 2.2$ \\
L10-082 & Field5 & 01:33:58.52 & +30:51:54.30 &            0.86 &       45 &              65 &    38 &          A &  $\rm 30.57\pm 1.37$ &    $\rm 524.9\pm 55.4$ &  $\rm 427.2\pm 45.1$ &  $\rm 286.0\pm 31.9$ &  $\rm 37.2\pm 5.3$ &  $\rm 53.1\pm 5.7$ & $\rm 32.5\pm 4.3$ \\
L10-084 & Field2 & 01:34:00.31 & +30:42:19.40 &            1.13 &       92 &              87 &   103 &          A &  $\rm 22.34\pm 0.84$ &    $\rm 859.4\pm 94.3$ &  $\rm 442.6\pm 48.9$ &  $\rm 286.0\pm 32.7$ &  $\rm 57.2\pm 3.9$ &  $\rm 62.9\pm 4.0$ & $\rm 50.1\pm 3.4$ \\
L10-085 & Field2 & 01:34:00.32 & +30:47:24.10 &            0.97 &       52 &              51 &    50 &          B &  $\rm 19.26\pm 0.50$ &    $\rm 922.6\pm 99.6$ &  $\rm 358.6\pm 39.9$ &  $\rm 286.0\pm 31.4$ &  $\rm 42.4\pm 2.7$ &  $\rm 55.9\pm 2.8$ & $\rm 41.0\pm 2.3$ \\
L10-086 & Field5 & 01:34:00.53 & +30:49:02.71 &            0.60 &       46 &              43 &    28 &          C &   $\rm 0.81\pm 0.08$ &    $\rm 375.9\pm 64.2$ &  $\rm 123.2\pm 25.0$ &  $\rm 286.0\pm 52.6$ &  $\rm 27.6\pm 6.9$ &  $\rm 27.7\pm 7.8$ & $\rm 32.8\pm 8.6$ \\
L10-087 & Field3 & 01:34:01.34 & +30:35:20.20 &            0.88 &       29 &              38 &    26 &          A &  $\rm 40.87\pm 1.14$ &    $\rm 852.4\pm 63.9$ &    $\rm 59.4\pm 7.7$ &  $\rm 286.0\pm 20.5$ &  $\rm 31.6\pm 3.2$ &  $\rm 50.4\pm 3.3$ & $\rm 37.6\pm 2.8$ \\
L10-088 & Field3 & 01:34:02.24 & +30:31:06.80 &            0.91 &       41 &              46 &    37 &          A &  $\rm 41.90\pm 1.28$ &   $\rm 1005.5\pm 61.7$ &   $\rm 132.7\pm 9.1$ &  $\rm 286.0\pm 17.9$ &  $\rm 31.4\pm 3.7$ &  $\rm 51.0\pm 3.4$ & $\rm 39.8\pm 3.0$ \\
L10-089 & Field3 & 01:34:03.31 & +30:36:22.90 &            0.58 &       28 &              25 &    16 &          A & $\rm 284.02\pm 6.09$ &    $\rm 499.4\pm 18.1$ &    $\rm 46.1\pm 3.1$ &  $\rm 286.0\pm 10.0$ &  $\rm 26.6\pm 2.2$ &  $\rm 32.9\pm 2.1$ & $\rm 24.6\pm 1.9$ \\
L10-090 & Field1 & 01:34:03.48 & +30:44:43.80 &            0.99 &       47 &              41 &    46 &          A &  $\rm 13.78\pm 0.79$ &    $\rm 786.6\pm 97.5$ &  $\rm 669.7\pm 83.2$ &  $\rm 286.0\pm 38.7$ &  $\rm 51.7\pm 5.8$ &  $\rm 59.3\pm 5.9$ & $\rm 39.9\pm 4.8$ \\
L10-091 & Field3 & 01:34:04.26 & +30:32:57.10 &            0.87 &       53 &              48 &    46 &          B &   $\rm 7.83\pm 0.60$ & $\rm 4220.7\pm 1032.1$ & $\rm 621.7\pm 149.8$ &  $\rm 286.0\pm 71.5$ &  $\rm 30.2\pm 5.9$ &  $\rm 50.2\pm 7.7$ & $\rm 37.2\pm 6.5$ \\
L10-092 & Field3 & 01:34:07.23 & +30:36:22.00 &            0.80 &       25 &              23 &    20 &          A & $\rm 162.24\pm 3.93$ &    $\rm 565.2\pm 23.9$ &    $\rm 56.1\pm 3.7$ &  $\rm 286.0\pm 11.5$ &  $\rm 28.3\pm 2.5$ &  $\rm 45.6\pm 2.5$ & $\rm 34.7\pm 2.2$ \\
L10-093 & Field3 & 01:34:07.50 & +30:37:08.00 &            0.95 &       29 &              33 &    27 &          B &   $\rm 3.94\pm 0.21$ &  $\rm 3612.9\pm 840.9$ & $\rm 595.8\pm 135.8$ &  $\rm 286.0\pm 66.0$ &  $\rm 33.8\pm 6.3$ &  $\rm 50.2\pm 5.8$ & $\rm 44.4\pm 5.4$ \\
L10-094 & Field1 & 01:34:08.37 & +30:46:33.20 &            0.69 &       24 &              43 &    17 &          C &   $\rm 7.88\pm 0.31$ &    $\rm 427.6\pm 41.8$ &  $\rm 377.9\pm 35.7$ &  $\rm 286.0\pm 28.5$ &  $\rm 39.9\pm 4.7$ &  $\rm 39.3\pm 4.0$ & $\rm 29.7\pm 3.6$ \\
L10-095 & Field1 & 01:34:10.02 & +30:47:14.90 &            0.90 &       65 &              71 &    58 &          A &   $\rm 8.26\pm 0.35$ &    $\rm 599.2\pm 48.9$ &  $\rm 452.5\pm 36.8$ &  $\rm 286.0\pm 25.7$ &  $\rm 37.4\pm 4.0$ &  $\rm 51.7\pm 4.4$ & $\rm 38.0\pm 3.8$ \\
L10-096 & Field1 & 01:34:10.70 & +30:42:24.00 &            1.29 &       57 &              53 &    74 &          A &  $\rm 49.83\pm 0.88$ &     $\rm 274.5\pm 6.9$ &   $\rm 126.5\pm 3.9$ &   $\rm 286.0\pm 8.6$ &  $\rm 42.9\pm 1.9$ &  $\rm 71.4\pm 1.9$ & $\rm 57.5\pm 1.7$ \\
L10-097 & Field3 & 01:34:11.04 & +30:38:59.90 &            0.81 &       47 &              55 &    38 &          B &   $\rm 8.90\pm 0.30$ &    $\rm 840.0\pm 91.3$ &  $\rm 166.2\pm 19.4$ &  $\rm 286.0\pm 30.0$ &  $\rm 33.5\pm 4.1$ &  $\rm 44.7\pm 3.8$ & $\rm 36.2\pm 3.4$ \\
L10-098 & Field3 & 01:34:12.69 & +30:35:12.00 &            0.44 &        0 &               0 &     0 &          A &  $\rm 90.76\pm 1.39$ &    $\rm 425.4\pm 18.3$ &    $\rm 96.0\pm 4.5$ &  $\rm 286.0\pm 10.3$ &  $\rm 18.7\pm 1.5$ &  $\rm 24.4\pm 1.6$ & $\rm 19.3\pm 1.6$ \\
L10-099 & Field5 & 01:34:13.11 & +30:48:33.47 &            0.64 &       27 &              28 &    17 &          C &  $\rm 91.26\pm 1.10$ &     $\rm 199.8\pm 7.8$ &    $\rm 38.2\pm 3.3$ &   $\rm 286.0\pm 9.3$ &  $\rm 26.6\pm 1.2$ &  $\rm 37.4\pm 1.2$ & $\rm 26.2\pm 1.0$ \\
L10-100 & Field1 & 01:34:13.65 & +30:43:27.00 &            0.86 &       56 &              64 &    48 &          A &   $\rm 5.11\pm 0.23$ &    $\rm 579.8\pm 95.7$ &  $\rm 509.7\pm 81.8$ &  $\rm 286.0\pm 46.8$ &  $\rm 38.1\pm 4.8$ &  $\rm 52.4\pm 4.8$ & $\rm 33.2\pm 3.9$ \\
L10-101 & Field1 & 01:34:13.71 & +30:48:17.50 &            0.74 &       27 &              24 &    20 &          B &  $\rm 54.32\pm 1.10$ &     $\rm 297.8\pm 8.3$ &    $\rm 34.3\pm 2.4$ &   $\rm 286.0\pm 8.7$ &  $\rm 30.3\pm 2.2$ &  $\rm 44.8\pm 2.0$ & $\rm 29.5\pm 1.7$ \\
L10-102 & Field3 & 01:34:14.10 & +30:34:30.90 &            0.50 &       32 &              30 &    16 &          B & $\rm 122.34\pm 2.45$ &    $\rm 357.3\pm 11.9$ &    $\rm 49.2\pm 3.1$ &   $\rm 286.0\pm 9.8$ &  $\rm 16.7\pm 1.9$ &  $\rm 28.2\pm 2.0$ & $\rm 22.0\pm 1.8$ \\
L10-103 & Field1 & 01:34:14.35 & +30:41:53.60 &            0.87 &       36 &              33 &    31 &          A &  $\rm 11.29\pm 0.32$ &    $\rm 436.1\pm 46.2$ &  $\rm 226.3\pm 25.8$ &  $\rm 286.0\pm 30.4$ &  $\rm 39.9\pm 2.9$ &  $\rm 50.7\pm 2.8$ & $\rm 35.9\pm 2.4$ \\
L10-104 & Field3 & 01:34:14.38 & +30:39:41.60 &            1.07 &       45 &              45 &    48 &          A &  $\rm 21.29\pm 0.70$ &  $\rm 1818.5\pm 178.0$ &  $\rm 143.3\pm 16.1$ &  $\rm 286.0\pm 27.9$ &  $\rm 32.3\pm 3.5$ &  $\rm 61.3\pm 3.7$ & $\rm 45.3\pm 3.1$ \\
L10-105 & Field5 & 01:34:14.36 & +30:53:51.90 &            0.98 &       45 &              56 &    44 &          A &  $\rm 57.29\pm 1.74$ &    $\rm 447.1\pm 25.2$ &   $\rm 159.5\pm 9.4$ &  $\rm 286.0\pm 16.6$ &  $\rm 30.6\pm 3.2$ &  $\rm 58.2\pm 3.6$ & $\rm 39.7\pm 2.9$ \\
L10-106 & Field3 & 01:34:14.67 & +30:31:50.90 &            0.93 &       33 &              36 &    30 &          A &  $\rm 18.77\pm 0.70$ &  $\rm 1063.0\pm 196.3$ &  $\rm 462.6\pm 83.4$ &  $\rm 286.0\pm 51.4$ &  $\rm 37.0\pm 4.2$ &  $\rm 50.5\pm 4.1$ & $\rm 42.0\pm 3.8$ \\
L10-107 & Field3 & 01:34:15.57 & +30:32:59.90 &            0.82 &       40 &              50 &    33 &          A &  $\rm 20.84\pm 0.60$ &   $\rm 847.6\pm 108.7$ &  $\rm 273.0\pm 36.6$ &  $\rm 286.0\pm 37.0$ &  $\rm 31.5\pm 3.6$ &  $\rm 46.2\pm 3.3$ & $\rm 35.8\pm 2.9$ \\
L10-108 & Field5 & 01:34:16.31 & +30:52:32.70 &            0.57 &       27 &              28 &    15 &          A & $\rm 150.10\pm 1.79$ &     $\rm 204.3\pm 8.5$ &    $\rm 68.7\pm 4.1$ &  $\rm 286.0\pm 10.2$ &  $\rm 22.1\pm 1.2$ &  $\rm 33.5\pm 1.2$ & $\rm 23.0\pm 1.0$ \\
L10-109 & Field5 & 01:34:17.00 & +30:51:47.10 &            0.39 &       23 &              28 &     9 &          B & $\rm 127.26\pm 1.31$ &     $\rm 173.5\pm 8.4$ &   $\rm 192.1\pm 9.3$ &  $\rm 286.0\pm 12.6$ &  $\rm 18.8\pm 1.0$ &  $\rm 22.5\pm 1.1$ & $\rm 16.1\pm 0.9$ \\
L10-110 & Field3 & 01:34:17.03 & +30:33:57.70 &            0.55 &       22 &              23 &    12 &          B & $\rm 139.04\pm 2.99$ &    $\rm 387.9\pm 13.1$ &    $\rm 46.6\pm 3.4$ &  $\rm 286.0\pm 10.5$ &  $\rm 16.7\pm 2.1$ &  $\rm 31.1\pm 2.2$ & $\rm 24.1\pm 2.0$ \\
L10-111 & Field1 & 01:34:17.61 & +30:41:23.30 &            1.01 &       66 &              71 &    67 &          A &  $\rm 24.42\pm 0.58$ &    $\rm 424.4\pm 26.5$ &  $\rm 354.7\pm 21.7$ &  $\rm 280.8\pm 17.7$ &  $\rm 42.5\pm 2.5$ &  $\rm 58.5\pm 2.6$ & $\rm 42.5\pm 2.2$ \\
L10-112 & Field5 & 01:34:18.32 & +30:54:05.80 &            0.88 &       21 &              25 &    19 &          A &  $\rm 30.34\pm 1.33$ &  $\rm 1249.3\pm 215.0$ &  $\rm 233.1\pm 42.3$ &  $\rm 286.0\pm 49.3$ &  $\rm 35.2\pm 4.5$ &  $\rm 50.9\pm 4.9$ & $\rm 36.6\pm 4.2$ \\
L10-113 & Field3 & 01:34:19.28 & +30:33:45.90 &            0.48 &       25 &              23 &    12 &          A & $\rm 153.91\pm 3.27$ &    $\rm 448.6\pm 32.5$ &  $\rm 298.3\pm 22.4$ &  $\rm 286.0\pm 21.3$ &  $\rm 15.7\pm 1.9$ &  $\rm 27.1\pm 2.1$ & $\rm 21.2\pm 1.9$ \\
L10-114 & Field3 & 01:34:19.87 & +30:33:56.00 &            0.69 &       29 &              29 &    20 &          A &  $\rm 60.52\pm 1.30$ &    $\rm 504.9\pm 41.0$ &  $\rm 336.4\pm 28.1$ &  $\rm 286.0\pm 23.8$ &  $\rm 20.6\pm 2.1$ &  $\rm 38.6\pm 2.2$ & $\rm 30.5\pm 2.0$ \\
L10-116 & Field5 & 01:34:23.54 & +30:54:23.23 &            0.83 &       32 &              48 &    27 &          A &   $\rm 1.89\pm 0.15$ &   $\rm 872.2\pm 261.8$ & $\rm 489.8\pm 149.0$ &  $\rm 286.0\pm 88.0$ &  $\rm 35.3\pm 7.2$ & $\rm 52.6\pm 10.8$ & $\rm 30.5\pm 8.0$ \\
L10-118 & Field1 & 01:34:25.33 & +30:48:30.30 &            0.78 &       47 &              40 &    37 &          A &  $\rm 10.51\pm 0.56$ &    $\rm 255.8\pm 21.1$ &   $\rm 87.2\pm 10.0$ &  $\rm 281.7\pm 25.9$ &  $\rm 32.2\pm 6.2$ &  $\rm 46.2\pm 5.2$ & $\rm 32.1\pm 4.4$ \\
L10-119 & Field3 & 01:34:25.87 & +30:33:16.80 &            0.45 &        0 &               0 &     0 &          A &   $\rm 4.29\pm 0.16$ &    $\rm 327.5\pm 56.9$ &   $\rm 56.1\pm 15.1$ &  $\rm 286.0\pm 39.0$ &  $\rm 12.3\pm 3.6$ &  $\rm 28.2\pm 4.1$ & $\rm 16.5\pm 4.0$ \\
L10-120 & Field1 & 01:34:29.61 & +30:41:33.40 &            0.62 &        0 &               0 &     0 &          C &  $\rm 14.21\pm 0.18$ &    $\rm 334.7\pm 22.8$ &  $\rm 358.8\pm 24.8$ &  $\rm 248.1\pm 16.8$ &  $\rm 32.2\pm 1.3$ &  $\rm 38.1\pm 1.4$ & $\rm 24.0\pm 1.4$ \\
L10-121 & Field3 & 01:34:30.29 & +30:35:44.80 &            0.97 &       54 &              54 &    52 &          A &  $\rm 30.03\pm 1.15$ &    $\rm 992.3\pm 98.2$ &  $\rm 311.6\pm 31.1$ &  $\rm 286.0\pm 29.6$ &  $\rm 21.9\pm 3.0$ &  $\rm 55.4\pm 4.1$ & $\rm 41.3\pm 3.5$ \\
\hline\\[-2ex]
\end{tabular}
\begin{tablenotes}
       \item Continued.
\end{tablenotes}
\end{table}
\end{landscape}

\newpage
\clearpage

\begin{landscape}
\begin{table}
\setlength{\tabcolsep}{2pt}
\scriptsize
\centering
\begin{tabular}{l c c c c c c c c c c c c c c c}
\hline\hline  \\
(1) & (2) & (3) & (4) & (5) & (6) & (7) & (8) & (9) & (10) & (11) & (12) & (13) & (14) & (15) & (16) \\
   Name &  Field &          RA &          Dec & [SII]/H$\alpha$ & $\sigma$ & $\sigma$([SII]) & $\xi$ & Morphology &       F(H$\alpha$) &            [OII]3727 &         [OIII]5007 &            H$\alpha$ &        [NII]6584 &        [SII]6717 &       [SII]6717 \\
 ID & & & & & km\, s$^{-1}$ & km\, s$^{-1}$ & & &  10$^{-15}$ erg cm$^{-2}$  s$^{-1}$ &  & & & & &\\
\hline\\[-2ex]
L10-123 & Field3 & 01:34:32.34 & +30:35:32.60 &            1.03 &       47 &              51 &    48 &          A &  $\rm 17.03\pm 0.59$ &   $\rm 1134.5\pm 73.4$ &   $\rm 135.0\pm 9.5$ &  $\rm 286.0\pm 18.9$ &  $\rm 23.9\pm 3.5$ &  $\rm 56.1\pm 3.8$ & $\rm 46.6\pm 3.4$ \\
L10-124 & Field1 & 01:34:33.02 & +30:46:39.02 &            0.72 &       48 &              55 &    34 &         A2 &  $\rm 37.22\pm 0.70$ &     $\rm 192.9\pm 3.4$ &   $\rm 137.7\pm 2.8$ &   $\rm 284.5\pm 7.1$ &  $\rm 26.5\pm 1.9$ &  $\rm 40.5\pm 1.9$ & $\rm 31.4\pm 1.7$ \\
L10-126 & Field3 & 01:34:36.22 & +30:36:23.60 &            0.69 &        0 &               0 &     0 &          A &  $\rm 12.52\pm 0.30$ &    $\rm 407.2\pm 66.4$ &  $\rm 508.0\pm 81.2$ &  $\rm 286.0\pm 45.7$ &  $\rm 32.9\pm 2.4$ &  $\rm 39.6\pm 2.7$ & $\rm 28.9\pm 2.6$ \\
L10-127 & Field3 & 01:34:39.00 & +30:37:59.80 &            0.38 &        0 &               0 &     0 &          A &  $\rm 79.16\pm 1.39$ &    $\rm 236.2\pm 12.7$ &    $\rm 36.7\pm 3.9$ &  $\rm 286.0\pm 12.3$ &  $\rm 12.9\pm 1.7$ &  $\rm 22.6\pm 1.9$ & $\rm 15.7\pm 1.9$ \\
L10-128 & Field1 & 01:34:40.73 & +30:43:36.40 &            0.77 &       45 &              43 &    34 &          A &  $\rm 43.28\pm 0.72$ &     $\rm 261.7\pm 7.8$ &   $\rm 141.7\pm 5.0$ &   $\rm 250.0\pm 8.3$ &  $\rm 26.6\pm 1.6$ &  $\rm 45.8\pm 1.6$ & $\rm 30.8\pm 1.4$ \\
L10-129 & Field1 & 01:34:41.10 & +30:43:28.30 &            0.96 &       58 &              60 &    55 &          A &  $\rm 89.42\pm 1.84$ &    $\rm 337.7\pm 12.6$ &   $\rm 247.0\pm 9.8$ &  $\rm 257.6\pm 10.9$ &  $\rm 35.1\pm 2.1$ &  $\rm 56.0\pm 2.2$ & $\rm 39.6\pm 1.8$ \\
L10-130 & Field1 & 01:34:41.16 & +30:43:55.40 &            0.64 &       41 &              40 &    26 &          A &   $\rm 9.02\pm 0.38$ &    $\rm 357.2\pm 32.7$ &  $\rm 387.1\pm 35.5$ &  $\rm 262.2\pm 25.8$ &  $\rm 25.0\pm 4.1$ &  $\rm 39.3\pm 4.1$ & $\rm 24.8\pm 3.4$ \\
L10-131 & Field3 & 01:34:41.89 & +30:37:35.20 &            0.39 &        0 &               0 &     0 &          A & $\rm 145.33\pm 2.92$ &    $\rm 209.9\pm 12.6$ &    $\rm 45.5\pm 4.7$ &  $\rm 286.0\pm 13.5$ &  $\rm 13.3\pm 1.9$ &  $\rm 23.2\pm 2.2$ & $\rm 15.3\pm 2.1$ \\
L10-132 & Field1 & 01:34:44.62 & +30:42:38.80 &            0.31 &        6 &               2 &     2 &          C &  $\rm 34.96\pm 0.91$ &     $\rm 130.0\pm 5.1$ &    $\rm 10.1\pm 2.1$ &   $\rm 278.1\pm 9.2$ &  $\rm 19.2\pm 2.9$ &  $\rm 18.4\pm 2.3$ & $\rm 13.0\pm 2.1$ \\
L14-002 & Field7 & 01:32:27.85 & +30:35:44.60 &            0.62 &       38 &              45 &    24 &        -- &  $\rm 29.55\pm 1.34$ &    $\rm 558.0\pm 62.2$ &  $\rm 139.5\pm 14.6$ &  $\rm 286.0\pm 28.3$ &  $\rm 19.0\pm 5.3$ &  $\rm 37.2\pm 5.0$ & $\rm 25.0\pm 4.0$ \\
L14-004 & Field7 & 01:32:35.36 & +30:35:19.80 &            0.30 &        0 &               0 &     0 &        -- &  $\rm 45.61\pm 0.58$ &    $\rm 281.7\pm 16.9$ &    $\rm 30.2\pm 5.1$ &  $\rm 286.0\pm 12.6$ &   $\rm 8.8\pm 1.2$ &  $\rm 17.4\pm 1.3$ & $\rm 13.0\pm 1.3$ \\
L14-005 & Field8 & 01:32:37.16 & +30:17:54.30 &            0.41 &        0 &               0 &     0 &        -- &  $\rm 12.62\pm 0.35$ &    $\rm 505.6\pm 47.1$ &  $\rm 434.6\pm 35.0$ &  $\rm 266.9\pm 22.2$ &  $\rm 12.7\pm 2.7$ &  $\rm 27.3\pm 3.0$ & $\rm 14.2\pm 2.9$ \\
L14-009 & Field7 & 01:32:40.94 & +30:31:50.00 &            0.28 &       16 &              17 &     5 &        -- &  $\rm 42.35\pm 0.96$ &    $\rm 548.5\pm 58.6$ &    $\rm 57.9\pm 4.1$ &  $\rm 286.0\pm 11.9$ &  $\rm 15.3\pm 2.2$ &  $\rm 18.2\pm 2.2$ & $\rm 10.3\pm 1.8$ \\
L14-010 & Field7 & 01:32:42.71 & +30:36:18.71 &            0.42 &       18 &              15 &     8 &        -- &  $\rm 27.40\pm 0.56$ &    $\rm 646.8\pm 67.8$ &    $\rm 27.1\pm 4.6$ &  $\rm 286.0\pm 13.8$ &  $\rm 12.0\pm 1.7$ &  $\rm 25.3\pm 1.9$ & $\rm 16.6\pm 1.6$ \\
L14-012 & Field8 & 01:32:45.47 & +30:23:14.20 &            0.42 &       14 &              10 &     6 &        -- &   $\rm 9.46\pm 0.27$ &    $\rm 296.1\pm 48.9$ &                   -- &  $\rm 281.4\pm 16.7$ &  $\rm 19.6\pm 3.1$ &  $\rm 25.2\pm 2.7$ & $\rm 16.3\pm 2.3$ \\
L14-016 & Field7 & 01:32:52.80 & +30:31:34.20 &            0.47 &       19 &              18 &     9 &        -- &  $\rm 51.25\pm 0.76$ &    $\rm 356.6\pm 17.2$ &    $\rm 29.6\pm 3.2$ &   $\rm 286.0\pm 9.6$ &  $\rm 12.7\pm 1.4$ &  $\rm 27.5\pm 1.4$ & $\rm 19.4\pm 1.2$ \\
L14-020 & Field7 & 01:32:56.12 & +30:33:30.40 &            0.85 &       28 &              24 &    24 &        -- &  $\rm 35.73\pm 1.14$ &  $\rm 1082.3\pm 109.7$ &    $\rm 57.4\pm 5.5$ &  $\rm 286.0\pm 16.4$ &  $\rm 22.8\pm 3.1$ &  $\rm 52.0\pm 3.2$ & $\rm 32.6\pm 2.5$ \\
L14-023 & Field7 & 01:32:57.18 & +30:39:14.70 &            0.41 &       18 &              16 &     7 &        -- &  $\rm 19.82\pm 0.36$ &    $\rm 531.3\pm 42.0$ &    $\rm 10.5\pm 3.0$ &   $\rm 286.0\pm 9.9$ &  $\rm 10.0\pm 1.5$ &  $\rm 23.0\pm 1.7$ & $\rm 17.7\pm 1.5$ \\
L14-038 & Field4 & 01:33:13.46 & +30:28:13.10 &            0.47 &       21 &              22 &    10 &        -- &  $\rm 40.92\pm 0.80$ &    $\rm 216.9\pm 13.4$ &    $\rm 67.0\pm 6.8$ &  $\rm 219.7\pm 13.6$ &  $\rm 24.4\pm 1.9$ &  $\rm 28.7\pm 2.1$ & $\rm 18.2\pm 1.7$ \\
L14-039 & Field7 & 01:33:13.81 & +30:39:44.00 &            0.42 &       30 &              26 &    13 &        -- &  $\rm 54.38\pm 1.71$ &    $\rm 546.6\pm 84.9$ &    $\rm 54.0\pm 4.7$ &  $\rm 286.0\pm 14.9$ &  $\rm 17.5\pm 3.7$ &  $\rm 23.5\pm 2.7$ & $\rm 18.7\pm 2.5$ \\
L14-040 & Field7 & 01:33:15.35 & +30:35:36.70 &            0.39 &        0 &               0 &     0 &        -- &  $\rm 11.90\pm 0.35$ &   $\rm 513.5\pm 127.7$ &                   -- &  $\rm 286.0\pm 29.6$ &  $\rm 21.2\pm 2.9$ &  $\rm 23.2\pm 3.2$ & $\rm 15.6\pm 3.1$ \\
L14-043 & Field6 & 01:33:17.52 & +30:46:43.66 &            0.50 &       23 &              22 &    11 &        -- &  $\rm 99.08\pm 1.26$ &    $\rm 333.8\pm 25.6$ &  $\rm 165.0\pm 11.1$ &  $\rm 286.0\pm 16.8$ &  $\rm 23.1\pm 1.2$ &  $\rm 30.2\pm 1.2$ & $\rm 19.9\pm 1.0$ \\
L14-044 & Field4 & 01:33:18.13 & +30:33:38.60 &            0.32 &        0 &               0 &     0 &        -- &   $\rm 4.87\pm 0.12$ &    $\rm 108.3\pm 40.0$ &                   -- &  $\rm 286.0\pm 37.9$ &   $\rm 9.9\pm 2.3$ &  $\rm 17.6\pm 2.5$ & $\rm 14.8\pm 2.5$ \\
L14-048 & Field9 & 01:33:21.19 & +30:19:20.63 &            0.28 &        0 &               0 &     0 &        -- &  $\rm 13.06\pm 0.27$ &    $\rm 314.4\pm 43.4$ &                   -- &  $\rm 286.0\pm 23.8$ &  $\rm 12.2\pm 2.0$ &  $\rm 15.7\pm 2.2$ & $\rm 12.3\pm 2.2$ \\
L14-049 & Field4 & 01:33:21.33 & +30:30:31.60 &            0.40 &       19 &              19 &     8 &        -- &  $\rm 22.55\pm 0.33$ &    $\rm 209.8\pm 11.9$ &                   -- &  $\rm 261.6\pm 14.2$ &  $\rm 18.9\pm 1.4$ &  $\rm 23.1\pm 1.5$ & $\rm 17.4\pm 1.3$ \\
L14-054 & Field4 & 01:33:24.01 & +30:36:56.80 &            0.41 &        0 &               0 &     0 &        -- &  $\rm 15.98\pm 0.29$ &    $\rm 195.8\pm 20.5$ &                   -- &  $\rm 263.9\pm 24.2$ &  $\rm 25.0\pm 1.8$ &  $\rm 23.1\pm 2.0$ & $\rm 17.8\pm 2.0$ \\
L14-055 & Field4 & 01:33:24.34 & +30:28:50.41 &            0.91 &        0 &               0 &     0 &        -- &   $\rm 2.52\pm 0.19$ &    $\rm 294.0\pm 63.6$ &  $\rm 277.3\pm 61.8$ &  $\rm 160.0\pm 35.6$ &  $\rm 36.8\pm 7.6$ &  $\rm 56.2\pm 8.9$ & $\rm 35.1\pm 8.3$ \\
L14-058 & Field9 & 01:33:27.94 & +30:18:17.04 &            0.86 &       49 &              50 &    42 &        -- &   $\rm 9.06\pm 0.54$ &   $\rm 966.2\pm 122.0$ &  $\rm 207.8\pm 28.2$ &  $\rm 286.0\pm 39.0$ &  $\rm 19.5\pm 4.6$ &  $\rm 51.8\pm 6.4$ & $\rm 34.0\pm 4.9$ \\
L14-059 & Field9 & 01:33:27.97 & +30:15:59.49 &            0.37 &        0 &               0 &     0 &        -- &   $\rm 2.40\pm 0.10$ &                     -- &                   -- &  $\rm 286.0\pm 49.2$ &                 -- &  $\rm 21.2\pm 4.5$ & $\rm 15.5\pm 4.5$ \\
L14-065 & Field9 & 01:33:30.64 & +30:21:01.50 &            0.65 &       29 &              30 &    19 &        -- &  $\rm 36.72\pm 1.05$ &    $\rm 568.7\pm 39.5$ &                   -- &  $\rm 286.0\pm 18.9$ &  $\rm 18.3\pm 3.3$ &  $\rm 37.5\pm 2.9$ & $\rm 27.6\pm 2.5$ \\
L14-066 & Field9 & 01:33:31.20 & +30:21:14.26 &            0.67 &       23 &              24 &    15 &        -- &  $\rm 21.16\pm 0.66$ &    $\rm 403.0\pm 48.8$ &                   -- &  $\rm 286.0\pm 34.0$ &  $\rm 15.4\pm 2.7$ &  $\rm 38.5\pm 3.2$ & $\rm 28.7\pm 2.8$ \\
L14-070 & Field9 & 01:33:35.05 & +30:19:24.64 &            0.59 &       15 &              14 &     9 &        -- &  $\rm 13.99\pm 0.62$ &    $\rm 614.7\pm 73.5$ &   $\rm 26.2\pm 10.6$ &  $\rm 286.0\pm 31.7$ &  $\rm 28.3\pm 7.6$ &  $\rm 32.9\pm 4.3$ & $\rm 26.2\pm 3.9$ \\
L14-072 & Field4 & 01:33:34.99 & +30:29:54.60 &            0.95 &       52 &              56 &    50 &        -- &   $\rm 7.72\pm 0.13$ &    $\rm 386.4\pm 19.8$ &  $\rm 202.5\pm 10.6$ &  $\rm 259.4\pm 13.0$ &  $\rm 31.5\pm 1.7$ &  $\rm 55.0\pm 2.0$ & $\rm 40.3\pm 1.6$ \\
L14-075 & Field4 & 01:33:37.02 & +30:33:10.00 &            0.60 &       27 &              30 &    16 &        -- &  $\rm 63.01\pm 0.66$ &    $\rm 298.1\pm 10.6$ &    $\rm 95.3\pm 4.5$ &   $\rm 286.0\pm 9.8$ &  $\rm 26.8\pm 1.1$ &  $\rm 36.2\pm 1.2$ & $\rm 23.8\pm 1.0$ \\
L14-080 & Field4 & 01:33:39.84 & +30:34:30.54 &            0.90 &       19 &              26 &    17 &        -- &  $\rm 10.05\pm 0.17$ &    $\rm 443.8\pm 31.2$ &   $\rm 103.0\pm 8.3$ &  $\rm 286.0\pm 17.3$ &  $\rm 39.2\pm 1.9$ &  $\rm 53.6\pm 2.0$ & $\rm 36.3\pm 1.7$ \\
L14-092 & Field9 & 01:33:47.51 & +30:17:13.77 &            0.42 &       13 &              11 &     6 &        -- &  $\rm 12.94\pm 0.39$ &   $\rm 792.1\pm 105.6$ &  $\rm 341.7\pm 45.3$ &  $\rm 286.0\pm 37.6$ &  $\rm 20.6\pm 3.2$ &  $\rm 23.3\pm 2.7$ & $\rm 19.2\pm 2.5$ \\
L14-093 & Field9 & 01:33:47.79 & +30:18:05.17 &            0.53 &       26 &              24 &    14 &        -- &  $\rm 34.81\pm 0.67$ &    $\rm 490.4\pm 20.9$ &                   -- &  $\rm 286.0\pm 10.2$ &  $\rm 13.1\pm 1.6$ &  $\rm 32.2\pm 1.8$ & $\rm 21.2\pm 1.5$ \\
L14-103 & Field4 & 01:33:52.56 & +30:28:38.40 &            0.40 &        0 &               0 &     0 &        -- &   $\rm 6.03\pm 0.19$ &    $\rm 109.5\pm 19.4$ &                   -- &  $\rm 230.2\pm 32.7$ &  $\rm 24.1\pm 3.1$ &  $\rm 25.3\pm 3.3$ & $\rm 15.2\pm 3.3$ \\
L14-106 & Field9 & 01:33:54.63 & +30:18:51.10 &            0.66 &       64 &              81 &    43 &        -- &  $\rm 17.60\pm 1.30$ &  $\rm 1160.6\pm 178.0$ &  $\rm 330.5\pm 43.6$ &  $\rm 286.0\pm 41.6$ &                 -- &  $\rm 43.9\pm 8.6$ & $\rm 22.6\pm 6.0$ \\
L14-122 & Field2 & 01:34:00.25 & +30:39:28.90 &            0.77 &        0 &               0 &     0 &        -- &  $\rm 34.01\pm 1.01$ &  $\rm 1636.8\pm 265.7$ &  $\rm 134.8\pm 26.5$ &  $\rm 286.0\pm 45.8$ &  $\rm 43.4\pm 3.1$ &  $\rm 44.1\pm 3.4$ & $\rm 32.9\pm 3.3$ \\
L14-125 & Field5 & 01:34:00.52 & +30:50:43.09 &            0.80 &       32 &              38 &    26 &        -- &   $\rm 3.94\pm 0.32$ &    $\rm 239.8\pm 52.7$ &  $\rm 156.0\pm 33.3$ &  $\rm 224.4\pm 47.5$ &  $\rm 40.2\pm 9.8$ &  $\rm 47.4\pm 9.5$ & $\rm 32.7\pm 7.8$ \\
L14-133 & Field5 & 01:34:04.83 & +30:58:32.35 &            0.29 &       14 &              15 &     4 &        -- &  $\rm 36.52\pm 0.91$ &    $\rm 254.6\pm 31.1$ &                   -- &  $\rm 286.0\pm 16.9$ &  $\rm 15.0\pm 3.7$ &  $\rm 16.5\pm 2.4$ & $\rm 12.5\pm 2.2$ \\
L14-148 & Field3 & 01:34:13.85 & +30:30:39.80 &            0.59 &       16 &              32 &     9 &        -- &   $\rm 3.34\pm 0.13$ &  $\rm 1164.4\pm 200.4$ &  $\rm 528.2\pm 80.6$ &  $\rm 286.0\pm 44.3$ &  $\rm 26.6\pm 4.0$ &  $\rm 30.4\pm 4.4$ & $\rm 28.2\pm 4.3$ \\
L14-153 & Field1 & 01:34:14.55 & +30:44:36.20 &            0.68 &        0 &               0 &     0 &        -- &  $\rm 14.32\pm 0.58$ &    $\rm 578.4\pm 57.4$ &  $\rm 120.4\pm 14.5$ &  $\rm 286.0\pm 28.8$ &  $\rm 32.1\pm 4.1$ &  $\rm 41.9\pm 4.6$ & $\rm 26.4\pm 4.4$ \\
L14-162 & Field5 & 01:34:19.45 & +30:52:48.90 &            0.57 &       24 &              30 &    14 &        -- &  $\rm 39.41\pm 1.27$ &    $\rm 243.5\pm 46.4$ &                   -- &  $\rm 286.0\pm 35.3$ &  $\rm 18.9\pm 3.6$ &  $\rm 34.4\pm 3.6$ & $\rm 23.1\pm 3.0$ \\
L14-163 & Field3 & 01:34:19.68 & +30:33:41.50 &            0.38 &       24 &              20 &     9 &        -- & $\rm 105.05\pm 2.16$ &    $\rm 294.8\pm 31.5$ &  $\rm 425.9\pm 46.2$ &  $\rm 286.0\pm 30.8$ &  $\rm 12.1\pm 1.8$ &  $\rm 21.2\pm 2.0$ & $\rm 16.6\pm 1.8$ \\
L14-167 & Field3 & 01:34:24.08 & +30:33:24.40 &            0.48 &        0 &               0 &     0 &        -- &  $\rm 32.59\pm 0.53$ &    $\rm 328.1\pm 21.7$ &  $\rm 194.6\pm 13.8$ &  $\rm 286.0\pm 18.5$ &  $\rm 17.5\pm 1.6$ &  $\rm 26.5\pm 1.8$ & $\rm 21.0\pm 1.7$ \\
L14-168 & Field1 & 01:34:24.48 & +30:48:58.30 &            0.41 &        0 &               0 &     0 &        -- &  $\rm 33.21\pm 0.61$ &     $\rm 109.4\pm 7.3$ &                   -- &  $\rm 286.0\pm 11.2$ &  $\rm 18.3\pm 1.8$ &  $\rm 24.6\pm 2.0$ & $\rm 16.6\pm 2.0$ \\
L14-173 & Field3 & 01:34:29.77 & +30:35:08.40 &            0.53 &       16 &              22 &     9 &        -- &  $\rm 29.22\pm 1.57$ &    $\rm 513.5\pm 74.0$ &    $\rm 26.9\pm 9.5$ &  $\rm 286.0\pm 30.3$ &  $\rm 14.1\pm 5.0$ &  $\rm 27.5\pm 5.4$ & $\rm 25.9\pm 5.3$ \\
L14-180 & Field1 & 01:34:37.40 & +30:44:11.00 &            0.41 &        0 &               0 &     0 &        -- &  $\rm 46.33\pm 0.79$ &     $\rm 132.1\pm 5.6$ &    $\rm 17.0\pm 3.0$ &   $\rm 286.0\pm 9.9$ &  $\rm 19.3\pm 1.7$ &  $\rm 24.7\pm 1.8$ & $\rm 16.4\pm 1.8$ \\
L14-182 & Field3 & 01:34:39.69 & +30:39:17.60 &            0.60 &       34 &              35 &    21 &        -- &  $\rm 20.38\pm 0.71$ &    $\rm 448.8\pm 46.5$ &                   -- &  $\rm 286.0\pm 25.4$ &   $\rm 7.9\pm 2.8$ &  $\rm 34.6\pm 3.6$ & $\rm 25.1\pm 3.1$ \\
L14-187 & Field1 & 01:34:42.68 & +30:40:51.50 &            0.48 &        0 &               0 &     0 &        -- &   $\rm 6.65\pm 0.24$ &    $\rm 222.7\pm 30.1$ &                   -- &  $\rm 286.0\pm 28.5$ &  $\rm 18.7\pm 3.6$ &  $\rm 24.6\pm 3.9$ & $\rm 23.1\pm 3.9$ \\
L14-189 & Field3 & 01:34:45.40 & +30:35:35.20 &            0.89 &       52 &              56 &    47 &        -- &  $\rm 10.18\pm 0.72$ &  $\rm 1265.3\pm 181.8$ &  $\rm 331.9\pm 46.8$ &  $\rm 286.0\pm 43.6$ &  $\rm 25.2\pm 7.8$ &  $\rm 54.3\pm 8.0$ & $\rm 34.9\pm 6.3$ \\
L14-191 & Field3 & 01:34:47.24 & +30:34:25.00 &            0.42 &        3 &              18 &     1 &        -- &   $\rm 8.30\pm 0.39$ &    $\rm 301.9\pm 48.9$ &                   -- &  $\rm 286.0\pm 25.4$ &  $\rm 16.8\pm 4.8$ &  $\rm 23.9\pm 5.0$ & $\rm 18.4\pm 4.5$ \\
\hline
\end{tabular}
\begin{tablenotes}
       \item ``$^\dagger$'' Line ratios of the regions where H$\beta$ is not measured; these line ratios are computed by estimating an H$\beta$ flux \citep[3$\sigma$, see][]{2021A&A...645A..57D} value and are not used in our analysis.
\end{tablenotes}
\end{table}
\end{landscape}

\begin{table*}
\begin{minipage}[t]{.48\linewidth}\vspace{0cm}
\setlength{\tabcolsep}{2pt}
\caption{Impact of the threshold $\xi$: Fraction of SNRs in the L10 and L14 samples which do not pass the given threshold.}
\centering
\begin{threeparttable}
        \label{tab:xivar}
        \begin{tabular}{ccccc}
                \hline
                 Sample & $\xi \leq 12$ & $\xi \leq 10$ & $\xi \leq 8$ & $\xi \leq 6$ \\
                \hline
                L10 (118 SNRs) & 18\% & 14\% & 10\% & 8\% \\
                L14 (45 SNRs) & 67\% & 62\% & 47\% & 40\% \\            
                \hline
        \end{tabular}
\end{threeparttable}
\end{minipage}
\end{table*}

\begin{figure}
	\includegraphics[width=0.8\columnwidth]{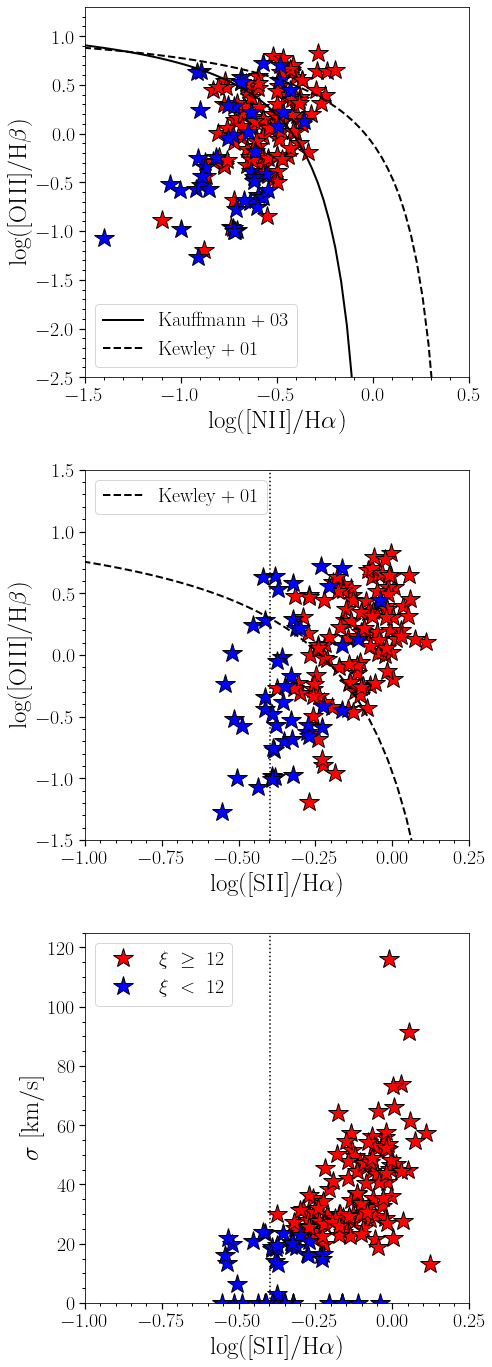}
    \caption{Diagnostic diagrams (upper and center panel) and scheme of identification (lower panel) for the sample of SNR candidates in M33. The red and blue stars correspond to those candidates that do and do not satisfy the condition $\xi\ \geq$12. The dashed, continuous, and vertical dotted lines show the \citet{2001ApJ...556..121K} curve, the \citet{2003MNRAS.346.1055K} demarcation, and the limit value of [SII]/H$\alpha$=0.4, respectively.}
    \label{fig:diagramaas}
\end{figure}

Finally, in Fig.~\ref{fig:diagramaas} three diagnostic plots are shown: [NII]/H$\alpha$ vs. [OIII]/H$\beta$ (BPTO3N2), [SII]/H$\alpha$ vs. [OIII]/H$\beta$ (BPTO3S2), and $\sigma$ vs. [SII]/H$\alpha$. In each of them, the SNR candidates have been colour coded by their $\xi$ value; red stars for candidates with $\xi\ \geq$ 12 and blue stars for those with $\xi\ <$ 12. As in Fig.~\ref{fig:N6822bpt}, in the case of the BPTO3N2 and BPTO3S2 diagrams there is a substantial population of candidates with low values of $\sigma$, which might be considered as older SNRs, for which the expansion has been considerably slowed down. When considering the [SII]/H$\alpha$ vs. $\sigma$ diagram it can be seen that the candidates with $\xi\ <$ 12 populate the lower zone of this diagram and those with higher values than 12 populate the upper part, as expected from Fig.~\ref{fig:N6822sigma} for Ho 12 and all HII regions in NGC6822. 

\section{Conclusions}
\label{sec:5_conclu}
Based on spectral datacubes obtained with SITELLE, we presented a detailed study of the optical properties of a large sample (164) of known supernova remnants in NGC 6822 and M33, with a particular emphasis on the measurement of their integrated properties including their velocity dispersion, $\sigma$. Our main conclusions are the following:

\begin{itemize}
\item  Ho 12 in NGC 6822 presents an intriguing spiral shape composed of multiple knots mixed with a more diffuse structure. Although its global morphology is similar at all wavelengths, we find noticeable differences in the line ratios, radial velocities and velocity dispersion from one knot to the next. We also note clear radial trends the properties of Ho12 as well as those of the compact and shell-shaped SNRs in M33 which set them apart from those of HII regions.

\item We introduce a diagram, $\sigma$ vs. [SII]/H$\alpha$, that can be used on a pixel-by-pixel basis for individual objects or to assess the integrated properties of a population of emission-line regions in a given galaxy. The shape of the pixel-to-pixel diagrams depends on the morphological type of the supernova remnants. 

\item M33 SNRs showing A, A2, or A3 morphological types present higher values of $\sigma$ in the central parts, decreasing as we move towards the outer parts. In the case of type B and C SNRs this dependence is less evident. Also, the scatter of [SII]/H$\alpha$ at a constant $\sigma$ value is lower in the types A2 and A3 than in the rest of the morphological types defined by \cite{2010ApJS..187..495L}. 

\item Given that (a) the [SII]/H$\alpha$ ratio and the velocity dispersion are generally spatially anti-correlated in HII regions; (b) the diffuse ionized gas shows high values of [SII]/H$\alpha$ but low values of velocity dispersion, and (c) supernova remnants are expected to show higher values of both [SII]/H$\alpha$ and velocity dispersion than HII regions, we defined a parameter that combines both properties, $\xi$=([SII]/H$\alpha$)$\times\ \sigma$. This parameter can be determined on a pixel-by-pixel basis in order to facilitate the detection of supernova remnants using maps extracted from datacubes, or as a parameter obtained from the integrated spectrum of emission-line regions.

\item While a threshold of $\xi = 12$ clearly separates the sole SNR in NGC 6822 from the large HII regions in the same galaxy, the situation is not as clear-cut in M33, which presents a wide variety of SNR morphologies and ISM structure. While neither the L10 SNR sample nor the new SNRs found by L14 were assembled using a measure of the velocity dispersion, we note a significant discrepancy between them: 82\% of the L10 SNRs satisfy the $\xi = 12$ threshold (and 90\% with $\xi = 8$), but only a third of the additional L14 SNRs satisfy the first threshold and half the second. These might be older objects for which the expansion has been significantly slowed down.

\item As expected, the detection rate of a significant velocity dispersion ($\sigma$ > 0) in known M33 SNRs is higher (92\%, or 60/65) in our SN3 datacubes  with the canonical SIGNALS spectral resolution R = 5000 (Fields 5 to 9) than it is (82\%, or 70/98) in the datacubes obtained during the proof of concept program (Fields 1 to 4, with R = 2200 - 2900).

\item We will make use of the results presented in this paper to study the SNR population in more distant galaxies of the SIGNALS survey, such as NGC 4214 \citep{2023MNRAS.524.3623V}.

\end{itemize}

\section*{Acknowledgements}
We are grateful to the anonymous referee for very constructive suggestions that have helped us to improve this manuscript. Based on observations obtained with SITELLE, a joint project of Universit\'e Laval, ABB, Universit\'e de Montr\'eal, and the Canada-France-Hawaii Telescope (CFHT) which is operated by the National Research Council of Canada, the Institut National des Sciences de l'Univers of the Centre National de la Recherche Scientifique of France, and the University of Hawaii. The authors wish to recognize and acknowledge the very significant cultural role that the summit of Mauna Kea has always had within the indigenous Hawaiian community. We are most grateful to have the opportunity to conduct observations from this mountain. 

SDP acknowledges financial support from Juan de la Cierva Formaci\'on fellowship (FJC2021-047523-I) financed by MCIN/AEI/10.13039/501100011033 and by the European Union `NextGenerationEU'/PRTR, Ministerio de Econom\'ia y Competitividad under grants PID2019-107408GB-C44, PID2020-113689GB-I00, and PID2020-114414GB-I00, PID2022-136598NB-C32, and from Junta de Andalucía Excellence Project P18-FR-2664 and FQM108.
SDP, LD, and CR are grateful to the Natural Sciences and Engineering Research Council of Canada, the Fonds de Recherche du Qu\'ebec, and the Canada Foundation for Innovation for funding. 
LRN is grateful to the National Science foundation NSF - 2109124 and the Natural Sciences and Engineering Research Council of Canada NSERC - RGPIN-2023-03487 for their support. This research made use of astropy, a community-developed core python \citep[http://www.python.org,][]{Van_Rossum_python} package for Astronomy \citep{2013A&A...558A..33A,2018AJ....156..123A,2022ApJ...935..167A}; IPython \citep{PER-GRA:2007}; Matplotlib \citep{Hunter:2007}; NumPy \citep{2011CSE....13b..22V}; SciPy \citep{2020SciPy-NMeth,scipy_11255513}; and Pandas \citep{mckinneyprocscipy2010}.

\section*{Data Availability}

SIGNALS data are part of the public domain already. The measurements underlying this article can be shared upon request to the first author.

\bibliographystyle{mnras}
\bibliography{biblio} 

\bsp	


\section*{SUPPORTING INFORMATION}

Supplementary data are available at MNRAS online. \\

\noindent Please note: Oxford University Press is not responsible for the content or functionality of any supporting materials supplied by the authors. Any queries (other than missing material) should be directed to the corresponding author for the article.



\bsp	
\label{lastpage}
\end{document}